\documentclass[a4paper,fleqn,usenatbib]{mnras}
\usepackage{newtxtext,newtxmath}

\usepackage[T1]{fontenc}
\usepackage{ae,aecompl}
\usepackage[czech]{babel} 
\usepackage[utf8]{inputenc}

\usepackage{graphicx}	
\usepackage{amsmath}	
\usepackage{amssymb}	

\newcommand{\Msun}{{\mathrm M}_{\odot}}
\newcommand{\NH}{N_{\mathrm{H}}}
\newcommand{\NJ}{N_{\mathrm{J}}}
\newcommand{\nH}{n_{\mathrm{H}}}

\newcommand{\nHtwo}{n_{\mathrm{H2}}}
\newcommand{\NHtwo}{N_{\mathrm{H2}}}

\newcommand{\scm}{\,\mathrm{cm}^{-2}}
\newcommand{\ccm}{\,\mathrm{cm}^{-3}}
\newcommand{\Zsol}{{\mathrm Z}_{\odot} }

\newcommand{\K}{\,\mathrm{K}}

\newcommand{\NS}{N_{\mathrm{sh}}}

\defcitealias{haardt_radiative_2012}{HM12}
\defcitealias{wiersma_effect_2009}{W09}
\defcitealias{rahmati_evolution_2013}{R13}
\defcitealias{wolfire_neutral_1995}{W95}
\defcitealias{faucher-giguere_cosmic_2020}{FG20}

\usepackage{multirow}
\usepackage{array}

\newcolumntype{L}[1]{>{\raggedright\let\newline\\\arraybackslash\hspace{0pt}}p{#1}}
\newcolumntype{C}[1]{>{\centering\let\newline\\\arraybackslash\hspace{0pt}}p{#1}}
\newcolumntype{R}[1]{>{\raggedleft\let\newline\\\arraybackslash\hspace{0pt}}p{#1}}

\title[Cooling rates, species abundances and emissivities]{Radiative cooling rates, ion fractions, molecule abundances and line emissivities including self-shielding and both local and metagalactic radiation fields}

\author[S. Ploeckinger \& J. Schaye]{
Sylvia Ploeckinger,$^{1,2,3}$\thanks{E-mail: ploeckinger@lorentz.leidenuniv.nl}
Joop Schaye$^{1}$
\\
$^{1}$Leiden Observatory, Leiden University, PO Box 9513, NL-2300 RA Leiden, the Netherlands\\
$^{2}$Lorentz Institute for theoretical physics, Leiden University, PO Box 9506, NL-2300 RA Leiden, the Netherlands\\
$^{3}$Institute for Computational Cosmology, Durham University, South Road, Durham DH1 3LE, UK\\
}

\date{Accepted XXX. Received YYY; in original form ZZZ}

\pubyear{2019}

\begin{document}
\label{firstpage}
\pagerange{\pageref{firstpage}--\pageref{lastpage}}
\maketitle

\begin{abstract}
We use the spectral synthesis code \textsc{cloudy} to tabulate the properties of gas for an extensive range in redshift ($z=0$ \-- $9$), temperature ($\log T [K] = 1$ \-- $9.5$), metallicity ($\log Z/\mathrm{Z}_{\odot} = -4$ \-- $+0.5$, $Z=0$), and density ($\log \nH [\ccm] = -8$ \-- $+6$). This therefore includes gas with properties characteristic of the interstellar, circumgalactic and intergalactic media.
The gas is exposed to a redshift-dependent UV/X-ray background, while for the self-shielded lower-temperature gas (i.e. ISM gas) an interstellar radiation field and cosmic rays are added. The radiation field is attenuated by a density- and temperature-dependent column of gas and dust. Motivated by the observed star formation law, this gas column density also determines the intensity of the interstellar radiation field and the cosmic ray density. The ionization balance, molecule fractions, cooling rates, line emissivities, and equilibrium temperatures are calculated self-consistently. We include dust, cosmic rays, and the interstellar radiation field step-by-step to study their relative impact.
These publicly available tables are ideal for hydrodynamical simulations. They can be used stand alone or coupled to a non-equilibrium network for a subset of elements. The release includes a C routine to read in and interpolate the tables, as well as an easy to use python graphical user interface to explore the tables. 
\end{abstract}

\begin{keywords}
radiative transfer -- ISM: general --  intergalactic medium -- galaxies: ISM
\end{keywords}



\section{Introduction}\label{sec:intro}
Radiative processes are a critical ingredient for all models that include baryons. Radiative losses are not only crucial for the formation of stars and therefore for the ignition of the cosmic baryon cycle, they are also heavily used in observational astronomy to classify the properties of gas within galaxies (interstellar medium, ISM), around galaxies (circumgalactic medium, CGM) and in between galaxies (intergalactic medium, IGM). The photons that are emitted at specific energies are important tracers for the composition, physical properties (e.g. density, temperature), and the movement of the gas (e.g. turbulence, inflow, outflows). 

Determining the radiative cooling rate of a parcel of gas relies on a series of intertwined chemical reactions of species that can also interact with the ambient radiation field. A full implementation of these processes in a simulation would therefore require solving a large chemical network together with a coupled radiative transfer code for each resolution element at each timestep. While this can currently be done for a limited set of species in small-scale simulations of galaxy formation and evolution (e.g. isolated galaxies, or patches of the ISM), it is computationally too expensive to do in a large cosmological volume.  A popular compromise to include radiative processes in simulations is to tabulate the cooling (and heating) rates. Therefore, the way these tabulated rates are calculated can be arbitrarily complicated without affecting the runtime of the simulation. 

Traditionally, radiative cooling functions were calculated under the assumption of an optically thin gas in collisional ionization equilibrium (CIE,  e.g. 
\citealt{cox_ionization_1969, dalgarno_heating_1972, raymond_radiative_1976, shull_ionization_1982, gaetz_line_1983, boehringer_metallicity-dependence_1989, sutherland_cooling_1993}). The ionization rates in CIE (and therefore the recombination rates and the radiative losses, $\Lambda$) are proportional to the number of collisions between particles (electrons and atoms). As hydrogen is the most abundant element in the Universe, the number density of particles scales with the hydrogen number density $\nH$ and the normalised cooling rate $\Lambda / \nH^2 \,[\mathrm{erg}\,\mathrm{s}^{-1}\,\mathrm{cm}^3]$ is only a function of temperature (case I in Table~\ref{tab:dims}). In this case, the contribution of individual elements scales linearly with their element abundance and the cooling rate for each chemical element can either be provided in terms of tables (e.g. \citealt{boehringer_metallicity-dependence_1989}) or fitting functions (e.g. \citealt{dalgarno_heating_1972}).

When including photo-ionisation, the gas is over-ionised compared to the CIE case. This leads to different cooling rates caused by the different ion fractions of each individual element \citep[e.g.][]{efstathiou_suppressing_1992, wiersma_effect_2009}. As the photo-ionization rate scales with the number of particles to ionise (again approximated by $\nH$), while collisional processes scale with $\nH^2$, the normalised cooling rate $\Lambda / \nH^2 \,[\mathrm{erg}\,\mathrm{s}^{-1}\,\mathrm{cm}^3]$ that combines both types of processes is a function of both density and temperature (case II in Table~\ref{tab:dims}).

As the radiation field is not constant throughout the Universe, additional dependences are introduced.  
In galaxy-scale or cosmological simulations, a redshift-dependent uniform UV background (UVB) is often used as radiation field, which adds a redshift dependence to the tabulated cooling rates (case IIa in Table~\ref{tab:dims}, e.g. \citealt{wiersma_effect_2009}, hereafter: \citetalias{wiersma_effect_2009}). \citetalias{wiersma_effect_2009} showed that at the peak of the cooling function, between temperatures of $10^4$ and $10^6\K$, the cooling rate is reduced by up to an order of magnitude compared to the CIE case when photoionization from the UVB is included.

\citet{gnedin_cooling_2012} tabulate the properties of photo-ionized gas for general radiation fields from stars or active galactic nuclei (AGN). They parametrize their incident spectrum with 4 independent values, one for the intensity and three that define its shape (case IIb in Table~\ref{tab:dims}), resulting in an equal number of additional dimensions (7 in total, including density, temperature, and metallicity) for their tabulated cooling rates of optically thin gas in ionization equilibrium. 

At temperatures between $10^4$ and $\approx 5 \times 10^6\,\K$ the cooling timescale can become shorter than the recombination timescale, leading to a ``recombination lag". This rapidly cooling gas is therefore over-ionized compared to ionization equilibrium (e.g.~\citealp{kafatos_time-dependent_1973}). \citet{gnat_time-dependent_2007} provide tables of radiative cooling rates including this non-equilibrium effect. Their calculations start with initially hot ($T > 5 \times 10^{6}\,\K$) gas that cools either at constant density or constant pressure in the absence of a radiation field. The resulting cooling functions are smoother with a less pronounced peak due to hydrogen recombination compared to the equilibrium rates. As metal line cooling is important, their tabulated cooling rates also depend on the gas metallicity (case N-I, Table~\ref{tab:dims}). 

\begin{table} 
	\caption{In collisional ionization equilibrium (CIE), the cooling rate $\Lambda$ scales with the hydrogen number density, $\nH^2$, and the term $\Lambda / \nH^2$ therefore only depends on the temperature $T$. Including extra processes (rows) induces additional dependences on $\nH$, redshift $z$, hydrogen column density $\NH$, gas metallicity $Z$, as well as the intensity, $I$, and shape of the radiation field (RF). }
	\begin{tabular}{lll}
		Case &								& $\Lambda / \nH^2$ dependence \\
		\hline
		\multicolumn{3}{c}{Ionization equilibrium}\\
		\hline
		I	&CIE 							& $T$\\
		II	&Photo-ionization 					& $T, \nH$\\
		IIa	&Photo-ionization (UVB) 				& $T, \nH, z$\\
		IIb	&Photo-ionization (general RF) 			& $T, \nH, Z, I, \mathrm{shape}$\\
		III	&UVB, self-shielded gas 				& $T, \nH, z, Z, \NH$ \\
		IV 	&UVB, ISRF, self-shielded gas 			& $T, \nH, z, Z, \NH, I, \mathrm{shape}$ \\
			&								&\\
		\multicolumn{3}{l}{this work:}\\		
		Va	&UVB, unshielded gas					& $T, \nH, z, Z$ \\
		Vb	&UVB, self-shielded gas					& $T, \nH, z, Z$ \\
		Vc	&UVB, ISRF, unshielded gas				& $T, \nH, z, Z$ \\
		Vd	&UVB, ISRF, self-shielded gas			& $T, \nH, z, Z$ \\
		\hline
		\multicolumn{3}{c}{Non-equilibrium}\\
		\hline
		N-I	&CIE								& $T, Z$ \\
		N-IIa	&Photo-ionization (UVB) 				& $T, \nH, z, Z$\\
		\hline
	\end{tabular}\label{tab:dims}
\end{table}

Similar to the approach from  \citet{gnat_time-dependent_2007}, \citet{oppenheimer_non-equilibirum_2013} use their non-equilibrium reaction network to provide tabulated cooling and photo-heating rates by following hot gas as it cools either at constant pressure or constant density. In contrast to \citet{gnat_time-dependent_2007},   \citet{oppenheimer_non-equilibirum_2013} include a redshift-dependent background radiation field and they tabulate their cooling rates for each ion of each considered element. As a hydrodynamic code typically does not trace the individual ion abundances, they also provide tabulated ion fractions of 11 elements, depending on the hydrogen number density, temperature, and metallicity of the gas, as well as the redshift for the background radiation field (case N-IIa in Table~\ref{tab:dims}). \citet{oppenheimer_non-equilibirum_2013}  conclude that non-equilibrium effects are reduced when the background radiation field is included. 

All above approaches focus on ionized gas with temperatures above $10^4\,\K$, as is typical for the CGM and IGM. At these temperatures, molecules cannot form, dust grains are quickly destroyed by thermal sputtering (e.g. \citealt{tielens_physics_1994}) and gas is optically thin to ionizing radiation. In the ISM, the incident radiation is however attenuated by gas and dust and can self-shield from photo-ionizing or photo-dissociating radiation. As the ion fractions depend on the photo-ionization rate, the cooling rate varies with the hydrogen column density $\NH$ (case III in Table~\ref{tab:dims}). This can lead to structures where the illuminated side of a gas parcel is ionized, but deeper into the cloud, the gas can remain neutral and cold.

In neutral gas, the formation of molecules and the existence of dust grains further complicates the picture. Dust grains and molecules shield different parts of the intrinsic radiation field and the reaction rates of various molecules have a non-trivial dependence on the shape of the spectrum. In addition, an important formation channel of $\mathrm{H}_{\mathrm{2}}$ depends on dust grains as catalysts, which results in a non-trivial relation between the gas metallicity and the $\mathrm{H}_{\mathrm{2}}$ abundance. 

The ISM gas resolution elements in hydrodynamic simulations can have a large variety of thermal and chemical histories: from gas that cools through thermal instabilities and is accreted from the galaxy halo to dense gas that is heated by nearby stellar feedback events. Tabulating non-equilibrium cooling rates and other chemical properties in a similar way (i.e. cooling from high temperatures) is therefore not a good approach. 
Chemical networks that solve a large number of differential equations to follow the thermo-chemical evolution of the gas particles are required to include non-equilibrium effects in the ISM. 

\textsc{grackle} \citep{smith_grackle:_2017} is an open-source chemistry and cooling library that allows users to include hydrogen and helium non-equilibrium chemistry directly in the simulation. The primordial network includes the self-shielding approximation from \citet{rahmati_evolution_2013} (hereafter, \citetalias{rahmati_evolution_2013}). The cooling and heating rates from metals are calculated with \textsc{cloudy} under the assumption of ionization equilibrium, and tabulated for the UV background from \cite{haardt_radiative_2012}. In the 2017 \textsc{grackle} release, the metal cooling and heating rates assume that the gas is optically thin to the UV background and solar abundances. Combining the primordial network with the tabulated metal rates leads to inconsistencies in their electron fractions if self-shielding is included. The free electron fraction from hydrogen and helium used in the primordial network that includes self-shielding can be much lower than the free electron fraction from hydrogen and helium used in the optically thin \textsc{cloudy} tables. The metal cooling rates are therefore based on gas that is too highly ionized, leading to too high cooling rates for self-shielded gas. This has been discussed in \citet{hu_variable_2017} and \citet{emerick_simulating_2019} produced metal cooling tables for \textsc{grackle} that include self-shielding of the UV background by hydrogen and helium, while shielding on metals and dust grains is neglected. Furthermore, as dust grains are not included in the \textsc{cloudy} calculations, \textsc{grackle}'s shielding column lacks H$_{\mathrm{2}}$. The rates are calculated using a shielding length based on the local Jeans length (limited to a physical size of 100~pc) and for a single metallicity with solar relative abundances. 

\citet{glover_modelling_2010} present an ISM model for CO formation in turbulent molecular clouds within the ISM, including, in addition to hydrogen and helium, both carbon and oxygen chemistry. 
Reduced chemical networks have also been used to study the dependence of the abundance of molecules, such as $\mathrm{H}_{\mathrm{2}}$ and CO, on the cosmic ray rate \citep{bisbas_effective_2015, bisbas_cosmic-ray_2017} or (in addition) on metallicity and the radiation field intensity (e.g.~\citealp{bialy_coh2_2015, bisbas_simulating_2019}). An extended chemical network with 272 chemical reactions that follow the evolution of 157 species (among these: 20 different molecules) is described in \citet{richings_non-equilibrium_2014-1} for optically thin gas, and \citet{richings_non-equilibrium_2014} for shielded gas. The effects of self-shielding compared to the assumption of optically thin gas, as well as the impact of the radiation field strength and gas metallicity on molecule abundances, are demonstrated in simulations of isolated galaxies by \citet{richings_effects_2016}. These large chemical networks that also calculate metals in non-equilibrium are computationally expensive, as the number of reactions increases rapidly with every included species.

The aim of this work is to provide tables of gas properties in ionization equilibrium that include radiation by a UV background and, optionally, from an interstellar radiation field (ISRF) for both unshielded (or optically thin to the radiation field) and self-shielded gas. 
The large number of dimensions would make the tables expensive to produce and would also lead to large memory requirements during any simulation run using the tables. More importantly, neither the shielding gas column density nor the local radiation field is generally known on the fly in large-scale simulations. 

We solve this issue by assuming that the gas is self-gravitating and hence that its coherence scale is the local Jeans length \citep{schaye_model-independent_2001,schaye_physical_2001}. Note that \citet{rahmati_evolution_2013} showed that this assumption reproduces the shielding lengths in their cosmological, radiative transfer simulations. By setting the shielding column density to half of the local Jeans column density we remove the $\NH$ dimension. 
In addition, we assume that the incident radiation field, which could be described by an arbitrary number of parameters, comes in two flavours: (1) the redshift-dependent model of the UVB by \citet{faucher-giguere_cosmic_2020} (hereafter: \citetalias{faucher-giguere_cosmic_2020}), modified to make the treatment of attenuation before \ion{H}{I} and \ion{He}{II} reionization more self-consistent, and (2) optionally, a local interstellar radiation field (ISRF) with a constant spectral shape and a normalization that depends on the local Jeans column density of the gas as suggested by the observed Kennicutt-Schmidt \citep{kennicutt_global_1998} star formation law (case Vc,d in Table~\ref{tab:dims}).  For comparison, we also include the unshielded versions of the same tables (case Va,c in Table~\ref{tab:dims}). We use the spectral synthesis code \textsc{cloudy}  v17.01, last described in \cite{ferland_2017_2017}, to transmit the spectrum through the gas (for self-shielded gas), calculate the equilibrium ionization states of all atomic ions, and follow the formation and destruction of molecules to obtain the resulting cooling and heating rates for all included processes. 

While we do not take non-equilibrium effects into account, the tables can easily be coupled to non-equilibrium networks that calculate the abundances of H and He, an approach that captures non-equilibrium effects on metals caused by the non-equilibrium free electron density (see Appendix~\ref{sec:noneq}).

The data products include a set of tables in hdf5 format of cooling and heating rates (also per element), ion and molecule fractions, and line emissivities, a routine that reads in and interpolates the cooling rates (written in C), as well as a python-based graphical user interface (gui) to visualise the table content. In equations that refer directly to individual hdf5 datasets, a line with the dataset name is added (e.g. dataset: \texttt{ShieldingColumnDensityRef} in Eq.~\ref{eq:Nnorm}). Figures that can be reproduced directly with the provided gui are labelled accordingly in the figure captions. The python and C routines and additional information on how to download the tables as well as examples can be found on dedicated web pages (\url{http://radcool.strw.leidenuniv.nl/} or \url{https://www.sylviaploeckinger.com/radcool}).

The methods are described in detail in Sec.~\ref{sec:method}, including how the shielding column (Sec.~\ref{sec:column}), the radiation field (Sec.~\ref{sec:radfield}), and the dust content (Sec.~\ref{sec:metdust}) vary within the four-dimensional grid in redshift, temperature, metallicity, and density. We present the data products in Sec.~\ref{sec:data}, where the individual tables of the full set are presented and their information content is listed and explained. In Sec.~\ref{sec:results} we highlight a few key results, such as the dependence of the thermal equilibrium temperatures and the most important phase transitions (ionized - neutral, neutral - molecular) on redshift, metallicity, and the radiation field. We summarise the findings in Sec.~\ref{sec:summary} and add sections on how to use (Appendix~\ref{sec:howto}) and reproduce (Appendix~\ref{sec:reprod}) the provided tables. 

The grid spacing in temperature, density, redshift, and metallicity as well as most of the tabulated properties are in $\log$. Throughout the paper $\log$ refers to $\log_{\mathrm{10}}$ and we use $10^{-50}$ as a floor value for properties that would be zero. All bins are generally equally spaced in $\log$ but the metallicity dimension has an additional bin: primordial abundances ($Z=0$, referred to as $\log Z/\mathrm{Z}_{\odot} = -50$). 

\section{Method}\label{sec:method}

For the calculation of cooling and heating rates as well as ionization stages and molecular fractions, we use version 17.01 of the photoionization code \textsc{cloudy} \citep{ferland_cloudy_1998, ferland_2017_2017}. \textsc{cloudy}  is a powerful and widely used open source code that can calculate the ionization, chemical, and thermal equilibrium state of gas exposed to an external radiation field. 

Our basic setup is a large grid of \textsc{cloudy} runs with dimensions $z$ (redshift), $T$ (temperature [K]), $Z$ (metallicity [$\Zsol$]), and $\nH$ (total hydrogen number density [$\ccm$]). The aim of this work is to provide tables that cover the full range of gas properties typically occurring in galaxy-scale or cosmological simulations (see Table~\ref{tab:size}). Throughout the paper we use the solar metallicity ($\Zsol = 0.0134$) and solar abundance ratios from \citet{asplund_chemical_2009} (see Table~\ref{tab:solarabundances}).

\begin{table} 
	\caption{All gas properties are tabulated on an equally spaced grid in the dimensions: redshift $z$, gas temperature $\log T$, metallicity $\log Z/\Zsol$, and gas density $\log \nH$. The grid points for each dimension range from their minimum (column 3) to their maximum (column 4) value, with the spacing $\Delta$ between the grid points listed in column 5. For metallicity an additional field is included for primordial abundances ($\log Z/\Zsol = -50$, i.e. $Z=0$). The resulting number of bins per dimension in column 6 leads to a total of 3,089,636 grid points for each table. }
	\begin{tabular}{lllllll}
				 	& unit			& min 		&max 		& $\Delta$  	& 	extra			&nbins \\
		\hline
		z	&				&  0			& 9			&	0.2	  	&   					& 46		\\
		$\log T$  	& $[\K]$		&  1			& 9.5			&	0.1		&					& 86	 	\\
		$\log Z/\Zsol$	& 		& -4			& 0.5			&	0.5		&   -50	($Z=0$)	& 11 		\\
		$\log \nH$& $[\ccm]$		& -8			& 6			&	0.2		&					& 71		\\
		\hline
	\end{tabular}\label{tab:size}
\end{table}

For every grid point in the tables that include self-shielding, the chosen incident radiation field is set to propagate through a slab of plane-parallel gas with the given properties ($T$, $Z/\Zsol$, $\nH$) until its shielding column density $\NS$ is reached. \textsc{cloudy} bins the gas column into individual zones of adaptive widths and in each zone the chemical properties and ionization states are calculated and the resulting output spectrum is passed as input into the next zone. Within each gas column, the total hydrogen number density as well as the gas temperature are set to remain constant, while the ion fractions, electron densities, and molecule fractions depend on the depth into the slab of gas. For unshielded gas, the properties of the first zone are tabulated (1-zone model), while for self-shielded gas the same properties are tabulated for the last zone, where $\NH = \NS$. 

For ionized gas in ionization equilibrium, the contribution of individual elements to the total cooling and heating rates can be calculated by running the same 1-zone \textsc{cloudy} models once with the full abundance set and once excluding individual elements. The difference in the rates can then be attributed to the excluded element (this is done e.g. in \citetalias{wiersma_effect_2009}). Here, this approach is not suitable, as a combination of elements is involved in the formation of molecules\footnote{Clearly the contribution of CO cannot be derived from the sum of two runs, where one is without C but with O, and the other one includes C but no O.} and the shielding contributions from one species can affect the abundance of another species. We therefore run \textsc{cloudy} for each grid point only once and output the values for each cooling and heating channel, as listed in Tables~\ref{tab:cool} and \ref{tab:heat}. For a full list of gas properties tabulated for each grid point, see Sec.~\ref{sec:data}.

\begin{table} 
	\caption{Solar abundances from \citet{asplund_chemical_2009} for individually tabulated elements with a total metallicity of Z = 0.0134. }
	\begin{center}
	\begin{tabular}{cc}
		Element	&$n_i/\nH $	\\
		\hline
		H		&		1\\
		He		&		$8.51\times10^{-2}$\\
		C		&		$2.69\times10^{-4}$\\
		N		&		$6.76\times10^{-5}$\\
		O		&		$4.90\times10^{-4}$\\
		Ne		&		$8.51\times10^{-5}$\\
		Mg		&		$3.98\times10^{-5}$\\
		Si		&		$3.24\times10^{-5}$\\
		S		&		$1.32\times10^{-5}$\\
		Ca		&		$2.19\times10^{-6}$\\
		Fe		&		$3.16\times10^{-5}$\\
		\hline
	\end{tabular}\label{tab:solarabundances}
	\end{center}
\end{table}

\subsection{Column density}\label{sec:column}
\begin{table*} 
	\caption{Overview of the Milky Way / solar neighbourhood values used to scale the interstellar radiation field, the CR rate and the D/G ratio. }
	\begin{tabular}{cccll}
											& &&& Description \\
			
		\hline
		$\log N_{\mathrm{H,0}}$					&	\multicolumn{2}{r}{20.56} 							&$\scm$									& Total hydrogen gas column density \\
		$\log N_{\mathrm{sh,0}}$					&	\multicolumn{2}{r}{20.26} 							&$\scm$									& Total hydrogen shielding column density \\
		$\Sigma_{\mathrm{g,0}}$					&	\multicolumn{2}{r}{3.85}							&$\Msun\,\mathrm{pc}^{-2}$					& Total gas surface density \\
		$(\mathrm{D/G})_0$						& 	\multicolumn{2}{r}{$5.623\times10^{-3}$}				&										& Total dust mass to gas mass ratio for $Z=\Zsol$ \\
		$\Sigma_{\mathrm{SFR,0}}$				&	 \multicolumn{2}{r}{$10^{-3}$}				&$\Msun\,\mathrm{yr}^{-1}\mathrm{kpc}^{-2}$ 		& SFR surface density \\
		$J_{\mathrm{0}}$					 	&       \multicolumn{2}{r}{$1.4\times10^{-3}$}		&$\mathrm{erg}\,\mathrm{cm}^{-2}\,\mathrm{s}^{-1}$& Radiation field from stars ($4 \pi \nu J_{\nu}$), defined at 1000 \AA \\
		$\Gamma_{\mathrm{Phot,0}} (\ion{H}{I})$    	&   	\multicolumn{2}{r}{$3.75\times10^{-11}$}	& $\mathrm{s}^{-1}$							& \ion{H}{I} ionization rate\\
		$\Gamma_{\mathrm{Phot,0}} (\ion{He}{I})$  	&      \multicolumn{2}{r}{$3.10\times10^{-12}$}	& $\mathrm{s}^{-1}$							& \ion{He}{I} ionization rate\\
		$\Gamma_{\mathrm{Phot,0}} (\ion{He}{II})$ 	&      \multicolumn{2}{r}{$1.43\times10^{-14}$}	& $\mathrm{s}^{-1}$							& \ion{He}{II} ionization rate\\
		$\log \zeta_0$ 							&  	\multicolumn{2}{r}{$-15.7$}				&$\mathrm{s}^{-1}$							& Cosmic ray hydrogen ionization rate \\
		\hline
	\end{tabular}\label{tab:MW}
\end{table*}

We have already established that the gas properties of self-shielded gas depend on the depth (i.e. gas column density) into the irradiated slab of gas. 
While algorithms to calculate the optical depths for each resolution element in a simulation have been developed (e.g \citealp{wunsch_tree-based_2018} for the grid code FLASH, \mbox{\citealp{fryxell_flash:_2000}}), the shielding column or optical depth can be expensive to calculate and is not generally known on the fly.

The shielding length is often approximated based on the gradient in the gas density $\rho$

\begin{equation}
	L_{\mathrm{Sob}} = \frac{\rho}{| 2 \nabla \rho |} \, ,
\end{equation}

\noindent 
referred to as Sobolov approximation (used e.g. in \citealp{richings_effects_2016, hopkins_fire-2_2018}), or is assumed to be close to the local Jeans length for self-gravitating gas \citep[e.g. in ][]{schaye_physical_2001}

\begin{equation}
	L_{\mathrm{J}} = \left ( \frac{\gamma X k_{\mathrm{B}}}{m^2_{\mathrm{H}} \mu G} \right ) ^{0.5} \left ( \frac{T}{\nH} \right )^{0.5}  \, ,
\end{equation}

\noindent
where $\gamma$ is the ratio of specific heats, $X$ is the hydrogen mass fraction, $k_{\mathrm{B}}$ is the Boltzmann constant, $m_{\mathrm{H}}$ is the proton mass, $\mu m_{\mathrm{H}}$ is the mean particle mass, $G$ is the gravitational constant and $\nH$ and $T$ are the total hydrogen density and temperature.
The resulting Jeans column density is

\begin{align}\label{eq:NJ}
N_{\mathrm{J}} &= \left ( \frac{\gamma X k_{\mathrm{B}}}{m^2_{\mathrm{H}} \mu G} \right ) ^{0.5} \left (n_{\mathrm{H}} T \right )^{0.5}  \, .
\end{align}

\noindent
The \textsc{cloudy} calculation stops when a specified total hydrogen column density is reached. Using the values for $\gamma$ and $\mu$ from the current \textsc{cloudy} zone leads to discontinuities in the shielding column densities at phase transitions as $\mu$ changes steeply (ionized - neutral: $\mu \approx  0.5 \rightarrow 1$, atomic - molecular: $\mu \approx  1 \rightarrow 2$). In addition, the last zone would preferentially be at the phase transition, as the shielding length suddenly decreases ($N_{\mathrm{J}} \propto \mu^{-0.5}$). We therefore use the values of primordial neutral gas (from \mbox{\citealp{ade_planck_2016}}) for the shielding column density: $X = 0.7563$,   $\gamma = 5/3$, and $\mu = 1.2328$, which results in

\begin{equation}
	L_{\mathrm{J}} = 0.9\,\mathrm{kpc}\, \left ( \frac{T}{10^4\K} \right )^{\mathrm{0.5}} \left ( \frac{\nH}{1 \ccm} \right )^{\mathrm{-0.5}}  \,
\end{equation}

\noindent
and

\noindent
\begin{equation}\label{eq:logNJ}
\log N_{\mathrm{J}} \,[\scm]= 19.44 + 0.5 \times (\log \nH [\ccm]+ \log T[K] ) \, .
\end{equation}

The Jeans column density is defined in \citet{schaye_model-independent_2001, schaye_physical_2001}, and \citetalias{rahmati_evolution_2013} showed that this approximation for the shielding column reproduces the results of full radiative transfer calculations for \ion{H}{I} in cosmological simulations that however did not include a cold ISM. We use the Jeans length approximation here, as it has the additional advantage that $N_{\mathrm{J}}$ depends only on gas temperature and density, so no extra dimension $\nabla \rho$ is required in the tables. 

We cover a large range of gas densities and temperatures (see Table~\ref{tab:size}) and the Jeans column density cannot be directly applied everywhere.
High temperature gas ($T\gtrsim10^5\,\K$) is mostly ionized and therefore optically thin, but as $N_{\mathrm{J}} \propto T^{0.5}$, its theoretical shielding length would continue to increase for a constant density. This leads to practical problems in \textsc{cloudy}, as the 1D shielding column starts to radiate thermal emission for high temperatures which results in an increasing radiation field with shielding column density. In addition, the shielding length can become very large for very low densities ($\nH < 10^{-6}\,\ccm$), leading to unrealistic coherence length scales corresponding to dynamical and sound crossing times that exceed the Hubble time. We therefore limit the length scale to $l_{\mathrm{max}} = 100 \,\mathrm{kpc}$, the total hydrogen column densities to $N_{\mathrm{max}} = 10^{24}\,\scm$ and use an asymptotic function for a smooth transition between low ($\lesssim10^{3}\,\K$, shielded) and high ($\gtrsim10^5\,\K$, optically thin) temperature gas. 

The reference column density, $N_{\mathrm{ref}}$, is defined as

\noindent
\begin{align}\label{eq:Nnorm}
	\log N_{\mathrm{ref}} =& \log N'_{\mathrm{ref}} - \frac{\log N'_{\mathrm{ref}} - \log N_{\mathrm{min}} }{1 + \left( \sqrt{T_{\mathrm{min}}T_{\mathrm{max}}}  / T\right ) ^{k}} \, ,\\
	&\mathrm{dataset:} \, \texttt{ShieldingColumnRef} \nonumber
\end{align}

\noindent 
where $k = 5 / \ln (10) = 2.17$ regulates the steepness of the transition between the asymptotes $N'_{\mathrm{ref}}$ reached at $T \ll  \sqrt{T_{\mathrm{min}}T_{\mathrm{max}}}$ and $N_{\mathrm{min}}$ at $T \gg  \sqrt{T_{\mathrm{min}}T_{\mathrm{max}}}$. We will use $T_{\mathrm{min}}  = 10^{3}\,\K$ and $T_{\mathrm{max}} = 10^5\,\K$. $N'_{\mathrm{ref}}$ is based on the length and column density limited Jeans column density $\NJ$:

\begin{equation}\label{eq:Nrefprime}
	\log N'_{\mathrm{ref}} \,[\scm] = \mathrm{min} \begin{cases}
	\log N_{\mathrm{J}} & \\
	\log l_{\mathrm{max}} \,\nH \quad .&  \\
	\log N_{\mathrm{max}}  & 
	\end{cases} 
\end{equation}

\noindent
$N_{\mathrm{min}} = l_{\mathrm{max}} \, n_{\mathrm{H,min}} = 3.08\times10^{15}\scm$ is the column density used for the minimum density in the tables $n_{\mathrm{H,min}} = 10^{-8}\,\scm$.

\begin{figure}
	\begin{center}
		\includegraphics*[width=\linewidth]{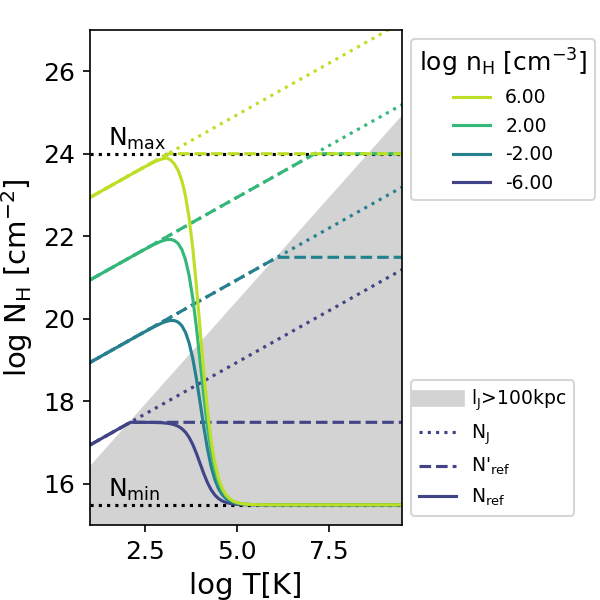}
		\caption{Temperature dependence of the discussed hydrogen column densities for selected gas densities $\nH$ (different colors). The Jeans column density  $\NJ$ (Eq.~\ref{eq:logNJ}, dotted lines) increases with temperature as $\NJ \propto T^{0.5}$ for a constant gas density. $N'_{\mathrm{ref}}$ (Eq.~\ref{eq:Nrefprime}, dashed lines) follows the Jeans column density for low temperatures but is limited by a maximum column density $N_{\mathrm{max}} = 10^{24}\scm$ and a maximum length scale of $l_{\mathrm{max}} = 100\,\mathrm{kpc}$ (the grey area indicates where $l_{\mathrm{J}} > 100\,\mathrm{kpc}$). The final reference column density $N_{\mathrm{ref}}$  (Eq.~\ref{eq:Nnorm}, solid lines) includes the asymptotic transition in temperature from self-shielded gas to optically thin gas with a minimum column density $N_{\mathrm{min}}$. The values for $N_{\mathrm{ref}}$, which corresponds to twice the shielding column density, over the full temperature-density parameter space are shown in Fig.~\ref{fig:Nref}. }
		\label{fig:Nrefexpl}
	\end{center}
\end{figure}

\begin{figure}
	\begin{center}
		\includegraphics*[width=\linewidth, trim = 1.7cm 0.5cm 0.5cm 1.5cm,clip]{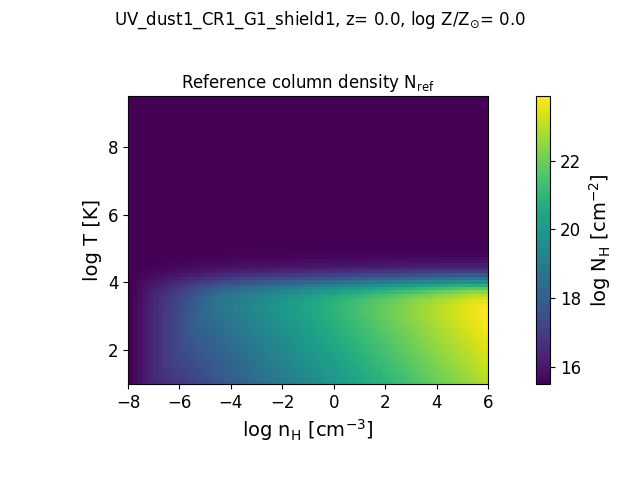}
		\caption{Reference column density $N_{\mathrm{ref}}$ (Eq.~\ref{eq:Nnorm}) for hydrogen number density $\nH$ and gas temperature $T$. $N_{\mathrm{ref}}$ corresponds to twice the shielding column density. Figure made with provided gui.}
		\label{fig:Nref}
	\end{center}
\end{figure}

Eqs.~\ref{eq:Nnorm} and \ref{eq:Nrefprime} are illustrated in Fig.~\ref{fig:Nrefexpl}. $N'_{\mathrm{ref}}$ (dashed line) follows the Jeans length $N_{\mathrm{J}}$ (dotted lines) for low temperatures but is limited by $l>l_{\mathrm{max}}=100\,\mathrm{kpc}$ (grey area) at low densities and by $N_{\mathrm{max}}$ at high densities. Adding the asymptotic transition to $N_{\mathrm{min}}$ for $T\gg\sqrt{T_{\mathrm{min}}T_{\mathrm{max}}}$ leads to the final $N_{\mathrm{ref}}$ (solid lines). A plot of $\log N_{\mathrm{ref}}$ as a function of density and temperature is presented in Fig.~\ref{fig:Nref}. 

For unshielded tables, the \textsc{cloudy} calculation is stopped for each grid point after the first zone, where the maximum size of the first zone\footnote{Setting a maximum zone size is important, as the zone size in \textsc{cloudy} is adaptive and for very low density gas, the zone sizes can become very large and vary significantly between adjacent grid points. As the radiation field is determined at the centre of each zone, this leads to noisy ion fractions. Limiting the zone size removes this artefact.} is set to $\Delta r = 10^{20}\,\mathrm{cm}$.
The code stops adding zones when the total hydrogen column density exceeds the shielding column density 
\begin{align}\label{eq:Nsh}
	\log N_{\mathrm{sh}} = &\begin{cases}
	\log (0.5 \times N_{\mathrm{ref}})& \mathrm{for\, shielded\, gas}\\
	\log (\Delta r \times \nH)&\mathrm{for\, optically\, thin\, gas}
	\end{cases} \\
	&\mathrm{dataset:} \, \texttt{ShieldingColumn} \nonumber
\end{align}

$N_{\mathrm{sh}}$ is therefore the column density from the edge of the self-gravitating gas clump/disc to its centre/midplane, while $N_{\mathrm{ref}}$ describes the total column column density through the gas clump. The shielding column density to the last zone of the \textsc{cloudy} column is tabulated as dataset \texttt{ShieldingColumn}. Depending on the thickness of the last zone, a small deviation from $0.5 N_{\mathrm{ref}}$ is possible. For unshielded runs, \texttt{ShieldingColumn} is the column density of the first zone, while the values for $N_{\mathrm{ref}}$ in dataset \texttt{ShieldingColumnRef} are the same for both unshielded and self-shielded runs.

\subsection{Radiation field}\label{sec:radfield}

The incident spectrum in all tables includes a redshift dependent UVB based on \citetalias{faucher-giguere_cosmic_2020} (Sec.~\ref{sec:UVB}), as well as the cosmic microwave background (CMB) as a black\-body spectrum with a redshift dependent temperature of $T_{\mathrm{CMB}} = T_0\,(1+z)$ with $T_0 = 2.725\,\mathrm{K}$. In some tables a local radiation field is added that models the contribution from nearby stars (Sec.~\ref{sec:ISRF}) in the ISM.

\subsubsection{UV background radiation field}\label{sec:UVB}

The spectrum for the UVB from distant galaxies and active galactic nuclei (AGN) is based on \citetalias{faucher-giguere_cosmic_2020}, which consists of three main contributions. First, the radiation field from stars (for $z\le8$) is from the BPASS \citep{eldridge_binary_2017,stanway_re-evaluating_2018} spectrum for a stellar population with a constant star formation rate at an age of 300~Myr and a metallicity of $0.1\mathrm{Z}_{\odot}$. This spectrum is dust attenuated with a constant $E(B-V) = 0.129$. The redshift-dependent normalization is chosen to match the observed total UV emissivity from GALEX for $z<2$ \citep{chiang_broadband_2019} and the rest-frame UV galaxy luminosity function from Bouwens et al. (in prep.) for $z>2$ at 8.3 eV (1500\AA). Second, the radiation from AGN is modelled by combining spectral templates of obscured (75 per cent) and unobscured AGN (25 per cent) and the spectrum is normalized to the AGN ionizing emissivity from \citet{shen_bolometric_2020} at 912\AA. Third, recombination emission from the photons absorbed by the ISM of star-forming galaxies as well as the recombination emission lines related to the ``sawtooth"~absorption by \ion{He}{II} Lyman series lines.

In addition to redshift-dependent spectra, \citetalias{faucher-giguere_cosmic_2020} also provide effective photo-ionization $\Gamma_{\mathrm{x,eff}}$ and photo-heating rates $\dot{q}_{\mathrm{x,eff}}$ (with $x = \left \{  \right . $\ion{H}{I}, \ion{He}{I}, \ion{He}{II} $ \left . \right \}$) that are constructed to produce a prescribed volume-averaged ionized fraction (\ion{H}{II} and \ion{He}{III}) for each redshift, assuming photoionization equilibrium. These effective rates are lower than the rates calculated directly from the spectra before reionization in order to match the electron scattering optical depth of $\tau_{\mathrm{e}} = 0.054$, the best fit value to the Planck 2018 CMB data \citep{planck_collaboration_planck_2018}. The best model in  \citetalias{faucher-giguere_cosmic_2020} uses reionization redshifts of $z_{\mathrm{rei,HI}}=7.8$ and $z_{\mathrm{rei,HeII}}=3.5$, while at $z < z_{\mathrm{rei,HI}} - \Delta z_{\mathrm{rei,HI}} = 7.2$ and $z < z_{\mathrm{rei,HeII}} - \Delta z_{\mathrm{rei,HeII}} = 3$, reionization is completed and the respective effective rates match the rates from the spectra. 
   
For the UVB used in this work, we modify the spectra from \citetalias{faucher-giguere_cosmic_2020} for $z>3$ so that they yield rates that match the effective photo-ionization and photo-heating rates before \ion{H}{I} and \ion{He}{II} reionization. The details of this method as well as the resulting rates are shown in Appendix~\ref{sec:fg20eff}. In short, we attenuate the UVB by \ion{H}{I} gas for $z>z_{\mathrm{rei,HI}} - \Delta z_{\mathrm{rei,HI}}$ (before \ion{H}{I} reionization is complete) and by \ion{He}{II} gas for $z>z_{\mathrm{rei,HeII}} - \Delta z_{\mathrm{rei,HeII}}$ (before \ion{He}{II} reionization is complete). The gas column density in both cases is chosen to match the respective effective photoionization rates from \citetalias{faucher-giguere_cosmic_2020}.

\begin{figure}
	\begin{center}
		\includegraphics*[width=\linewidth, trim = 0.2cm 0cm 1.2cm 0.7cm,clip]{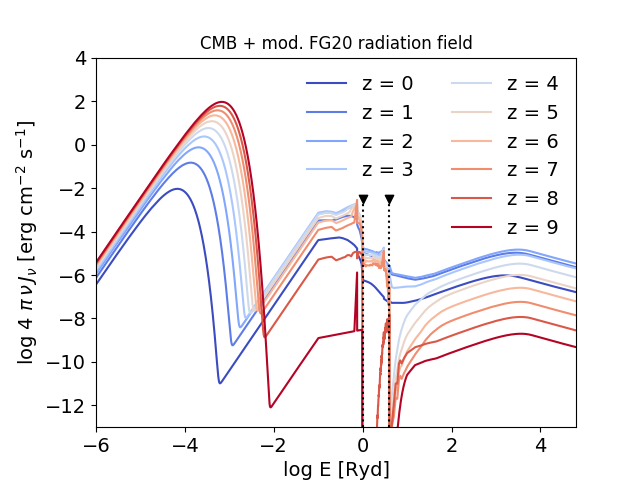}
		\caption{The incident background radiation field present in all tables for different redshifts consists of the CMB (lowest energy bump) and the modified \citetalias{faucher-giguere_cosmic_2020} UV / X-ray background (see Sec.~\ref{sec:UVB} and Appendix~\ref{sec:fg20eff} for details). The vertical dotted lines indicate the \ion{H}{I} (left) and \ion{He}{II} (right) ionization energies. Before reionization is complete, the UVB is reduced for $ z > 3$ ($z > 7.2$) at energies above the \ion{He}{II}  (\ion{H}{I}) ionization energy. See Fig.~\ref{fig:SP20spectrum} for a comparison to the original \citetalias{faucher-giguere_cosmic_2020} UVB.}
		\label{fig:RF}
	\end{center}
\end{figure}

The resulting spectrum (including the CMB) is shown in Fig.~\ref{fig:RF} for redshifts between 0 and 9. The CMB contribution dominates for energies $\lesssim 10^{-3}\,\mathrm{Ryd}$ while photons with higher energies originate from the UVB. For $z=9$ the \citetalias{faucher-giguere_cosmic_2020} UVB is greatly reduced at energies $-2 <\log E \,[ \mathrm{Ryd} ] < 0$ as the contribution from star forming galaxies is only included for $z\le 8$. The vertical dotted lines indicate the \ion{H}{I} and \ion{He}{II} ionization energies to illustrate the onsets of the above mentioned attenuation.

\subsubsection{Interstellar radiation field}\label{sec:ISRF}

For gas in the ISM, the diffuse ISRF from nearby stars can be stronger than the UVB. As discussed in the introduction, this type of radiation field can be described by its intensity as well as an undefined number of parameters that model its spectral shape. 

Here, we fix the shape of the ISRF to that described in \citet{black_heating_1987} for the Milky Way Galaxy and normalize it based on the reference column density $N_{\mathrm{ref}}$, which only depends on the gas density and temperature (Eq.~\ref{eq:Nnorm}). This is motivated by the observed star formation law and it avoids an extension of the tables to additional dimensions describing the ISRF and allows their use in simulations where the strength of the ISRF is not directly traced. 

\begin{figure}
	\begin{center}
		\includegraphics*[width=\linewidth, trim = 0.2cm 0cm 1.2cm 0.7cm,clip]{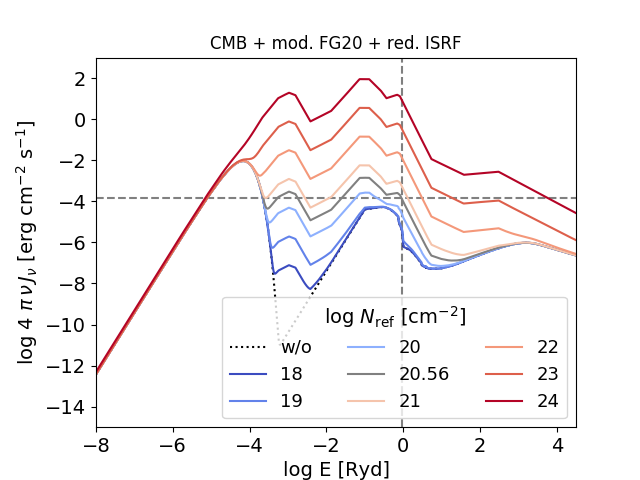}
		\caption{Total incident radiation field at $z=0$, including the CMB, the UVB and the local ISRF for different reference column densities, $N_{\mathrm{ref}}$. The shape of the ISRF is from \citet{black_heating_1987} and its intensity scales with $N_{\mathrm{ref}}^{1.4}$ as suggested by the Kennicutt-Schmidt law (see Eq.~\ref{eq:radG0}). The intensity of the total radiation field is defined at 1000 \AA \,(vertical dashed line), with $4 \pi \nu J_{\nu} = 1.4 \times10^{-4}\,\mathrm{erg}\,\mathrm{cm}^{-2}\,\mathrm{s}^{-1}$ (horizontal dashed line) for $\log N_{\mathrm{ref}} = \log N_{\mathrm{H,0}} [\scm]= 20.56$ (scaling as in Eq.~\ref{eq:radG0} with $\log R = -1$). For $\log N_{\mathrm{ref}} [\scm ]< 18$, the total radiation field converges to the \citetalias{faucher-giguere_cosmic_2020} plus CMB radiation field (black dotted line, see Fig.~\ref{fig:RF} for other redshifts) for most energies. }
		\label{fig:ISRF}
	\end{center}
\end{figure}

In the case of a universal initial mass function (IMF), the rate at which OB stars are formed is proportional to the local SFR and the normalization of the \citet{black_heating_1987} radiation field is therefore assumed to scale with the SFR surface density $\Sigma_{\mathrm{SFR}}$ as

\begin{equation} \label{eq:radG0basic}
	\log  \frac{J}{J_{\mathrm{0}}}=  \log \Sigma_{\mathrm{SFR}}- \log \Sigma_{\mathrm{SFR,0}}  + \log R\, ,
\end{equation}

\noindent
where $R$ is a renormalization factor.
The solar neighbourhood values $J_0 = 4.4\times10^5\,\mathrm{photons}\,\mathrm{cm}^{-2}\,\mathrm{s}^{-1}\,\mathrm{sr}^{-1}\,\mathrm{eV}^{-1}$ (or $4 \pi \nu J_{\nu,0} = 1.4 \times10^{-3}\,\mathrm{erg}\,\mathrm{cm}^{-2}\,\mathrm{s}^{-1}$) defined at a wavelength of 1000 \AA~\citep{black_heating_1987}  and $\Sigma_{\mathrm{SFR,0}} = 10^{-3}\, \mathrm{M}_{\odot} \, \mathrm{yr}^{-1} \, \mathrm{kpc}^{-2}$ \citep{bonatto_constraining_2011} are used to normalise the ISRF (see Table~\ref{tab:MW} for an overview of solar neighbourhood quantities used in this work). 

\begin{figure}
	\begin{center}
		\includegraphics*[width=\linewidth, trim = 1.8cm 0.5cm 0.5cm 1.2cm,clip]{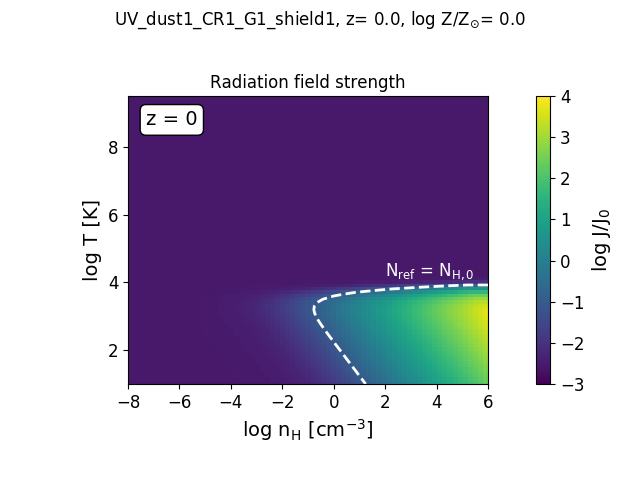}
		\includegraphics*[width=\linewidth, trim = 1.8cm 0.5cm 0.5cm 1.2cm,clip]{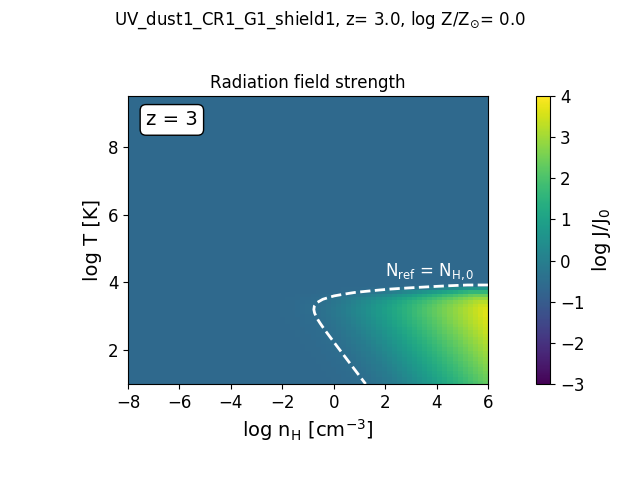}
		\caption{Total radiation field intensity defined at 1000 \AA \, relative to the MW value of $J_{\mathrm{0}}$ (see Table~\ref{tab:MW}) at $z=0$ (top panel) and $z=3$ (bottom panel)  for table UVB\_dust1\_CR1\_G1. The radiation field is dominated by the ISRF at high densities and temperatures $\lesssim 10^4\,\K$  (with $\log J/J_{\mathrm{0}} \propto 1.4 \times \log  \Sigma_{\mathrm{SFR}}$) and smoothly transitions to the redshift-dependent UVB for higher temperatures and towards low densities. For reference, the dashed line indicates where $N_{\mathrm{ref}} = N_{\mathrm{H,0}}$ (MW value, see Table~\ref{tab:MW}). Figures made with provided gui.}
		\label{fig:ISRF2D}
	\end{center}
\end{figure}

We assume that the SFR surface density $\Sigma_{\mathrm{SFR}}$ follows the observed Kennicutt-Schmidt relation \mbox{\citep{kennicutt_global_1998}}:

\begin{equation} \label{eq:KS}
	\Sigma_{\mathrm{SFR}} = A \left (  \frac{\Sigma_g} {\mathrm{M}_{\odot}\, \mathrm{pc}^{-2}} \right )^n \, ,
\end{equation}

\noindent
with $A = 1.515\times 10^{-4}\,\Msun\, \mathrm{yr}^{-1}\,\mathrm{kpc}^{-2}$ and $n=1.4$. The amplitude $A$ has been decreased to account for a \citet{chabrier_galactic_2003} IMF. These are the same values as used in the EAGLE simulations \citep{schaye_eagle_2015} for which a similar scaling of the ISRF was used in \citet{lagos_molecular_2015} to compute molecular fractions.
The gas surface density for a SFR of $\Sigma_{\mathrm{SFR,0}}$ is therefore 

\begin{equation}
\Sigma_{\mathrm{g,0}} = 3.85 \,\Msun \, \mathrm{pc}^{-2}\, .
\end{equation}

\noindent
As the gas surface density is expressed as a total hydrogen column density in the tables, $\Sigma_{\mathrm{g,0}}$ is re-written as   

\begin{align}
N_{\mathrm{H,0}} &= 3.65 \times 10^{20} \,\scm \quad \mathrm{or} \quad \log N_{\mathrm{H,0}} [\scm] = 20.56 \, ,
\end{align}

\noindent
using $X=0.7563$ as for the column density in Sec.~\ref{sec:column}. The final scaling of the ISRF based on $N_{\mathrm{ref}}$ is

\begin{align} \label{eq:radG0}
	\log  \frac{J}{J_{\mathrm{0}}}=  & 1.4 \times \left (\log N_{\mathrm{ref}}-\log N_{\mathrm{H,0}} \right ) + \log R  \\
	&\mathrm{dataset:} \, \texttt{RadField} \nonumber
\end{align}

\noindent
with a photoionization rate $\Gamma_{\mathrm{phot}}$ of 

\begin{equation}\label{eq:radphoto}
	\log \frac{\Gamma_{\mathrm{phot}}}{\Gamma_{\mathrm{phot,0}}} = 1.4 \times \left (\log N_{\mathrm{ref}}-\log N_{\mathrm{H,0}} \right ) + \log R  \, .
\end{equation}

\noindent 
where $R$ is the renormalization constant as in Eq.~\ref{eq:radG0basic}, $\Gamma_{\mathrm{phot,0}} (\ion{H}{I}) = 3.75\times10^{-11}\,\mathrm{s}^{-1}$ for \ion{H}{I} ionization, $\Gamma_{\mathrm{phot,0}} (\ion{He}{I}) = 3.10\times10^{-12}\,\mathrm{s}^{-1}$ for \ion{He}{I} and $\Gamma_{\mathrm{phot,0}} (\ion{He}{II}) = 1.43\times10^{-14}\,\mathrm{s}^{-1}$ for \ion{He}{II} ionization.

For an unscaled \citet{black_heating_1987} radiation field ($\log R = 0$), the molecular hydrogen fraction $f_{\mathrm{H2}}$ rises above 10 per cent only for neutral hydrogen column densities of $N_{\ion{H}{I}} + 2\NHtwo \gtrsim 10^{21}\,\scm$ for the thermal equilibrium temperature, which is inconsistent with observations. In addition, low-metallicity gas does not significantly cool below $10^4\K$ at any redshift. Renormalising the ISRF and the CR rate by a factor of 1/10 ($\log R = -1$) alleviates these tensions and the thermal equilibrium properties are in much better agreement with observations. This mismatch between the ISRF and the observed H$_{\mathrm 2}$ fractions has been found in previous work, where the H$_{\mathrm 2}$ formation rate on dust grains has to be boosted for a radiation field of 2.3 $J_0$ to reproduce the observations (e.g. \mbox{\citealp{gnedin_modeling_2009}}, \mbox{\citealp{gnedin_environmental_2011}}, \mbox{\citealp{gnedin_line_2014}}). They relate this boost factor to the clumping factor $C_{\rho}$ and calibrate their model to this free parameter. This would account for unresolved higher density gas with higher molecule fractions. They find that clumping factors between $C_{\rho} \approx$ 10 and 30 are necessary to reproduce the observed \ion{H}{I} - H$_{\mathrm 2}$ in the MW. 

As \textsc{cloudy} solves a chemical network including several hundred molecules and atomic species, artificially increasing the formation rate of $H_{\mathrm 2}$ by a factor that is a free parameter is neither practical nor self-consistent. In this case, decreasing the radiation field is a more reasonable approach.
Note also that the assumed scaling of the ISRF with the star formation surface density may overestimate the radiation field. A higher $\Sigma_{\mathrm{SFR}}$ is linked to stronger local shielding of the stellar sources and may lead to a larger star formation scale height, both of which would reduce the local ISRF.

Fig.~\ref{fig:ISRF} shows the fiducial input spectrum ($\log R = -1$) at $z=0$ for different column densities $\log N_{\mathrm{ref}}$.  
For very small values of $N_{\mathrm{ref}}$, the UVB/CMB radiation field (black dotted line) dominates for $\log E\, [\mathrm{Ryd}] \lesssim -4$ and $\log E\, [\mathrm{Ryd}] \gtrsim -2$ and the intensity of the total radiation field in this energy range therefore becomes independent of the density and temperature. This transition depends on the intensity of the UVB and therefore on the redshift (see Fig.~\ref{fig:ISRF2D}).

\subsubsection{Before reionization}\label{sec:befre}
We apply the ISRF scaling $J\propto N_{\mathrm{ref}}^{1.4}$ at all column densities, so also at column densities lower than expected for the ISM. After reionization this is not a problem, because the UVB dominates over the ISRF at low densities. However, before reionization the ionizing radiation of distant galaxies is absorbed close to the source, while local radiation fields, such as the ionizing radiation from nearby stars (Sec.~\ref{sec:ISRF}) remain unaffected. As the local radiation field is not attenuated, its contribution can even become dominant at low column / volume densities ($\log \nH [\ccm] \lesssim -5$), which would not be part of the ISM and should therefore not be exposed to the ISRF.  In order to avoid artificial gas heating, the ISRF normalization is drastically reduced compared to the reference value from $\log \nH [\ccm] = -2$ to $-6$ with a similar asymptotic function as in Eq.~\ref{eq:Nnorm}:

\begin{align}
	\log (J/J_{\mathrm{0}})_{z>\mathrm{z_{reioniz}} } =&\log (J/J_{\mathrm{0}})_{\mathrm{min}}  - \frac{ \log (J/J_{\mathrm{0}})_{\mathrm{min}}  - \log J/J_{\mathrm{0}} }{1 + e^{-2 (\log \nH[\ccm] + 4)} } \\
	\mathrm{or:} & \nonumber \\
	\log (J/J_{\mathrm{0}})_{z>\mathrm{z_{reioniz}} } =&\log (J/J_{\mathrm{0}})_{\mathrm{min}}  - \frac{ \log (J/J_{\mathrm{0}})_{\mathrm{min}}  - \log J/J_{\mathrm{0}} }{1 +   c_{\mathrm{J}} \nH^{k_{\mathrm{J}}} } \nonumber 
\end{align}

\noindent 
where $k_{\mathrm{J}} = -2/\ln(10)$ and $c_{\mathrm{J}} = e^{-8}$ regulate the steepness and position of the asymptotic transition to $\log (J/J_{\mathrm{0}})_{\mathrm{min}} = -20$. Without this extra scaling the radiation field would not converge to the (modified) \citetalias{faucher-giguere_cosmic_2020} background radiation for very low densities. Note, that this additional scaling is only necessary for $z>7.5$ as the photo-ionization rate of the UVB dominates that of the ISRF for lower redshifts at these densities. 

\begin{figure}
	\begin{center}
		\includegraphics*[width=\linewidth, trim = 0cm 0cm 0cm 0cm,clip]{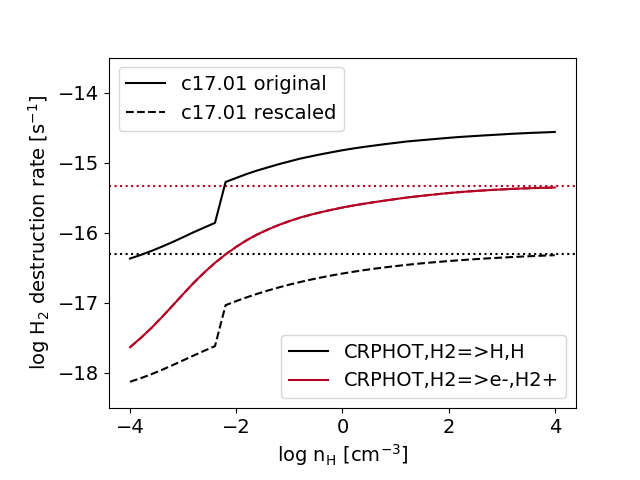}
		\caption{The H$_{\mathrm{2}}$ destruction rates by cosmic rays for H$_{\mathrm{2}}$ dissociation (``CRPHOT,H2=$>$H,H", black lines) and ionization (``CRPHOT,H2=$>$e-,H2+", red lines). In \textsc{cloudy} v17.01 (solid lines), the ionization rate converges to the input secondary ionization rate of $4.6 \times 10^{-16}\,\mathrm{s}^{-1}$ (red, dotted line) but the dissociation rate is a factor of $\approx 6$ higher. In this work we rescale the rate for the reaction ``CRPHOT,H2=$>$H,H" (dashed line) so that it converges to $5 \times 10^{-17}\,\mathrm{s}^{-1}$ (black, dotted line) for high densities, the value expected from the UMIST database (see text for details). }
		\label{fig:CRresc}
	\end{center}
\end{figure}

\subsection{Cosmic rays}\label{sec:CR}

Cosmic rays (CRs) are charged particles accelerated to high energies by diffuse shock acceleration in supernova remnants \citep{bell_turbulent_2004}. Their production rate is therefore expected to be proportional to the rate of SNe and furthermore, for a constant IMF, to the SFR. This is supported by observations of starburst galaxies, where the CR rate is several orders of magnitude higher than locally in the MW Galaxy \citep[e.g.][]{suchkov_cosmic-ray--dominated_1993, collaboration_connection_2009}.

Analogously to the scaling of the ISRF intensity (Sec.~\ref{sec:ISRF}), the \ion{H}{I} CR ionization rate $\zeta_{\mathrm{CR}}$ is set to

\begin{align}\label{eq:CR}
	\log \zeta_{\mathrm{CR}} \, [\mathrm{s}^{-1}]=& \log \zeta_{\mathrm{CR,0}} +1.4 \times \left (\log N_{\mathrm{ref}} - \log N_{\mathrm{H,0}} \right ) + \log R\\
	&\mathrm{dataset:} \, \texttt{CosmicRayRate} \nonumber	
\end{align}

\noindent 
where $\zeta_{\mathrm{CR,0}} = 2 \times 10^{-16}\,\mathrm{s}^{-1}$ is the mean CR ionization rate of neutral hydrogen in local, diffuse clouds \citep{indriolo_h+3_2007} and $\log R$ is the same renormalization constant as in Eq.~\ref{eq:radG0} (we use $\log R = -1$ for the fiducial model). 

We use the large \textsc{cloudy} model for the H$_{\mathrm{2}}$ molecule (described in \citealp{shaw_molecular_2005}) which includes several thousand levels for the H$_{\mathrm{2}}$ molecule. The H$_{\mathrm{2}}$ ionization rate is therefore not constant but depends for example on the gas density (see the red solid line in Fig.~\ref{fig:CRresc}). For high densities the H$_{\mathrm{2}}$ destruction rate by cosmic rays due to H$_{\mathrm{2}}$ ionization (labelled ``CRPHOT,H2=$>$e-,H2+"~in \textsc{cloudy}) converges to the input value of $4.6 \times 10^{-16}\,\mathrm{s}^{-1}$ (red dotted horizontal line), following the ratio between \ion{H}{I} and $\mathrm{H}_{\mathrm{2}}$ ionization rates from \citet{glassgold_model_1974}. According to the \mbox{UMIST database\footnote{\url{http://udfa.net}}} \citep{mcelroy_umist_2013}, the ratio between the rates for H$_{\mathrm{2}}$ destruction by ionization and dissociation (labelled ``CRPHOT,H2=$>$H,H"~in \textsc{cloudy}) is 0.108, and the dissociation rate is therefore expected to converge towards $5 \times 10^{-17}\,\mathrm{s}^{-1}$ (black, dotted horizontal line). In \textsc{cloudy} 17.01, the dissociation rate is a factor of 6 higher than the ionization rate (black solid line). We therefore rescale this rate to converge towards the expected value of $5 \times 10^{-17}\,\mathrm{s}^{-1}$ (black dashed line) for high densities. This is more consistent with other chemical network codes, such as \textsc{chimes} \citep{richings_non-equilibrium_2014-1,richings_non-equilibrium_2014} that use the UMIST values and leads to higher $\mathrm{H}_{\mathrm{2}}$ fractions, in better agreement with observations (see Sec.~\ref{sec:phasetransitions}).

The high H$_{\mathrm{2}}$ dissociation rate in \textsc{cloudy} is caused by the assumed mean kinetic energy of the secondary electrons of 20~eV. For this energy the cross section for H$_{\mathrm{2}}$ dissociation is at a maximum \citep{dalgarno_electron_1999}. Future versions of \textsc{cloudy} will use a mean kinetic energy of 36~eV, more representative for the broad range of energies of secondary electrons, and this rescaling will not be necessary anymore (see \citealp{shaw_cosmic_2020} for details).

\begin{figure} 
	\begin{center}
		\includegraphics*[width=\linewidth, trim = 1.7cm 0.5cm 0.3cm 1.4cm, clip]{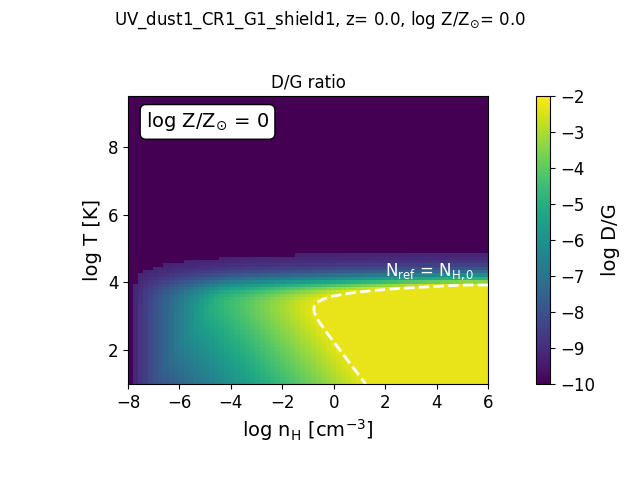}
		\caption{Assumed dust-to-gas mass ratio for solar metallicity. For $N_{\mathrm{ref}}>N_{\mathrm{H,0}}$ (higher density, lower temperature than the white dashed contour that indicates $N_{\mathrm{ref}}=N_{\mathrm{H,0}}$) the dust-to-metal mass ratio is constant while for lower densities and higher temperatures the dust content is reduced (Eq.~\ref{eq:DGratio}). Figure made with provided gui.}
		\label{fig:DG}
	\end{center}
\end{figure}

\begin{figure*} 
	\begin{center}
		\includegraphics*[width=\linewidth, trim = 1cm 0.2cm 1.7cm 0.8cm,clip]{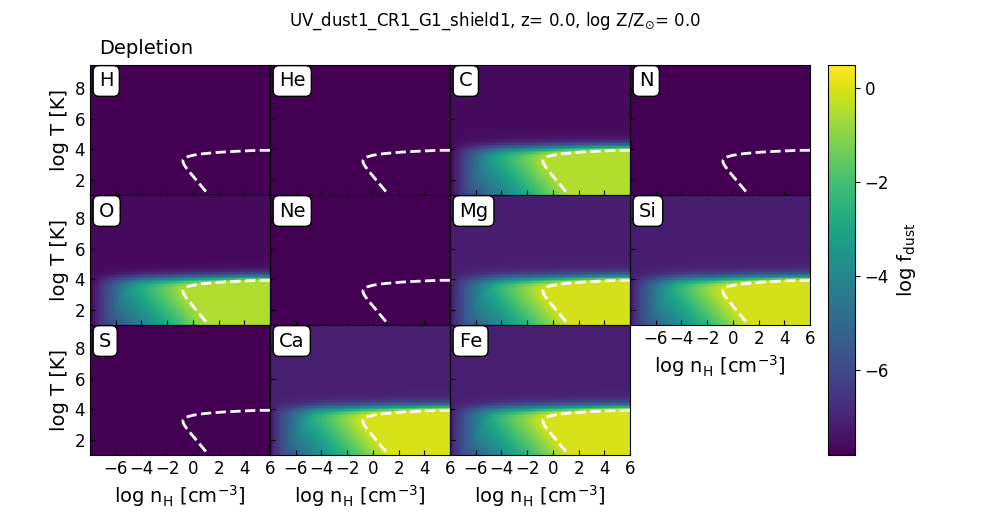}
		\caption{Metal depletion for all 11 tabulated elements. The color code shows the fraction of atoms of a given element that is locked into dust grains. Note the constant depletion for $N_{\mathrm{ref}} >  N_{\mathrm{H,0}}$ (white dashed contour indicates $N_{\mathrm{ref}} =  N_{\mathrm{H,0}}$) and the scaling with $N_{\mathrm{ref}}^{1.4}$ towards lower densities (Eq.~\ref{eq:depletion}). Figure made with provided gui.}
		\label{fig:depl}
	\end{center}
\end{figure*}

\subsection{Metallicity and dust}\label{sec:metdust}

One of the table dimensions is the gas metallicity $Z$ in units of solar metallicity $\Zsol$, which ranges from $\log Z/\Zsol = -4$ to $0.5$ in steps of 0.5 dex (see Table~\ref{tab:size}). The abundances of all elements other than hydrogen and helium are scales with metallicity relative to their solar values (Table~\ref{tab:solarabundances}). For primordial abundances, an additional metallicity bin is added ($\log Z/\Zsol = -50$), using the primordial helium abundance $(n_{\mathrm{He}}/\nH)_{\mathrm{prim}} = 0.08246$ \citep{ade_planck_2016}. To obtain a smooth transition between zero and solar metallicities, the helium abundance is linearly interpolated in between 

\begin{equation}\label{eq:helium}
	\left ( \frac{n_{\mathrm{He}}}{\nH} \right )_Z =\left [  \left ( \frac{n_{\mathrm{He}}}{\nH} \right )_{\Zsol} - \left ( \frac{n_{\mathrm{He}}}{\nH} \right )_{\mathrm{prim}}\right ] \left ( \frac{Z}{\Zsol} \right ) +  \left ( \frac{n_{\mathrm{He}}}{\nH} \right )_{\mathrm{prim}}
\end{equation}

\noindent
with $(n_{\mathrm{He}}/\nH)_{\odot} = 0.0851$ \citep{asplund_chemical_2009}.
The tables contain various datasets with information about the used abundance ratios (see Sec.~\ref{sec:additional}).

The primordial abundance set in \textsc{cloudy} does not only include hydrogen and helium, but also traces of lithium and beryllium. Their minor cooling contribution explains why the metal cooling rate can be non-zero for primordial abundances.

\subsubsection{Dust}

We include the ``Orion'' grain set from \textsc{cloudy}. For solar metallicity, this grain set has a default dust-to-gas mass ratio of $(D/G)_0 = 5.6\times10^{-3}$ and the relation between the extinction $A_V$ and the total hydrogen column density $\NH$ is $\left( A_V/\NH \right) = 5.54 \times10^{-22} \,\mathrm{mag\,cm}^{2}$. The grain distribution with sizes between 300 and 2500 \AA \,consists of both graphites and silicates and matches the extinction observed in Orion. In addition to dust grains, also polycyclic aromatic hydrocarbons (PAHs, with a power-law size distribution from \citealp{abel_sensitivity_2008}) are included as they contribute to photoelectric heating, collisional processes, and stochastic heating. As PAHs are destroyed in ionised gas by ionising photons and get depleted onto larger grains in molecular gas, their abundance is set to scale with the atomic neutral hydrogen fraction $n_{{\ion{H}{I}}}/\nH$ in \textsc{cloudy}.

A constant dust-to-metal ratio for ISM gas with $N_{\mathrm{ref}}>N_{\mathrm{H,0}}$ is assumed. As for this solar neighbourhood dust-to-metal ratio, some elements (e.g. Mg, Si, Ca, Fe) are already almost completely depleted on dust grains, higher dust-to-metal ratios would not be consistent. For lower reference column densities, a similar scaling as for the ISRF and the CR rate is used. The dust-to-gas mass ratio $D/G$ is then

\begin{align}\label{eq:DGratio}
	\log D/G = &\log Z/\Zsol + \log (D/G)_0 \,+\\ \nonumber
			&+ \mathrm{min} \begin{cases}
		1.4 \times \left ( \log N_{\mathrm{ref}} - \log N_{\mathrm{H,0}} \right )  &\\
		0 \\
	\end{cases}\\
	&\mathrm{dataset:} \, \texttt{DGratio} \nonumber		
\end{align}

For $N_{\mathrm{ref}} = N_{\mathrm{min}}$ grain physics is completely disabled, as even extremely low grain abundances can cause \textsc{cloudy} to become unstable for high temperatures (i.e. $T > 10^6\,\K$). Here, dust grains would be destroyed by thermal sputtering on very short timescales \citep[e.g.][]{tielens_physics_1994}, but this process is not included in \textsc{cloudy}.
Fig.~\ref{fig:DG} summarises the dust-to-gas ratio dependence on density and temperature. 
The option \texttt{no qheat} is a standard \textsc{cloudy} command to disable quantum heating. This was necessary for code stability.

\subsubsection{Metal depletion}

The abundances of the Sun, which define the solar abundances, differ from those of the local ISM. For some elements a large fraction of atoms are depleted onto dust grains and their abundance ratios measured in the gas phase are correspondingly reduced. If the gas phase abundance were set to solar and dust grains were added to the mix without depleting the gas phase, the total (dust + gas) abundances would become super-solar.

For each element $i$, the number fraction of atoms that are depleted on dust grains is $f_{\mathrm{dust,i}}$, while $f_{\mathrm{gas,i}} = 1 - f_{\mathrm{dust,i}}$ is the number fraction of atoms of element $i$ in the gas phase. The reference values for solar neighbourhood conditions $\log f_{\mathrm{dust,i}}'$ are taken from table 4 of \citet{jenkins_unified_2009}, corrected for the different solar abundances used in their work\footnote{For the corrected abundances, nitrogen would have $f_{\mathrm{gas, N}} > 1$ and is therefore limited to $1$ (i.e. no dust depeletion).}. 

As we add dust grains with constant dust-to-metal mass ratio for $N_{\mathrm{ref}} \ge N_{\mathrm{H,0}}$ and scale their abundance with the star formation activity for $N_{\mathrm{ref}} < N_{\mathrm{H,0}}$ (Eq.~\ref{eq:DGratio}), the dust depletion for each grid point with $N_{\mathrm{ref}}$ and for each element $i$ is  

\begin{align}\label{eq:depletion}
	\log f_{\mathrm{dust,i}} =& \log f_{\mathrm{dust,i}}'  + \mathrm{min} \begin{cases}
		1.4 \times \left ( \log N_{\mathrm{ref}} - \log N_{\mathrm{H,0}} \right )  &\\
		0 \\
	\end{cases}\\
	&\mathrm{dataset:} \, \texttt{Depletion} \nonumber		
\end{align}

\noindent
and Fig.~\ref{fig:depl} illustrates this scaling.

\subsection{HD cooling}\label{sec:HD}

In the \textsc{cloudy} version used (17.01), deuterium chemistry is disabled by default, contrary to the information in the user manual\footnote{See discussion in the \textsc{cloudy} user group: \url{https://cloudyastrophysics.groups.io/g/Main/message/3934}}. Tests with a coarser grid where the deuterium chemistry was specifically added through the command line input, revealed that \textsc{cloudy} aborts for an increased number of grid points due to non-convergence issues in the chemistry solver. 
We therefore do not use the HD cooling from \textsc{cloudy} but add it analytically, following the \textsc{cloudy} prescription as closely as possible. 

As in \textsc{cloudy} v17, the cooling rate per HD molecule, $W_{\mathrm{HD}} [\mathrm{erg\,s^{-1}}]$, is taken from the fitting function provided by \citet{flower_cooling_2000}. $W_{\mathrm{HD}}$ depends on the gas density and temperature and was calculated by \citet{flower_cooling_2000} for temperatures  between $30$ and $3000\K$ and densities $\nH \ge 1\,\ccm$. The cooling rate from HD per unit volume is

\begin{equation}
	\Lambda_{\mathrm{cool,HD}} = n_{\mathrm{HD}} W_{\mathrm{HD}} \,\mathrm{erg\,cm^{-3}\,s^{-1}}
\end{equation}

\noindent 
where $n_{\mathrm{HD}} [\ccm]$ is the number density of HD molecules. As done in earlier \textsc{cloudy} versions, the HD abundance is assumed to follow the abundance of H$_{\mathrm{2}}$ with $n_{\mathrm{HD}} / n_{\mathrm{H2}} = D/H$, with the primordial deuterium abundance of $D/H = 1.65\times10^{-5}$ \citep{pettini_new_2001}. 
The cooling rate $\Lambda_{\mathrm{cool,HD}}$ is extrapolated to lower densities to avoid a sharp transition at $\nH = 1 \,\ccm$. As both $n_{\mathrm{HD}}$ and $W_{\mathrm{HD}}$ are lower for lower densities, $\Lambda_{\mathrm{cool,HD}}$ is negligible for these densities. For temperatures $> 10^4\K$, $\Lambda_{\mathrm{cool,HD}}$ is set to zero, as $W_{\mathrm{HD}}$ increases super-linearly with temperature, which can lead to non-negligible HD cooling rates, even if the HD abundance is very low. 

At low redshift the HD cooling rate is only important for very low metallicities and very high densities (it contributes between 10 and 50 per cent of the total cooling rate for $\log Z/\Zsol < -3$ and $\log \nH [\ccm] > 3$). For redshifts $z > 8$ HD cooling dominates for temperatures below the CMB temperature ($T_{\mathrm{CMB}} (z=9) = 27.3 \K$). This is an artefact from the analytic fitting function that does not include a redshift dependence. It only affects the lowest temperature bins at the highest redshifts in the table, but the HD cooling rate should be used with care in this region of parameter space.

\begin{table} 
	\caption{Overview of the models and the included processes. For the column ``dust", an entry ``yes"~means that dust grains with a dust-to-gas ratio according to Eq.~\ref{eq:DGratio} are included and that individual elements are depleted onto dust grains as described by Eq.~\ref{eq:depletion}. Cosmic rays with a rate following Eq.~\ref{eq:CR} are included for runs labelled ``yes"~in the column ``CR".  In addition to the radiation from the CMB and the modified UVB from \citetalias{faucher-giguere_cosmic_2020} (all tables), an interstellar radiation field (Eq.~\ref{eq:radG0}) is added to the runs with ``yes"~in the column ``ISRF". The shielding column density of Eq.~\ref{eq:Nsh} is used for self-shielded gas (``yes"~in column ``shielding"). The fiducial model ``UVB\_dust1\_CR1\_G1\_shield1"~is highlighted in italic. The symbol $\star$ indicates that the corresponding rate/intensity is uncalibrated ($\log R = 0$ in Eqs.~\ref{eq:radG0} and \ref{eq:CR}).}
	\begin{tabular}{lllll}
		Model name 					& dust	& CR			& ISRF		&shielding \\		
		\hline
		\multicolumn{5}{c}{Main tables, results discussed in Sec.~\ref{sec:results}} \\
		\hline
		UVB\_dust1\_CR0\_G0\_shield0	& yes	&	no			& no			& no\\
		UVB\_dust1\_CR0\_G0\_shield1	& yes	&	no			& no			& yes\\
		UVB\_dust1\_CR1\_G1\_shield0	& yes	&	yes			& yes		& no\\
		{\emph{UVB\_dust1\_CR1\_G1\_shield1}}	& {\emph{yes}}	&	{\emph{yes}}			& {\emph{yes}}		& {\emph{yes}}\\
		\hline
		\multicolumn{5}{c}{Additional tables available (not discussed here)} \\
		\hline
		UVB\_dust0\_CR0\_G0\_shield0	& no		&	no			& no			& no\\
		UVB\_dust1\_CR1\_G0\_shield0	& yes	&	yes			& no			& no\\
		UVB\_dust1\_CR1\_G0\_shield1	& yes	&	yes			& no			& yes\\
		UVB\_dust1\_CR2\_G2\_shield0	& yes	&	yes$^{\star}$	& yes$^{\star}$	& no\\
		UVB\_dust1\_CR2\_G2\_shield1	& yes	&	yes$^{\star}$	& yes$^{\star}$	& yes\\
		\hline
	\end{tabular}\label{tab:tables}
\end{table}

\section{Models}\label{sec:models}

The fiducial model that we recommend for use in simulations is ``UVB\_dust1\_CR1\_G1\_shield1"~as it includes all processes discussed in Sec.~\ref{sec:method} (a redshift-dependent metagalactic radiation field, Sec.~\ref{sec:radfield}, as well as a density- and temperature-dependent shielding column density, Sec.~\ref{sec:column}, interstellar radiation field, Sec.~\ref{sec:radfield}, cosmic ray rate, Sec.~\ref{sec:CR}, and dust abundance, Sec.~\ref{sec:metdust}). In this model, both the CR rate as well the ISRF intensity scale with the pressure as expected for a self-gravitating disk following the Kennicutt-Schmidt star formation law, with a normalization that is reduced by 1 dex relative to the \citet{black_heating_1987} value for the Milky Way ($\log R = -1$ in Eqs.~\ref{eq:radG0} and \ref{eq:CR}) in order to match the observed transition from atomic to molecular hydrogen (as will be discussed in Sec.~\ref{sec:phasetransitions}). The unshielded counterpart of this model is ``UVB\_dust1\_CR1\_G1\_shield0"~where all other \textsc{cloudy} inputs, beside the shielding column density, are identical to model ``UVB\_dust1\_CR1\_G1\_shield1".

For comparison of the fiducial model to models without CRs and ISRF, we include ``UVB\_dust1\_CR0\_G0\_shield0"~and  ``UVB\_dust1\_CR0\_G0\_shield1"~the unshielded (``shield0") and self-shielded (``shield1" ) models, where the radiation field only depends on redshift and is the sum of the contributions from the CMB and the modified \citetalias{faucher-giguere_cosmic_2020} UVB (see Fig.~\ref{fig:RF}). The unshielded table ``UVB\_dust1\_CR0\_G0\_shield0"~is conceptually comparable to the approach in \citetalias{wiersma_effect_2009}, and can serve as an update to the \citetalias{wiersma_effect_2009} tables. The main differences are that we use a more recent \textsc{cloudy} version  (here: v17.01, \citetalias{wiersma_effect_2009}: v07.02), a different UV background (here: \citetalias{faucher-giguere_cosmic_2020}, \citetalias{wiersma_effect_2009}: \citealt{haardt_modelling_2001}), and include dust grains for $\log T [K] \lesssim 5$. By construction, all other tables only differ from ``UVB\_dust1\_CR0\_G0\_shield0"~for $\log T [\K] \lesssim 5$. At higher temperatures (and also very low densities) all tables converge to this one, as all additionally included processes are relevant only for neutral and molecular gas. 

Table~\ref{tab:tables} presents an overview of the models for which hdf5 files are provided. In addition to the above mentioned tables whose results will be further discussed in Sec.~\ref{sec:results}, additional tables are listed separately. We do not discuss them in detail here, as we do not recommend them for use in simulations, but they can be used to explore the impact of individual processes (e.g. a model with CRs but without ISRF: ``UVB\_dust1\_CR1\_G0\_shield1"~or a model without dust: ``UVB\_dust0\_CR0\_G0\_shield0"). The model with the full ISRF intensity and CR rate ($\log R = 0$ in Eqs.~\ref{eq:radG0} and \ref{eq:CR}, ``UVB\_dust1\_CR2\_G2\_shield1") is included as it motivates the renormalization of these two quantities. 

For self-shielded models (``shield1") without CRs  the individual \textsc{cloudy} runs can easily become unstable and crash as CRs are the only source of ionization in highly molecular regions. This is known behaviour, it is mentioned clearly in the \textsc{cloudy} user manual that: ``the chemistry network will probably collapse if the gas becomes molecular but cosmic rays are not present".  Grid points for which \textsc{cloudy} crashes (7,223 out of 3,089,636 or 0.2~per~cent for model ``UVB\_dust1\_CR0\_G0\_shield1"~compared to 315 or 0.01~per~cent for model ``UVB\_dust1\_CR1\_G0\_shield1") can be identified with the hdf5 dataset \texttt{GridFails} (see Sec.~\ref{sec:datasets}) while for all other datasets the values for the crashed runs are interpolated in 2D (density, temperature) between their neighbouring grid points with successful runs.

\section{Data products}\label{sec:data}

The data products released with this publication are hdf5 files for the different models listed in Table~\ref{tab:tables}. The main datafile of each model is available as ``$<$modelname$>$.hdf5"~(e.g.~``UVB\_dust1\_CR1\_G1\_shield1.hdf5") and contains both input datasets (e.g. CR rate, radiation field intensity) as well as \textsc{cloudy} results (e.g. cooling and heating rates, ion fractions, $\mathrm{H}_{\mathrm{2}}$ and CO abundances). The content of these files is described in Sec.~\ref{sec:datasets} and listed in Table~\ref{tab:alldatasets}.

For the model ``UVB\_dust1\_CR1\_G1\_shield0"~(unshielded) and the fiducial model ``UVB\_dust1\_CR1\_G1\_shield1"~(self-shielded), additional files are available with emissivities of 183 emission lines (``$<$modelname$>$\_lines.hdf5"). These files are described in Sec.~\ref{sec:lines}.

\begin{figure*} 
	\begin{center}
		\includegraphics*[width=0.9\linewidth, trim = 0.6cm .0cm 1.2cm 0.3cm,clip]{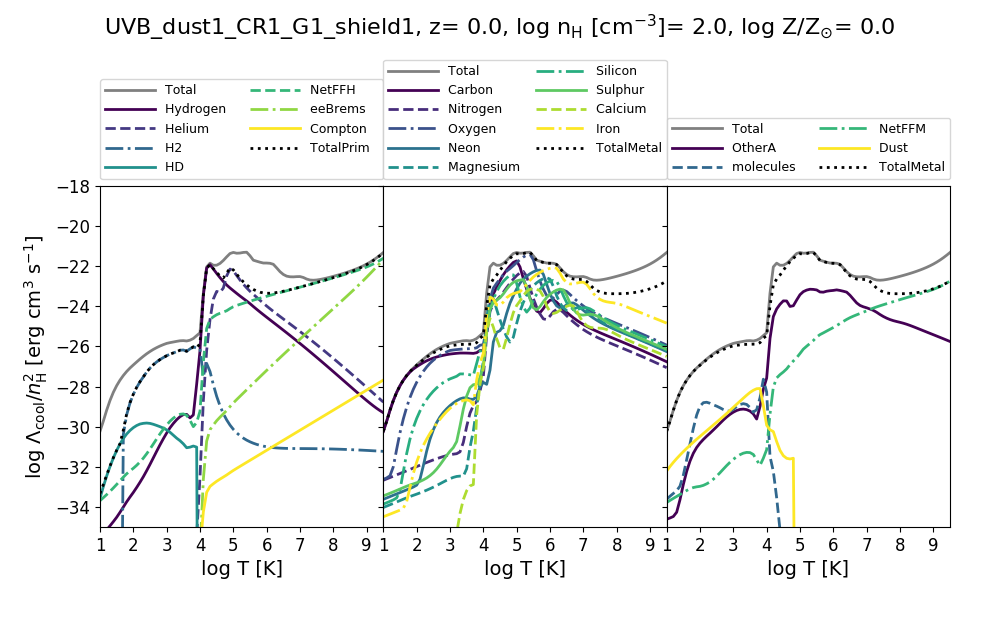}
		\includegraphics*[width=0.9\linewidth, trim = 0.6cm 1.0cm 1.2cm 0.3cm,clip]{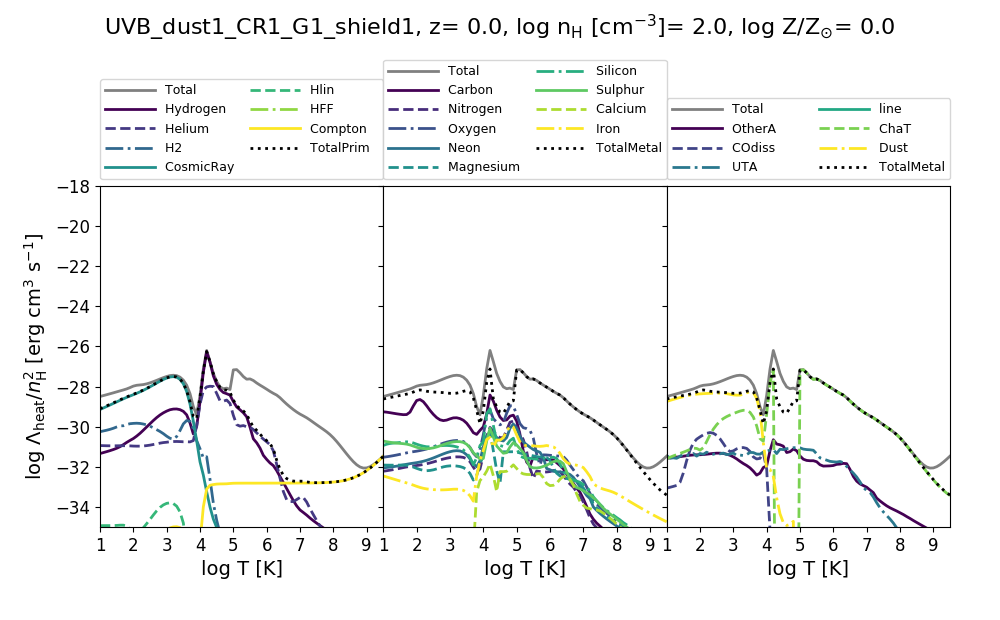}		
		\caption{The top (bottom) panels show an example of the cooling (heating) contributions as listed in Table~\ref{tab:cool} (Table~\ref{tab:heat}) from model UVB\_dust1\_CR1\_G1\_shield1 (see Table~\ref{tab:tables}) for a gas density of $\nH = 100\,\ccm$, solar metallicity and at redshift 0. The components are split into primordial (left panel), atomic metal (middle panel) and the remaining metal processes (right panel). The black dotted line in each panel shows the total rate of the contribution of H and He (left panel) or from metals (middle and right panel) and the solid grey line in each panel indicates the total rate. The individual cooling (heating) processes are explained in Table~\ref{tab:cool} (Table~\ref{tab:heat}). Figure made with provided gui.}
		\label{fig:coolheat}
	\end{center}
\end{figure*}

\subsection{Main hdf5 files: rates and fractions}\label{sec:datasets}

The results for each model listed in Table~\ref{tab:tables} are stored in an hdf5 file that contains both the data as well as documentation. 
An overview of all datasets is given in Table~\ref{tab:alldatasets} and additional information can be found in the attributes \texttt{Dimension}, \texttt{Info} and \texttt{Unit} of each dataset. We discuss a selection of entries that require more explanation below. 

\subsubsection{General structure}\label{sec:tablebins}

The hdf5 group \texttt{/TableBins/} contains the values for each dimension already mentioned in Table~\ref{tab:size} and the number of grid points in each dimension is stored in the hdf5 group \texttt{/NumberOfBins/}. A \textsc{cloudy} run was started for each redshift, temperature, metallicity, and density, which results in $N_z \times N_T \times N_Z \times N_{\nH} = 46 \times 86 \times 11 \times 71 = 3,089,636$ individual \textsc{cloudy} runs per table. 

Many hydrodynamic solvers do not use the gas temperature, $T$, but the internal energy per unit mass, $U$, as their main hydro variable. Hence, to use cooling rates tabulated as a function of temperature, the internal energy first has to be converted into the gas temperature. The relation between $U$ and $T$ is non-linear at phase transitions (mainly $\mathrm{H}_{\mathrm{2}}$ - \ion{H}{I} and \ion{H}{I}  - \ion{H}{II}), where both the mean particle mass $\mu$ and the ratio of specific heats $\gamma$ change rapidly. To improve the usability of the code, we provide all full data hypercuboids (Sec.~\ref{sec:fulldata}) both as a function of temperature (hdf5 group: \texttt{/Tdep/}) and internal energy (hdf5 group: \texttt{/Udep/}). The dimension of the full grid in \texttt{/Udep/} is $N_z \times N_U \times N_Z \times N_{\nH}$ (with the internal energy replacing the temperature dimension) and the internal energy bins can be found in dataset: \texttt{InternalEnergyBins}.  
The hdf5 group \texttt{/ThermEq/} does not include a temperature or internal energy dimension (therefore the dimensions are: $N_z \times N_Z \times N_{\nH}$), as it contains the gas properties at their thermal equilibrium temperature (dataset: \texttt{ThermEq/Temperature}) defined at the last zone of the \textsc{cloudy} calculation. Note that by definition the net cooling rate (cooling - heating) is zero for the thermal equilibrium temperature.
\begin{table*} 
	\caption{Overview of all datasets stored in each hdf5 table. A dataset is labelled with the prefix $^{\mathrm{(in)}}$ if it contains \textsc{cloudy} input, and with $^{\mathrm{(out)}}$ if it contains results of the \textsc{cloudy} calculations. Note that \textsc{cloudy} assumes chemical and ionization equilibrium. Column 2 shows the size and dimensions of each dataset: z (redshift), T (gas temperature), Z (gas metallicity), $\nH$ (gas density), U (gas internal energy), E (individually tabulated elements), C (cooling channels), and H (heating channels). In general, all resulting properties refer to those in the last zone in the \textsc{cloudy} calculation. For hydrogen and CO, average fractions over the full column density are also stored. The suffix ``Col"~refers to the column density fraction and the suffix ``Vol"~refers to the volume density fraction at the last zone. }
	\begin{tabular}{llll}
	\hline
		Dataset name 				& dimension 				& comment\\
	\hline
	\multicolumn{3}{r}{Group: /TableBins/, table dimensions, see also Sec.~\ref{sec:tablebins}} \\
	\hline
            	RedshiftBins    				&       z (46)				& Redshifts for the $z$ dimension \\
            	TemperatureBins         		&       T (86)				& Temperatures for the $\log T [\K]$ dimension (group: /Tdep/) \\		
            	MetallicityBins         			&       Z (11)				& Metallicities for the $\log Z/\Zsol$ dimension \\
            	DensityBins     				&       $\nH$ (71)			& Densities for the $\log \nH [\ccm]$ dimension \\
            	InternalEnergyBins      		&       U (191)				& Internal energies for the $\log U [\mathrm{erg\,g}^{-1}]$ dimension (group: /Udep/)\\
	\hline
	\multicolumn{3}{r}{Additional information, see also Sec.~\ref{sec:additional}} \\
	\hline
            	ElementNamesShort              	&       E (11)				& Short identifier for individually traced elements (e.g. H, He, C,...)\\
            	ElementNames                    	&       E+1					& Names of elements  (incl. one entry for all other atoms)\\
            	ElementMasses                   	&       E+1					& Masses in $u$ ($1u=1.66054\times10^{-24}\,\mathrm{g}$) of elements listed in dataset: \texttt{ElementNames}\\
            	NumberOfIons                   	&       E					& Number of ions for each of the individually traced elements\\
            	AbundanceHe                    	&       Z					& He abundance for each metallicity 	(Eq.~\ref{eq:helium})	\\
            	TotalExactMetallicity           	&       Z					& Exact metal mass fraction, differs slightly from MetallicityBins (see Sec.~\ref{sec:additional})\\
            	TotalAbundances                	&       Z, E+1				& Total (dust + gas) abundances $n_i/\nH$ for each element $i$ at each metallicity		\\
            	TotalMassFractions              	&       Z, E+1				& Total (dust + gas) mass fractions $M_i / M_{\mathrm{tot}}$ for each element $i$ at each metallicity\\	
            	SolarMetallicity                		&       -					& Solar metallicity, $\Zsol$ = 0.0134 \citep{asplund_chemical_2009}\\
            	JMW                             		&       -					& Flux of the ISRF in the solar neighbourhood at 1000 \AA~($1.36\times10^{-3}$ erg cm$^{-2}$ s$^{-1}$) \\	
            	IdentifierCooling               		&       C (22)				& String identifier for 	the different cooling channels\\
            	IdentifierHeating               		&       H (24)				& String identifier for 	the different heating channels\\
	\hline
	\multicolumn{3}{r}{Groups: /Tdep/ ($\star = T$), /Udep/ ($\star = U$), /Thermeq/ ($\star$ dimension does not exist), see also Sec.~\ref{sec:fulldata}}\\
	\hline
	       $^{\mathrm{(in)}}$CosmicRayRate	&       z, ${\star}$, Z, $\nH$ 		& Cosmic ray hydrogen ionization rate $\log \zeta_{\mathrm{CR}} [\mathrm{s}^{-1}]$\\
                $^{\mathrm{(in)}}$DGratio			&       z, ${\star}$, Z, $\nH$ 		& Dust to gas mass ratio	\\
                $^{\mathrm{(in)}}$Depletion		&       z, ${\star}$, Z, $\nH$, E	& For each element the fraction $f_{\mathrm{dust}}$ that is depleted onto dust \\
                $^{\mathrm{(in)}}$RadField		&       z, ${\star}$, Z, $\nH$ 		& Strength of the incident radiation field relative to the MW value (dataset: \texttt{JMW})\\
                $^{\mathrm{(in)}}$ShieldingColumn&       z, ${\star}$, Z, $\nH$ 		& Exact shielding column $\log \NH$ used in the \textsc{cloudy} run\\                
                $^{\mathrm{(in)}}$ShieldingColumnRef&   z, ${\star}$, Z, $\nH$ 		& Reference shielding column $\log N_{\mathrm{ref}}$ (Eq.~\ref{eq:Nnorm})\\
                $^{\mathrm{(out)}}$Cooling                 			&       z, ${\star}$, Z, $\nH$ , C	& Cooling rate $\log \Lambda_{\mathrm{cool}} / \nH^2$ for different cooling channels (see Table~\ref{tab:cool})\\
                $^{\mathrm{(out)}}$Heating                 			&       z, ${\star}$, Z, $\nH$ , H	& Heating rate $\log \Lambda_{\mathrm{heat}}  / \nH^2$ for different heating channels (see Table~\ref{tab:heat})\\           
                $^{\mathrm{(out)}}$AVextend             &       z, ${\star}$, Z, $\nH$ 		& Extended-source extinction,	total visual extinction in mag at 5500 \AA\\
                $^{\mathrm{(out)}}$AVpoint                &       z, ${\star}$, Z, $\nH$ 		& Point-source extinction, total	visual extinction in mag at 5500 \AA \\
                $^{\mathrm{(out)}}$COFractionCol    	&       z, ${\star}$, Z, $\nH$ 		& CO column density fraction $\log N_{\mathrm{CO}}/\NH$\\
                $^{\mathrm{(out)}}$COFractionVol	&       z, ${\star}$, Z, $\nH$ 		& CO volume density fraction $\log n_{\mathrm{CO}}/\nH$		\\
                $^{\mathrm{(out)}}$ColumnDensitiesC&       z, ${\star}$, Z, $\nH$ , 3	& Selected carbon column densities: $\log N_{\ion{C}{I}}$, $\log N_{\ion{C}{II}}$, $\log N_{\mathrm{CO}}$		\\
                $^{\mathrm{(out)}}$ColumnDensitiesH &       z, ${\star}$, Z, $\nH$ , 3	& Selected hydrogen column densities: $\log N_{\ion{H}{I}}$, $\log N_{\ion{H}{II}}$, $\log N_{\mathrm{H2}}$		\\
                $^{\mathrm{(out)}}$ElectronFractions    		&       z, ${\star}$, Z, $\nH$ , E + 3& Free electron fraction $\log n_{\mathrm{e}}/\nH$ from each individual element + prim, metal, total	\\
                $^{\mathrm{(out)}}$GammaHeat               		&       z, ${\star}$, Z, $\nH$		& Ratio of specific heats $\gamma$, varies between 7/5 and 5/3 \\
                $^{\mathrm{(out)}}$GridFails               			&       z, ${\star}$, Z, $\nH$ 		& 0 if \textsc{cloudy} finished successfully, 1 if warnings or errors were present\\
                $^{\mathrm{(out)}}$HydrogenFractionsCol    		&       z, ${\star}$, Z, $\nH$, 3	& Column density H mass fractions: $\log N_{\ion{H}{I}}/\NH$, $\log N_{\ion{H}{II}}/\NH$, $\log 2N_{\mathrm{H2}}/\NH$\\
                $^{\mathrm{(out)}}$HydrogenFractionsVol   	 	&       z, ${\star}$, Z,$\nH$, 3	& Volume density H mass fractions: $\log n_{\ion{H}{I}}/\nH$, $\log n_{\ion{H}{II}}/\nH$, $\log 2n_{\mathrm{H2}}/\nH$\\
                $^{\mathrm{(out)}}$HydrogenFractionsVolExtended	&       z, ${\star}$, Z, $\nH$, 6	& As HydrogenFractionsVol but also: $\log 2n_{\mathrm{H2^+}}/\nH$, $\log 3n_{\mathrm{H3^+}}/\nH$, $\log n_{\mathrm{H^-}}/\nH$\\
                $^{\mathrm{(out)}}$IonFractions/00hydrogen      	&       z, ${\star}$, Z, $\nH$, 2	& Fractions of hydrogen atoms in each ionization state (gas-phase, see Eq.~\ref{eq:iongas})\\
                $^{\mathrm{(out)}}$IonFractions/01helium        	&       z, ${\star}$, Z, $\nH$, 3	& As 	IonFractionsVol/00hydrogen, but for helium \\
                $^{\mathrm{(out)}}$IonFractions/02carbon        	&       z, ${\star}$, Z, $\nH$, 7	& As 	IonFractionsVol/00hydrogen, but for carbon\\
                $^{\mathrm{(out)}}$IonFractions/03nitrogen      	&       z, ${\star}$, Z, $\nH$, 8	& As 	IonFractionsVol/00hydrogen, but for nitrogen\\
                $^{\mathrm{(out)}}$IonFractions/04oxygen        	&       z, ${\star}$, Z, $\nH$, 9	& As 	IonFractionsVol/00hydrogen, but for oxygen\\
                $^{\mathrm{(out)}}$IonFractions/05neon  		&       z, ${\star}$, Z, $\nH$, 11	& As 	IonFractionsVol/00hydrogen, but for neon\\
                $^{\mathrm{(out)}}$IonFractions/06magnesium  	&       z, ${\star}$, Z, $\nH$, 13	& As 	IonFractionsVol/00hydrogen, but for magnesium\\
                $^{\mathrm{(out)}}$IonFractions/07silicon       	&       z, ${\star}$, Z, $\nH$, 15	& As 	IonFractionsVol/00hydrogen, but for silicon\\
                $^{\mathrm{(out)}}$IonFractions/08sulphur      	&       z, ${\star}$, Z, $\nH$, 17	& As 	IonFractionsVol/00hydrogen, but for sulphur\\
                $^{\mathrm{(out)}}$IonFractions/09calcium      	&       z, ${\star}$, Z, $\nH$, 21	& As 	IonFractionsVol/00hydrogen, but for calcium\\
                $^{\mathrm{(out)}}$IonFractions/10iron  		&       z, ${\star}$, Z, $\nH$, 27	& As 	IonFractionsVol/00hydrogen, but for iron\\
                $^{\mathrm{(out)}}$MeanParticleMass        		&       z, ${\star}$, Z, $\nH$ 		& Gas-phase mean particle mass $\mu$ in units of atomic mass unit $u$ \\
	\hline
            	$^{\mathrm{(out)}}$Tdep/U\_from\_T                   	&       z, T, Z, $\nH$			& Internal energy $\log U [\mathrm{erg\,g}^{-1}]$ for group /Tdep/\\
            	$^{\mathrm{(out)}}$ThermEq/U\_from\_T          		&       z, Z, $\nH$			& Thermal equilibrium internal energy $\log U [\mathrm{erg\,g}^{-1}]$ for group /Thermeq/\\
            	$^{\mathrm{(out)}}$Udep/T\_from\_U                   	&       z, U, Z, $\nH$			& Temperature $\log T [\K]$ for group /Udep/ \\
		$^{\mathrm{(out)}}$ThermEq/Temperature 		&	z, Z, $\nH$			& Thermal equilibrium temperature\\
	\hline
	\end{tabular}\label{tab:alldatasets}
\end{table*}

\subsubsection{Miscellaneous information}\label{sec:additional}
We select the 11 elements that contribute most to the thermal state of the gas (i.e. the dominant cooling contributions) and tabulate their properties individually and in more detail (listed in Table~\ref{tab:solarabundances} and in dataset \texttt{ElementNames}). 
The remaining elements are combined into one entry where this is relevant (e.g. the cooling channel ``OtherA", see Table~\ref{tab:cool}). Several datasets contain information about these elements (i.e. their names, masses and number of ions) and their abundances.

As mentioned in Sec.~\ref{sec:metdust} (Eq.~\ref{eq:helium}), the helium abundance varies with metallicity to yield a smooth transition between the solar and primordial values. 
The values used for $n_{\mathrm{He}}/\nH$ at each metallicity are stored in dataset \texttt{AbundanceHe}. The extra scaling of helium changes the helium mass Y which results in a 
slightly different metal mass fraction $Z = Z / (X + Y + Z)$ compared to the values from the dataset \texttt{MetallicityBins}, which were used to scale the metal abundances independently.
The difference is very small ($<$ 0.03 dex), but for reference the exact metallicity values are stored in \texttt{TotalExactMetallicity}. This array can be used instead of \texttt{MetallicityBins},
but has the practical disadvantage that it is not exactly uniformly spaced in $\log Z$.

The dataset \texttt{ElementMasses} $m_i$ is added to allow the conversion between element abundances ($n_i/\nH$) and mass fractions, $M_i/M_{\mathrm{tot}} = (n_i/\nH) \times m_i / [\sum_{i} (n_i/\nH) \times m_i]$. 
The last entry of \texttt{ElementMasses} is the average atomic mass of the remaining elements for solar abundances from \citet{asplund_chemical_2009}.

All metallicities are tabulated relative to the solar metallicity $\Zsol$ and can be converted to absolute metallicities by multiplying them with the entry from the dataset \texttt{SolarMetallicity}. Similarly, the radiation field dataset \texttt{RadField} is normalised to the local value $J_{0}$, which is stored in the dataset \texttt{JMW}. 

The datasets \texttt{IdentifierCooling} and \texttt{IdentifierHeating} contain the labels listed in columns 2 of Tables~\ref{tab:cool} and \ref{tab:heat}. They correspond to the final dimension of the datasets \texttt{Cooling} and \texttt{Heating} (see Sec.~\ref{sec:fulldata}).

\subsubsection{Full data hypercuboid}\label{sec:fulldata}

\begin{figure*} 
	\begin{center}
		\includegraphics*[width=\linewidth, trim = 0.6cm 0.2cm 1.2cm 0.3cm,clip]{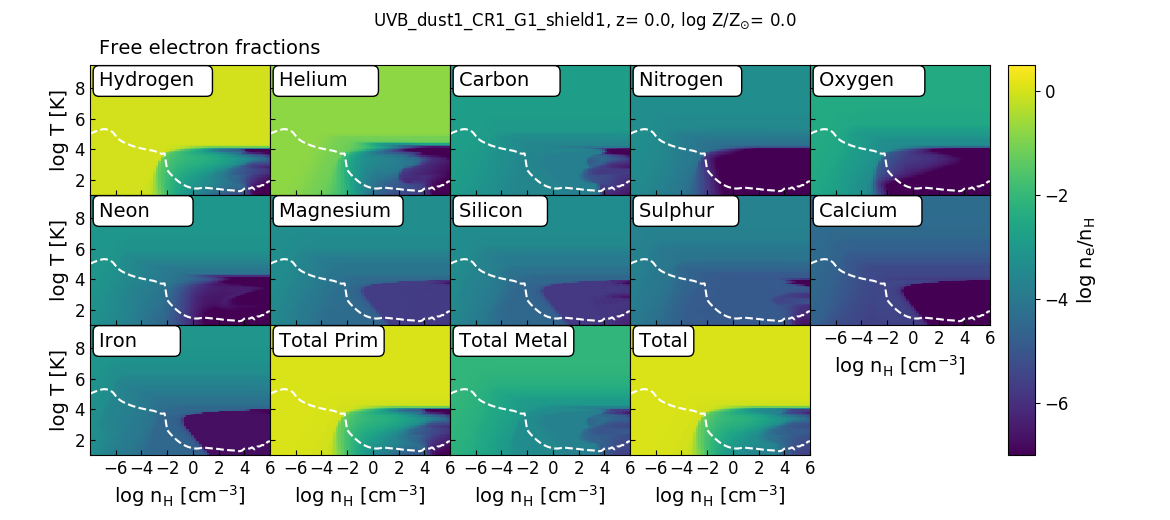}	
		\caption{An example of the dataset \texttt{ElectronFractions} for table UVB\_dust1\_CR1\_G1\_shield1 for solar metallicity gas at redshift~0. The colour coding shows the free electron fraction $n_{\mathrm{e}}/\nH$ in the last zone of the shielding column split up into contributions from individual elements, as well as the total contributions from hydrogen and helium (including electrons from hydrogen molecules, \texttt{TotalPrim}), from metals (\texttt{TotalMetal}) and the total electron fraction (including electrons from all molecules, \texttt{Total}). Figure made with provided gui.} 
		\label{fig:elecfrac}
	\end{center}
\end{figure*}

There are three big groups in each hdf5 file that contain the same datasets but are tabulated in slightly different ways. In group \texttt{/Tdep/} each grid point corresponds directly to a \textsc{cloudy} run with redshift, temperature, metallicity and density as inputs that remain constant throughout each 1D \textsc{cloudy} simulation. 
As already mentioned in Sec.~\ref{sec:tablebins}, some hydrodynamic codes use the internal energy as the thermal variable instead of the gas temperature. 
For this case, using the datasets from group \texttt{/Udep/} saves having to convert from $T$ to $U$, as the datasets now use the internal energy instead of the temperature dimension.

To create the datasets in group \texttt{/Udep/}, we first calculated the internal energy with

\begin{align}\label{eq:U}
	U [\mathrm{erg\, g}^{-1}] =&\frac{1}{\gamma - 1} \frac{k_B T }{\mu m_{u}}\\
		&\mathrm{dataset:} \, \texttt{Tdep/U\_from\_T} \nonumber		
\end{align}

\noindent
with the Boltzmann constant $k_B = 1.3806\times10^{-16}\,\mathrm{erg\,K}^{-1}$ and the atomic mass unit $m_u = 1.6605\times10^{-24}\, \mathrm{g}$.
Both the ratio of specific heats $\gamma$ and the mean particle mass $\mu$ vary within the table range. For $\mu$ we use the values from the dataset \texttt{MeanParticleMass}, which is a direct \textsc{cloudy} output. We approximate $\gamma$ by assuming it is dominated by the contributions from hydrogen and helium as

\begin{align}
	\gamma = &\frac{ 5 \left ( n_{\ion{H}{I}} +  n_{\ion{H}{II}} + n_{\mathrm{He}} + n_{\mathrm{e}} \right ) + 7 n_{\mathrm{H2}} }{3 \left ( n_{\ion{H}{I}} +  n_{\ion{H}{II}} + n_{\mathrm{He}} + n_{\mathrm{e}} \right ) + 5 n_{\mathrm{H2}} }\\
			&\mathrm{dataset:} \, \texttt{GammaHeat} \nonumber		
\end{align}

\noindent
assuring a smooth transition between $\gamma = 5/3$ for monoatomic gas and $\gamma = 7/5$ for molecular hydrogen (as done in the \textsc{krome} package, \citealp{grassi_krome_2014}).
In a second step, the properties of all datasets are interpolated onto fine, uniformly spaced bins in internal energy (dataset: \texttt{TableBins/InternalEnergyBins}). 

The large phase-space covered by the tables is not uniformly populated by the gas resolution elements in a simulation. Typically gas will spend a disproportionally large fraction of its time close to thermal equilibrium. For all datasets, the group \texttt{/ThermEq/} contains the properties at the thermal equilibrium temperature at the last \textsc{cloudy} zone (dataset: \texttt{/ThermEq/Temperature}). Occasionally, there are multiple equilibrium temperatures for a given density (at constant $z$ and $Z$). This occurs for example if an individual cooling contribution (e.g. molecular hydrogen) peaks around a given temperature while the heating rate stays roughly constant, which results in two thermal equilibrium tracks (at temperatures below and above the cooling peak). In this case (usually only for a narrow range in densities) we select the track that changes the equilibrium temperature the least, if the gas is moving to higher densities. The regions with multiple thermal equilibrium temperatures can be easily identified in a temperature - density plot of the net cooling rate (e.g with the provided gui). The minimum temperature in all tables is 10 K. If the cooling rate exceeds the heating rate at 10 K, we set $T_{\mathrm{eq}} = 10\,\K$.

\paragraph*{Input datasets:} Some of the datasets in Table~\ref{tab:alldatasets} have the prefix $^{\mathrm{(in)}}$. These are properties that are not calculated by \textsc{cloudy} but are used as inputs. As they vary over different dimensions, we output their values during runtime as an independent check to verify that the scaling is implemented correctly.

\paragraph*{Cooling and heating:}

\textsc{cloudy} follows a large number of different cooling and heating channels. We combined them in the 20 cooling and 22 heating groups listed in Tables~\ref{tab:cool} and \ref{tab:heat}. In addition, the last two entries in this dimension are the total rates from processes including hydrogen and helium (labelled as: TotalPrim) and from processes including metals (TotalMetal). This allows one to calculate the rates faster for solar relative abundances by adding only the last two entries (for more details on how to use the tables, see Appendix~\ref{sec:howto}). 

The heating and cooling labels from \textsc{cloudy} are listed in column 4 of Tables~\ref{tab:cool} and \ref{tab:heat} and we refer to the \textsc{cloudy} documentation for more details on the individual processes. An example of the cooling and heating rates can be found in Fig.~\ref{fig:coolheat}.

\paragraph*{Vol and Col:} 
A \textsc{cloudy} calculation that includes self-shielding splits the gas shielding column into a varying number of zones with adaptive sizes.  Each output property (labelled with $^{\mathrm{(out)}}$ in Table~\ref{tab:alldatasets}) is defined at each zone throughout the gas column. In this work, every tabulated property refers to its value calculated at the last zone of the \textsc{cloudy} simulation. For self-shielded gas the last zone is the zone at the shielding column density and for unshielded gas the last zone is the only zone in the calculation. 
 
If a phase transition occurs just before the last zone, the species fraction in the last zone can be very different from the species fraction integrated over the full shielding column. For the hydrogen and CO fractions we therefore store both values: $n_{\mathrm{x}}/\nH$ (suffix: ``Vol") at the last zone as well as $N_{\mathrm{x}}/\NH$ (suffix: ``Col") for the full shielding column. 

In simulations, the shielding length $l_{\mathrm{sh}} = N_{\mathrm{sh}} / \nH$ can be compared to the length scale of the resolution element $l_{\mathrm{sim}}$ to decide whether $n_{\mathrm{x}}/\nH$ (for $l_{\mathrm{sim}} > l_{\mathrm{sh}}$) or $N_{\mathrm{x}}/\NH$ (for $l_{\mathrm{sim}} < l_{\mathrm{sh}}$) is a better approximation. $n_{\mathrm{x}}/\nH$ will only differ significantly from $N_{\mathrm{x}}/\NH$ in small parts of the full table parameter space. For example for UVB\_dust1\_CR1\_G1\_shield1 $n_{\mathrm{H2}}/\nH$ ($n_{\ion{H}{I}}/\nH$) differs by more than a factor of 2 from $N_{\mathrm{H2}}/\NH$ ($N_{\ion{H}{I}}/\NH$)  in less than 10 (1) per cent of the 3,089,636 entries. 
 
\paragraph*{Free electron fractions:}

This dataset contains the number of free electrons per hydrogen atom $n_{\mathrm{e,X}}/\nH$ for each individually traced element $X$. 
The number of electrons per gas phase atom of element $X$ is calculated from the fractions of atoms in ion stage $i$ with         
   
\begin{align}       
	\frac{n_{\mathrm{e,X,gas}}}{n_{\mathrm{X,gas}}} = &\sum_{i=1}^{N_{\mathrm{ion,X}}} (i-1) \frac{n_{X,i}}{n_X} 
\end{align}

\noindent
 (e.g. for He: $n_{\mathrm{e,He,gas}}/ n_{\mathrm{He,gas}}= n_{\ion{He}{II}}/n_{\mathrm{He}} + 2n_{\ion{He}{III}}/n_{\mathrm{He}}$). Accounting for the total abundance of element $X$ and depletion onto dust (for tables that include grains), the number of electrons per hydrogen atom from element $X$ is

\begin{align}
         \frac{n_{\mathrm{e,X}}}{\nH} = & \frac{n_{\mathrm{e,X,gas}}}{n_{\mathrm{X,gas}}}  \underbrace{\frac{n_{\mathrm{X,gas}}}{n_{\mathrm{X}}}}_{\mathrm{1 -  f_{\mathrm{dust,X}}}}  \frac{n_{\mathrm{X}}}{n_{\mathrm{H}}}          
\end{align}

\noindent
where $n_{\mathrm{X,gas}}$ is the gas phase number density and $n_{\mathrm{X}}$ the total (gas + dust) number density of element $X$. The contributions from ionized molecules are not included here, but for metals these are negligible. The last three entries in the element dimension are the total primordial electron fractions (including species $n_{\mathrm{H2}^+}$, $n_{\mathrm{H3}^+}$ and $n_{\mathrm{H}^-}$), the total fractions of free electrons from metals $n_{\mathrm{e,met}}/\nH$ without molecules, and $n_{\mathrm{e,tot}}/\nH$ from all atoms and molecules included in \textsc{cloudy}. Fig.~\ref{fig:elecfrac} illustrates the individual entries for solar metallicity gas at redshift 0 (using table UVB\_dust1\_CR1\_G1\_shield1). 

Splitting the electron fractions into individual elements allows users to combine these tables with a chemical network that follows the non-equilibrium evolution of a subset of elements (e.g. H and He). The free electrons from metals in self-shielded gas can be added to the network individually (see Appendix~\ref{sec:noneq}).

\paragraph*{Ion fractions:}
An individual dataset containing the ion fractions of each element listed in Table~\ref{tab:solarabundances} is included in every hdf5 file. The size of the last dimension in each ion fraction dataset (see Table~\ref{tab:alldatasets}) is the number of ions, given in the additional dataset \texttt{NumberOfIons}. For each element the ion fractions are defined as the number of atoms in each ionization state relative to the total number of atoms of this element in the gas phase (e.g.~$n_{\ion{H}{I}}/\nH$). The sum of all ion fractions for an element does not add up to one if a non-negligible fraction of atoms are in molecules.

\begin{figure} 
	\begin{center}
		\includegraphics*[width=\linewidth, trim = 0cm 0.7cm 0cm 0cm,clip]{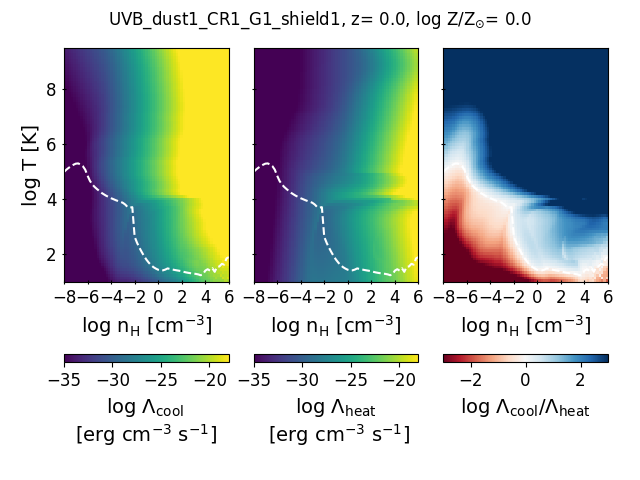}	
		\caption{Cooling ($\Lambda_{\mathrm{cool}}$, left panel) and heating ($\Lambda_{\mathrm{heat}}$, middle panel) rate, as well as their ratio ($\Lambda_{\mathrm{cool}}/\Lambda_{\mathrm{heat}}$, right panel) for solar metallicity gas at $z=0$ from table UVB\_dust1\_CR1\_G1\_shield1. The white dashed line in each panel indicates the thermal equilibrium temperature (i.e. $\Lambda_{\mathrm{cool}}/\Lambda_{\mathrm{heat}} =1$). Figure made with provided gui.}
		\label{fig:netcool2d}
	\end{center}
\end{figure}

\subsection{Line emissivity hdf5 files:}\label{sec:lines}

For tables UVB\_dust1\_CR1\_G1\_shield0 and UVB\_dust1\_CR1\_G1\_shield1 we provide additional information on the emissivity of selected lines (see Table~\ref{tab:lines} and dataset: \texttt{IdentifierLines}) in a separate hdf5 file (``$<$modelname$>$\_lines.hdf5"). The general file structure is the same as in the hdf5 files that contain the rates and species fractions (Sec.~\ref{sec:datasets}) as the emissivities are calculated for the same grid in redshift, metallicity, density and temperature. 

For each line the hdf5 file contains the line emissivity in units of $\mathrm{erg}\, \mathrm{cm}^{-3}\, \mathrm{s}^{-1}$ that is leaving the illuminated side of the cloud (using the \textsc{cloudy} keyword ``emergent"). The luminosity of a gas particle (or gas cell) in a simulation can be calculated with 

\begin{equation}
	L [\mathrm{erg}\,\mathrm{s}^{-1}] = \epsilon \frac{m_{\mathrm{gas}}}{\rho_{\mathrm{gas}}} \, ,
\end{equation}

\noindent
following e.g. \citet{bertone_metal-line_2010}, where $m_{\mathrm{gas}}$ is the mass of the resolution element, $\rho_{\mathrm{gas}}$ is the gas density and $\epsilon$ is either the emissivity at the last zone ($\epsilon_{\mathrm{Vol}}$, dataset: \texttt{EmissivitiesVol}) or the average emissivity of the shielding column ($\epsilon_{\mathrm{Col}}$, dataset: \texttt{EmissivitiesCol"}). The emergent emissivities of the last zone ($\epsilon_{\mathrm{Vol}}$) are calculated by \textsc{cloudy} with the command ``\texttt{save lines emissivity}"~and the column averaged emissivity is calculated by dividing the emergent line intensity of the full shielding column (``\texttt{save line list absolute}") by the length of the shielding column.
Analogous to the ``Vol"~and ``Col"~fractions from the main hdf5 file (discussed in Sec.~\ref{sec:fulldata}), comparing the shielding length to the length scale of the resolution element can help decide which emissivity to choose.

\begin{table*} 
	\caption{Line identifiers and wavelengths for the lines in the line emission files (see Sec.~\ref{sec:lines}). The wavelengths are in units of \AA~(A), cm (c), and $\mu$m (m). ``Blnd"~refers to a blended line with multiple components and is listed directly after its main component(s). The line with the \textsc{cloudy} identifier ``Cool"~at 1215\AA~is the contribution of collisional excitation to Ly$\alpha$ cooling. }
	\begin{tabular}{llllllllll}
	 	  Line 	& Wavelength & Line & Wavelength & Line & Wavelength & Line & Wavelength  & Line & Wavelength  \\
		\hline
H  1    &1025.72A       &C  5   &40.2678A       &O 3C   &4363.00A       &Mg 9   &62.7511A       &Fe16   &63.7110A\\
H  1    &1215.67A       &C  5   &41.4721A       &Blnd   &4363.00A       &Mg 9   &72.0276A       &Fe16   &66.3570A\\
Cool    &1215.67A       &C  6   &33.7372A       &O  3   &51.8004m       &Mg10   &63.2953A       &Fe17   &17.0510A\\
H  1    &21.1207c       &N  1   &1200.00A       &O  3   &88.3323m       &Mg11   &9.16875A       &Fe17   &15.2620A \\
H  1    &4861.33A       &N  1   &1200.22A       &O  4   &25.8832m       &Mg12   &8.42141A       &Fe17   &16.7760A \\
H  1    &6562.81A       &N  2   &1084.56A       &O  5   &1218.34A       &Si 2   &1816.93A       &Fe17   &17.0960A \\
H  1    &4340.46A       &N  2   &1084.58A       &Blnd   &1218.00A       &Si 2   &1817.45A       &Fe18   &16.0720A\\
H  1    &4101.73A       &N  2   &1085.10A       &O  6   &1031.91A       &Si 2   &34.8046m       &Fe18   &93.9322A \\
H  1    &920.963A       &Blnd   &1085.00A       &O  6   &1037.62A       &Si 3   &1206.50A       &Fe22   &11.7695A\\
H  1    &923.150A       &N  2   &121.767m       &Blnd   &1035.00A       &Si 3   &1882.71A       &Fe23   &11.7370A \\
H  1    &926.226A       &N  2   &205.244m       &O  7   &21.6020A       &Blnd   &1888.00A       &Fe25   &1.85040A \\
H  1    &930.748A       &N  2   &6583.45A       &O  7   &21.8044A       &Si 4   &1393.75A       &Fe26   &1.78177A \\
H  1    &937.804A       &N  3   &57.3238m       &O  7   &21.8070A       &Si 4   &1402.77A       &CO     &1300.05m\\
H  1    &949.743A       &N  3   &991.000A       &O  7   &22.1012A       &Blnd   &1397.00A       &CO     &260.169m\\
H  1    &972.537A       &N  4   &1486.50A       &O  7   &18.6270A       &Si11   &43.7501A       &CO     &2600.05m\\
He 2    &1640.43A       &Blnd   &1486.00A       &O  8   &16.0067A       &Si11   &52.2913A       &CO     &289.041m\\
He 2    &303.784A       &N  5   &1238.82A       &O  8   &18.9709A       &Si12   &40.9111A       &CO     &325.137m\\
C  1    &1561.34A       &N  5   &1242.80A       &Ne 2   &12.8101m       &Si12   &44.1650A       &CO     &371.549m\\
C  1    &1561.37A       &Blnd   &1240.00A       &Ne 3   &15.5509m       &Si13   &6.64803A       &CO     &433.438m\\
C  1    &1561.44A       &N  6   &29.5343A       &Ne 3   &3868.76A       &Si14   &6.18452A       &CO     &520.089m\\
C  1    &1656.27A       &N  6   &28.7870A       &Ne 3   &3967.47A       &S  2   &4068.60A       &CO     &650.074m\\
C  1    &370.269m       &N  7   &24.7807A       &Ne 4   &2424.28A       &Blnd   &4074.00A       &CO     &866.727m\\
C  1    &609.590m       &O  1   &145.495m       &Blnd   &2424.00A       &S  3   &18.7078m       &H2     &12.2752m\\
C  2    &1335.00A       &O  1   &5577.34A       &Ne 5   &14.3228m       &S  3   &33.4704m       &H2     &17.0300m\\
Blnd    &1335.00A       &O  1   &63.1679m       &Ne 7   &97.4960A       &S  4   &10.5076m       &H2     &28.2106m\\
C  2    &157.636m       &O  1   &6300.30A       &Ne 8   &88.0817A       &S  5   &1199.14A       &H2O    &269.199m\\
C  2    &2328.12A       &O  1   &6363.78A       &Ne 8   &98.1156A       &Blnd   &1199.00A       &H2O    &303.374m\\
Blnd    &2326.00A       &Blnd   &6300.00A       &Ne 8   &98.2601A       &S  6   &933.380A       &H2O    &398.534m\\
C  3    &1906.68A       &O  2   &3726.03A       &Ne 8   &770.410A       &S  7   &72.0290A       &HCO+   &480.147m\\
C  3    &1908.73A       &Blnd   &3726.00A       &Blnd   &774.000A       &S  7   &72.6640A       &HCO+   &560.140m\\
Blnd    &1909.00A       &O  2   &3728.81A       &Ne 9   &13.4471A       &S 15   &5.03873A       &HCO+   &840.150m\\
C  3    &977.000A       &Blnd   &3729.00A       &Ne 9   &13.6987A       &Ar 2   &6.98337m       &HNC    &661.221m\\
C  3    &977.020A       &O  3   &4958.91A       &Ne10   &12.1375A       &Ar 9   &49.1850A       &HNC    &826.492m\\
Blnd    &977.000A       &O  3   &5006.84A       &Mg 1   &2852.13A       &Ca 1   &4226.73A       &OH     &119.202m\\
C  4    &1548.19A       &Blnd   &5007.00A       &Mg 2   &2795.53A       &Fe 2   &2399.24A       &OH     &119.410m\\
C  4    &1550.78A       &O  3   &4363.21A       &Mg 2   &2802.71A       &Fe 2   &25.9811m       &       &\\
Blnd    &1549.00A       &O 3R   &4363.00A       &Blnd   &2798.00A       &Fe 5   &3891.28A       &       &\\
		\hline
	\end{tabular}\label{tab:lines}
\end{table*}

\section{Example results: Thermal equilibrium}\label{sec:results}

The full tables can be explored with the help of the provided gui. Here, we focus on the properties of gas close to the thermal equilibrium temperature at the last zone of the gas column.

\begin{figure} 
	\begin{center}
		\includegraphics*[width=\linewidth, trim = 0cm 0cm 0cm 0cm,clip]{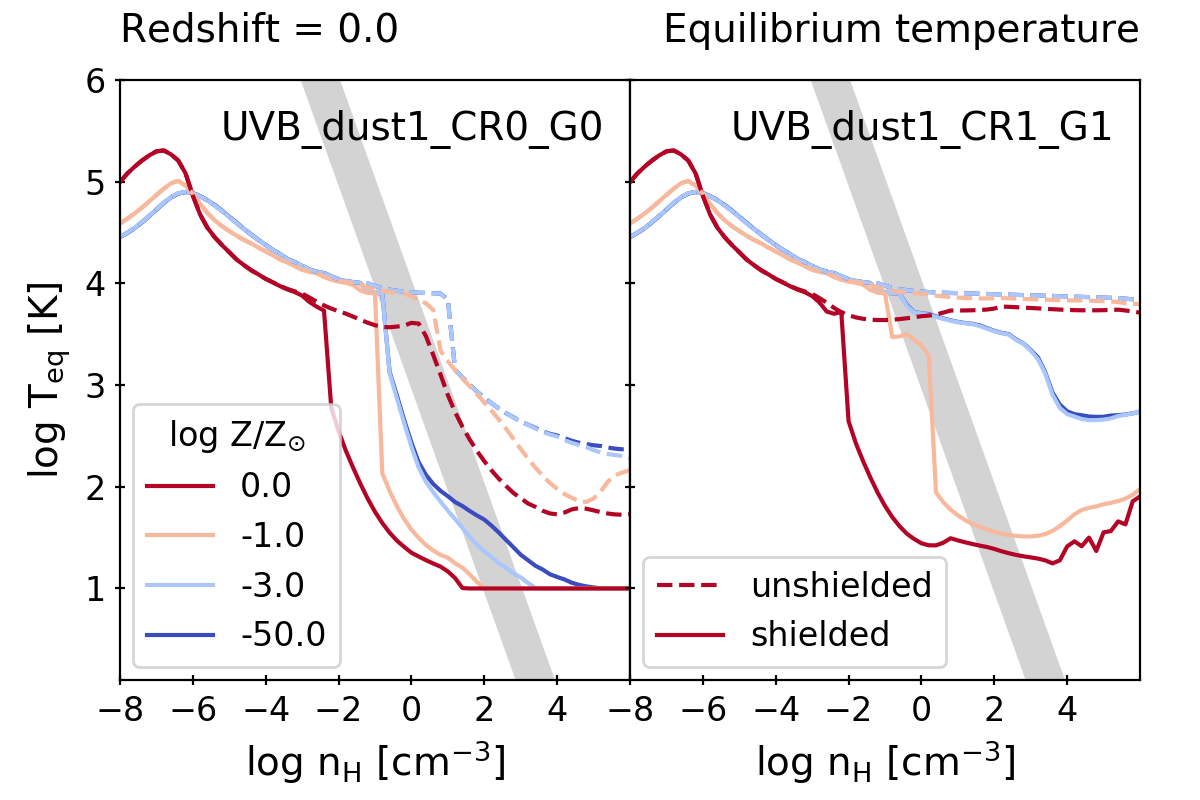}	
		\caption{Thermal equilibrium temperatures for two different radiation fields, only CMB and UVB (UVB\_dust1\_CR0\_G0, left panel) and including CRs and ISRF (UVB\_dust1\_CR1\_G1, right panel) at redshift 0. The colours in each panel refer to different gas metallicities and the line styles indicate if the thermal equilibrium temperature is for unshielded (dashed lines, ``\_shield0") or self-shielded (solid lines, ``\_shield1") gas. The shaded area indicates ISM pressures of $3< \log P/k_B [\K \ccm] <4$ as observed in the solar neighbourhood by \citet{jenkins_distribution_2011}. }
		\label{fig:Teqoverview}
	\end{center}
\end{figure}

An example for a density-temperature slice through the dataset hypercuboid for solar metallicity and $z=0$ is shown in Fig.~\ref{fig:netcool2d}. The plotted properties are the cooling rate ($\Lambda_{\mathrm{cool}}$, left panel), heating rate ($\Lambda_{\mathrm{heat}}$, middle panel), as well as the ratio between the two ($\Lambda_{\mathrm{cool}}/\Lambda_{\mathrm{heat}}$, right panel). The thermal equilibrium temperature, $T_{\mathrm{eq}}$, is defined as the temperature where $\Lambda_{\mathrm{cool}}/\Lambda_{\mathrm{heat}} =1$\footnote{In a few cases, multiple thermal equilibrium temperatures exist, see Sec. \ref{sec:fulldata} for details.} (dashed, white line). This temperature is stored in dataset \texttt{/ThermEq/Temperature} and all properties in the group \texttt{/ThermEq/} correspond to the relevant (for each $z$ and metallicity bin) thermal equilibrium temperatures.

$T_{\mathrm{eq}}$ typically serves as a lower temperature limit for ISM gas. Gas can have higher temperatures due to hydrodynamic processes (e.g. shocks) but temperatures below $T_{\mathrm{eq}}$ are rare as the heating rate typically dominates adiabatic losses at ISM densities. For densities around and below the cosmic mean\footnote{$\log \overline{\nH} \approx 1.7 \log (1+z) -6.2$ for $z\le1.3$ and $\log \overline{\nH} \approx 3.0 \log (1+z) -6.7$ for $z>1.3$, using \citealp{ade_planck_2016} cosmology} $T_{\mathrm{eq}}$ is not a good proxy for the actual \mbox{(quasi-)equilibrium} temperature because here adiabatic cooling due to the Hubble expansion typically dominates over radiative cooling. More generally, when the radiative time scales exceed the Hubble time, the gas will not be able to achieve thermal equilibrium. 

Fig.~\ref{fig:Teqoverview} displays an overview of the thermal equilibrium temperatures at $z=0$ for a selection of metallicities. Each panel shows two tables whose names start with the labels indicated in each panel. The dashed lines represent $T_{\mathrm{eq}}$ from the unshielded version (``\_shield0"), while the solid lines are for $T_{\mathrm{eq}}$ including self-shielding (``\_shield1"). For tables without an ISRF (left panel) even unshielded gas can cool down to temperatures below $10^4\,\K$ as soon as it becomes dense enough ($\nH \gtrsim 1\ccm$). Including self-shielding moves the density of the transition from the warm ($T \approx 10^4\,\K$) to the cold ($T \ll 10^4\,\K$) gas phase to lower densities, depending on the gas metallicity.

Most of the ISM gas in the solar neighbourhood has thermal pressures of $3< \log P/k_{\mathrm{B}} [\K \ccm] <4$ \mbox{\citep{jenkins_distribution_2011}}.
The shaded areas in Fig.~\ref{fig:Teqoverview} indicate this range of pressures. For solar metallicity, the minimum equilibrium temperature of the cold phase is $\approx 30\,\K$ in table UVB\_dust1\_CR1\_G1 while the warm phase (unshielded gas) has an equilibrium temperature of $\approx 10^4\,\K$. For tables including the reduced ISRF (right panel of Fig.~\ref{fig:Teqoverview}), a multi-phase medium can be established between the unshielded and the self-shielded gas for high metallicity ($Z \ge 0.1\,\Zsol$) gas over a large range of densities. 

\begin{figure*} 
	\begin{center}
		\includegraphics*[width=0.48\linewidth, trim = 0cm 0cm 0cm 0cm,clip]{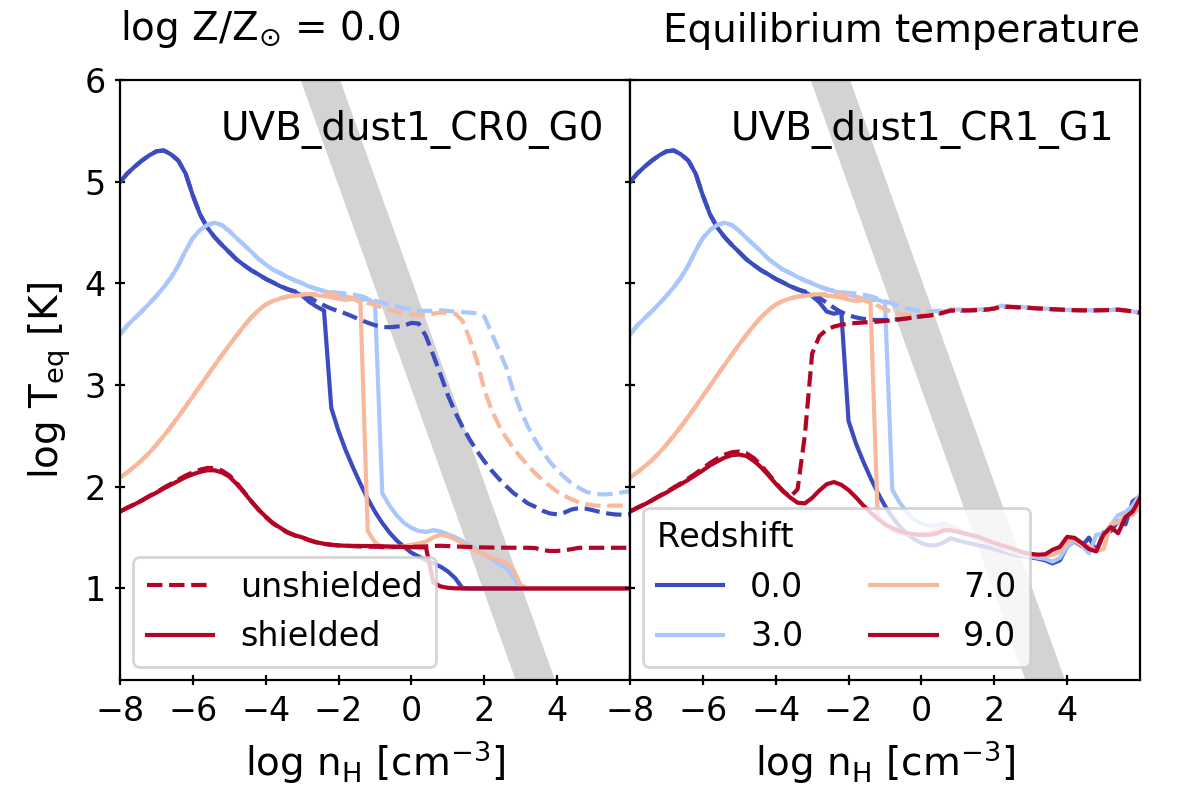}	
		\includegraphics*[width=0.48\linewidth, trim = 0cm 0cm 0cm 0cm,clip]{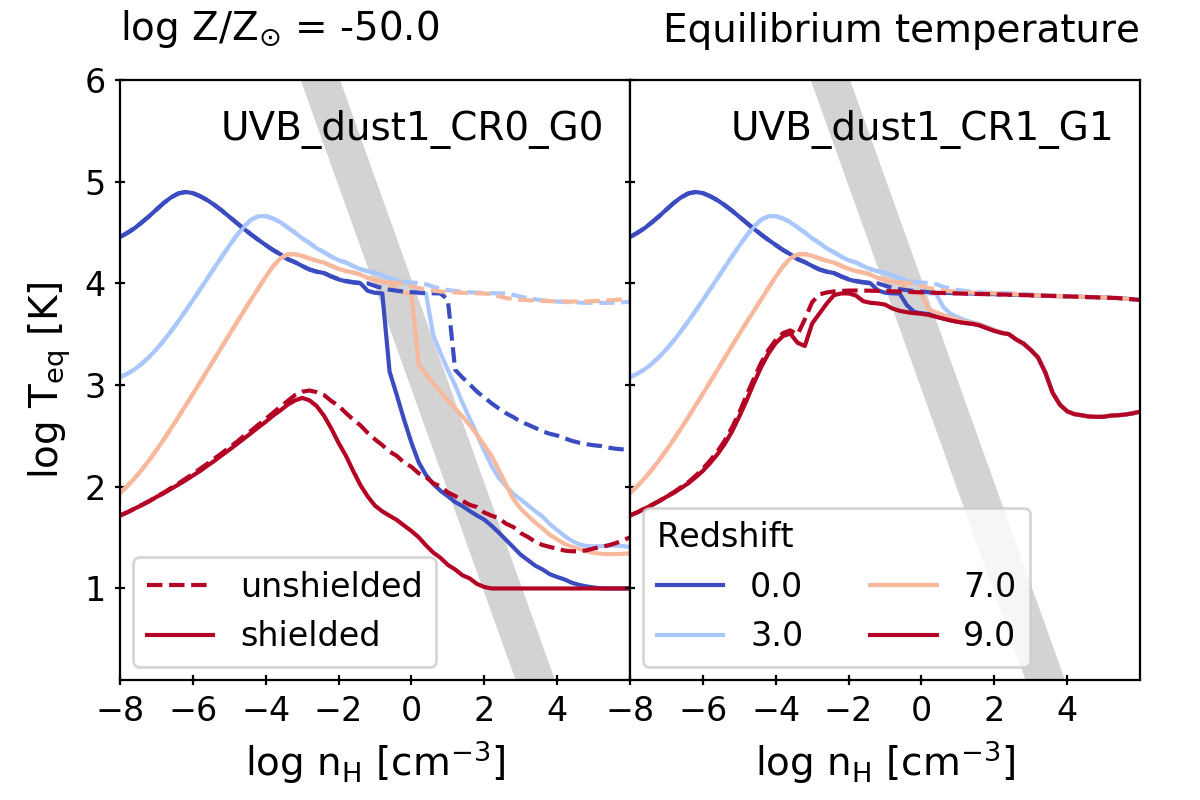}	
    		\caption{As. Fig.~\ref{fig:Teqoverview}, but where each figure is for constant metallicity (left figure, solar metallicity; right figure, primordial abundances) and each line colour represents a different redshift.}
		\label{fig:Teqoverview2}
	\end{center}
\end{figure*}

The redshift dependence of $T_{\mathrm{Eq}}$ is shown in Fig.~\ref{fig:Teqoverview2} for solar (left set of panels) and primordial abundances (right set of panels). As in Fig.~\ref{fig:Teqoverview}, the left panel does not include the ISRF and $T_{\mathrm{Eq}}$ depends on redshift over the full density range. In the right panels of Fig.~\ref{fig:Teqoverview2}, which include the redshift-independent ISRF, the thermal equilibrium temperatures of different redshifts converge at high densities.

$T_{\mathrm{Eq}}$ increases with density for the lowest densities in the tables (see Figs.~\ref{fig:Teqoverview} and \ref{fig:Teqoverview2}). This relation is caused by the combination of a photo heating rate  $\Lambda_{\mathrm{heat}}/n^2_{\mathrm{H}}$ that is independent of density for highly (photo-)ionized gas and Compton cooling rate $\Lambda_{\mathrm{cool,Compton}}/n^2_{\mathrm{H}}$ that increases with decreasing density. The redshift dependence at these densities is dominated by the steep dependence of Compton cooling with redshift. In practice the thermal equilibrium temperature from radiative processes is irrelevant for the lowest densities as the cooling time is longer than the Hubble time and adiabatic cooling (Hubble expansion) dominates over radiative cooling. At higher densities ($\log \nH [\ccm] \gtrsim -5$ for $z=0$) hydrogen and helium recombination cooling dominates between $\log T [\K] \approx 4-6$ and $T_{\mathrm{Eq}}$ decreases with density.

The thermal equilibrium curves in Figs.~\ref{fig:Teqoverview} and \ref{fig:Teqoverview2} are reminiscent of the classical two phase structure of the neutral ISM, as described in \citet{wolfire_neutral_1995} (hereafter:  \citetalias{wolfire_neutral_1995}).  As it can be tempting to directly compare the thermal equilibrium curves from this work to those of \citetalias{wolfire_neutral_1995}, we discuss here the key differences in both the general approach as well as the interpretation of the results. 

\citetalias{wolfire_neutral_1995} describe the ISM densities where warm neutral ISM gas can be in pressure equilibrium with cold neutral gas. Based on the idea of a dense, cold gas cloud embedded in more tenuous, warm ($10^4\K$) gas, the pressure equilibrium is determined at the edge of the cold gas cloud. They assume that the ISRF is shielded by a constant column density ($N_w = 10^{19}\scm$ in their fiducial model), where $N_w$ is the typical column density of the warm medium. 
In this work, the shielding from the ISRF corresponds to the self-shielding of the cold gas cloud (i.e. from its edge to its centre). The shielding column $\NS$ therefore varies with the assumed size of the gas cloud, which is a function of gas density and temperature (or pressure, as $\NS \propto (\nH T)^{0.5}$, see Sec.~\ref{sec:column}).  

In addition to the varying shielding column, also the radiation field intensity and the CR rate change along the thermal equilibrium curves. Higher gas pressures lead to larger column densities which are observed to result in higher SFR densities and subsequently a stronger ISRF (see Sec.~\ref{sec:ISRF}). 
While \citet{wolfire_neutral_2003} vary the ISRF to account for radial variations within the MW disk, the ISRF is constant with density for each of their equilibrium functions.

\begin{figure} 
	\begin{center}
		\includegraphics*[width=\linewidth, trim = 0cm -0.25cm 0cm 0cm,clip]{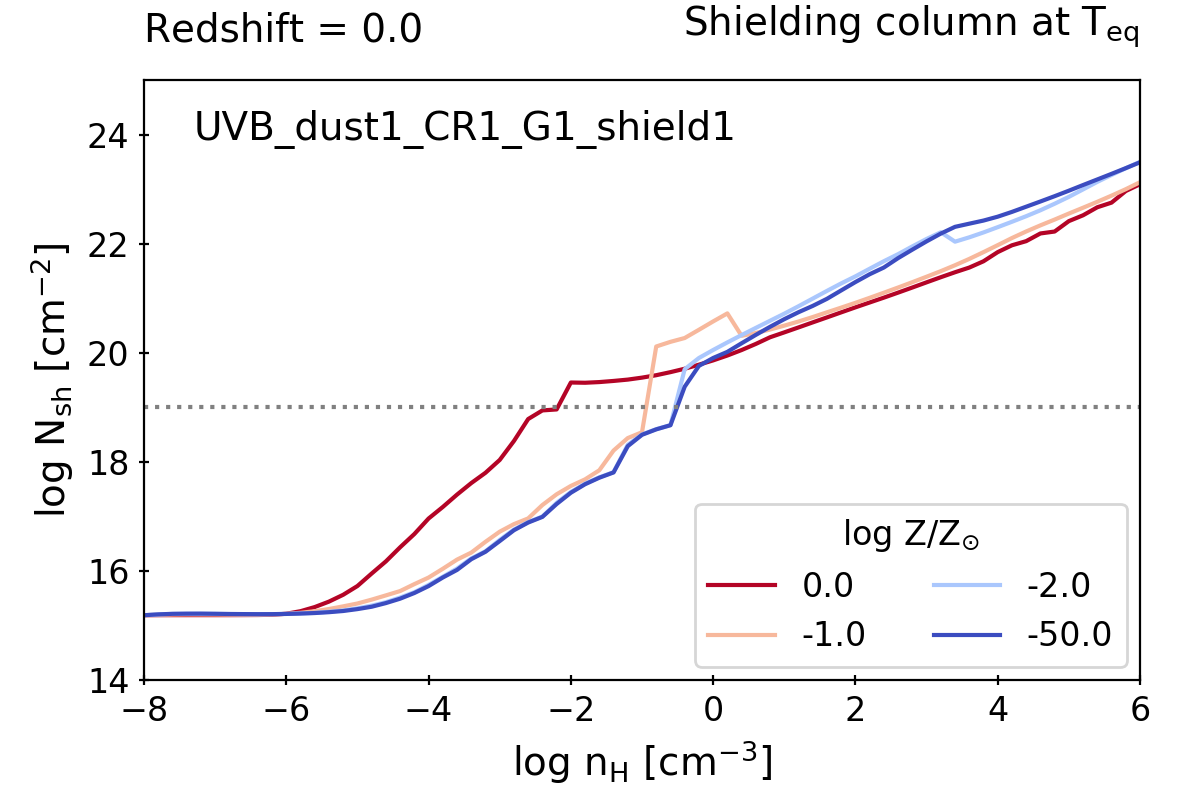}	
		\includegraphics*[width=\linewidth, trim = 0cm 0cm 0cm -0.25cm,clip]{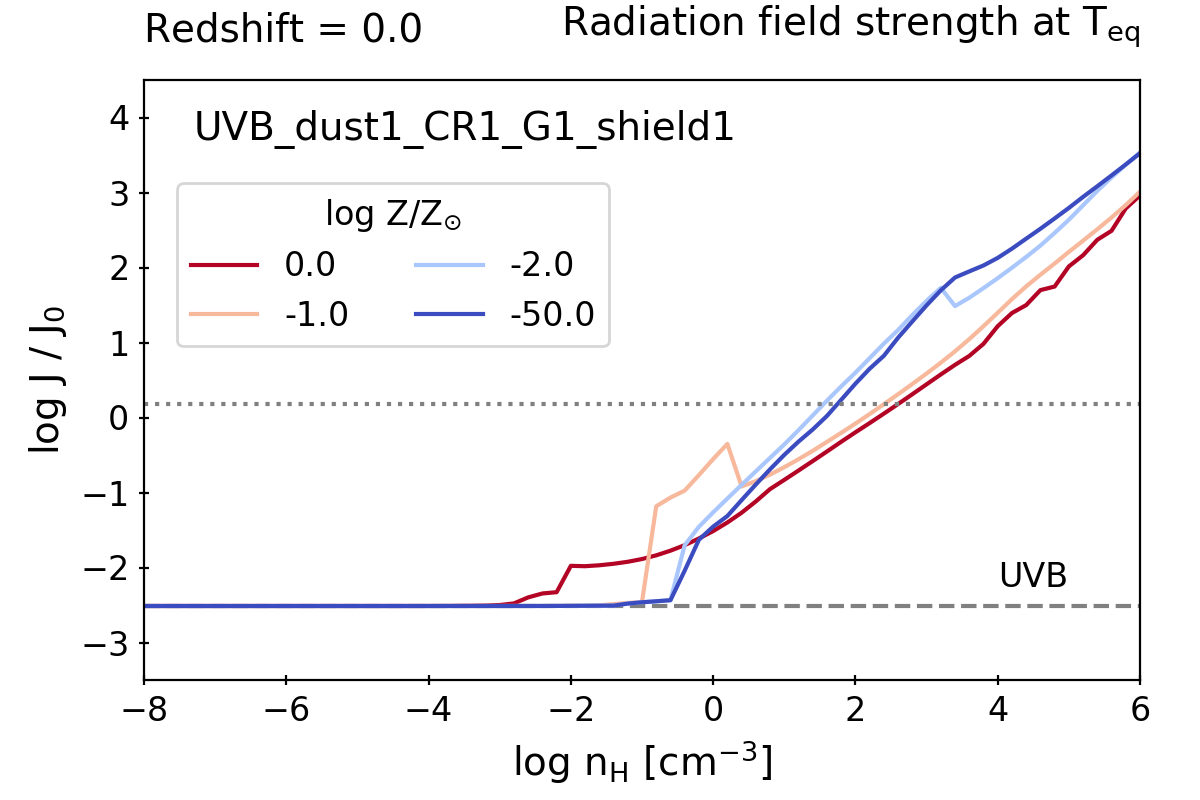}	
		\caption{Shielding column (top panels) and ISRF strength (bottom panels) assuming the density-dependent thermal equilibrium temperature displayed in the right panel of Fig.~\ref{fig:Teqoverview} for self-shielded gas at $z=0$. The horizontal dotted lines refer to a constant column density of $\log \NS [\scm] = 19$ (top panel) and a constant radiation field intensity of $J_{\mathrm{D78}} = 1.5 J_{\mathrm{0}}$ (bottom panel), the default values used in \citetalias{wolfire_neutral_1995}. At low densities, where the UVB (dashed line, bottom panel) dominates the ISRF at 1000\AA, the radiation field becomes independent of the density.}
		\label{fig:RadFieldOverTeq}
	\end{center}
\end{figure}

Fig.~\ref{fig:RadFieldOverTeq} illustrates the variations in the shielding column (top panels) and ISRF intensity (bottom panels). The coloured lines refer to the values at the thermal equilibrium temperatures for selected metallicities at $z=0$ and the grey horizontal lines show the fiducial values of the \citetalias{wolfire_neutral_1995} model, with $\log \NS [\scm] = 19$ and the radiation field from \citet{draine_photoelectric_1978} with $J_{\mathrm{D78}} = 6.6\times10^5\,\mathrm{photons}\,\mathrm{cm}^{-2}\,\mathrm{s}^{-1}\,\mathrm{sr}^{-1}\,\mathrm{eV}^{-1}$ and therefore $J_{\mathrm{D78}}/J_0 = 1.5$. 

The column densities for the tables that include self-shielding span several orders of magnitudes for typical ISM densities ($\nH \gtrsim 10^{-2}\,\ccm$). The radiation field spans an even larger range as $J\propto\Sigma_{\mathrm{SFR}}\propto{\NS}^{1.4}$. Higher pressure environments are therefore self-consistently exposed to more intense radiation fields.

\begin{figure} 
	\begin{center}
		\includegraphics*[width=\linewidth, trim = 1.5cm 0cm 1.5cm 0.5cm,clip]{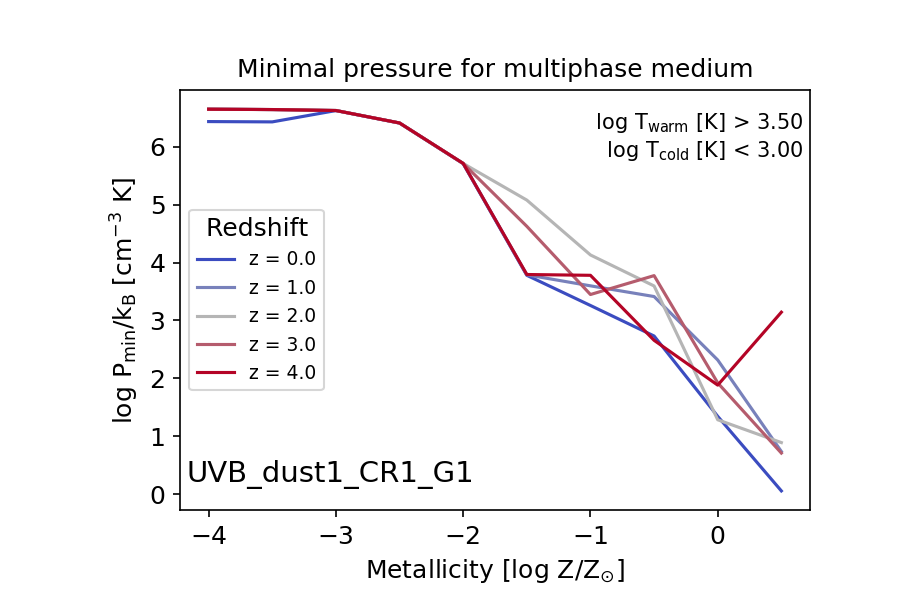}
		\caption{Minimum pressure above which a multiphase medium of unshielded and shielded gas in pressure equilibrium, with $\log T_{\mathrm{warm}} [\K] > 3.5$ and $\log T_{\mathrm{cold}} [\K] < 3$, is possible in model ``UVB\_dust1\_CR1\_G1"~for selected redshifts (different line colors). Note that these are the minimum possible pressures for a multiphase medium, based on the thermal equilibrium temperatures (see Figs.~\ref{fig:Teqoverview} and ~\ref{fig:Teqoverview2}). Hydrodynamical simulations are necessary to determine the density (and therefore pressure) distribution of gas along these thermal equilibrium temperature tracks.}
		\label{fig:multiphase}
	\end{center}
\end{figure}

\citetalias{wolfire_neutral_1995} and \citet{wolfire_neutral_2003} focus on the MW ISM and therefore assume solar metallicity. Recently, \citet{bialy_thermal_2019} extended their model to lower metallicity gas and included molecular hydrogen heating and cooling processes. They found that a multi-phase medium requires higher pressure for lower metallicities if the ISRF is the same as that assumed in solar metallicity galaxies. For extremely low metallicities ($Z\lesssim10^{-5}\Zsol$) and for a total CR ionization rate per hydrogen atom of $10^{-16}\,\mathrm{s}^{-1}$ they found that the ISM would be single-phase, as the equilibrium temperature changes smoothly from $\approx 10^4\K$ to $\approx 600\K$. Despite the above discussed differences, we find qualitatively similar behaviour for the thermal equilibrium temperatures of the recommended table UVB\_dust1\_CR1\_G1\_shield1. The colour scale in Fig.~\ref{fig:multiphase} shows the minimum pressure for unshielded (``\_shield0") and shielded (``\_shield1") gas to be in pressure equilibrium (where the former corresponds to the warm phase: $\log T_{\mathrm{warm}} [\K] > 3.5$ and the latter to the cold phase: $\log T_{\mathrm{cold}} [\K] < 3$). An overall trend with metallicity (i.e. lower metallicity - higher minimum pressure) is clearly visible for every redshift. 

Note that for very low-metallicity gas ($\log Z/\Zsol \lesssim -2$) the lack of metal-line cooling keeps the temperature at $\gtrsim 800\K$, even for very high density gas ($\nH = 10^6\,\ccm$, see Fig.~\ref{fig:Teqoverview}). This gas is mainly heated by CRs and by the vibrational and rotational energy of molecular hydrogen, as it absorbs UV photons.

\subsection{Phase transitions}\label{sec:phasetransitions}

\paragraph*{Ionized - neutral hydrogen: }

 \begin{figure} 
	\begin{center}
		\includegraphics*[width=\linewidth, trim = 0cm 0cm 0cm 0cm,clip]{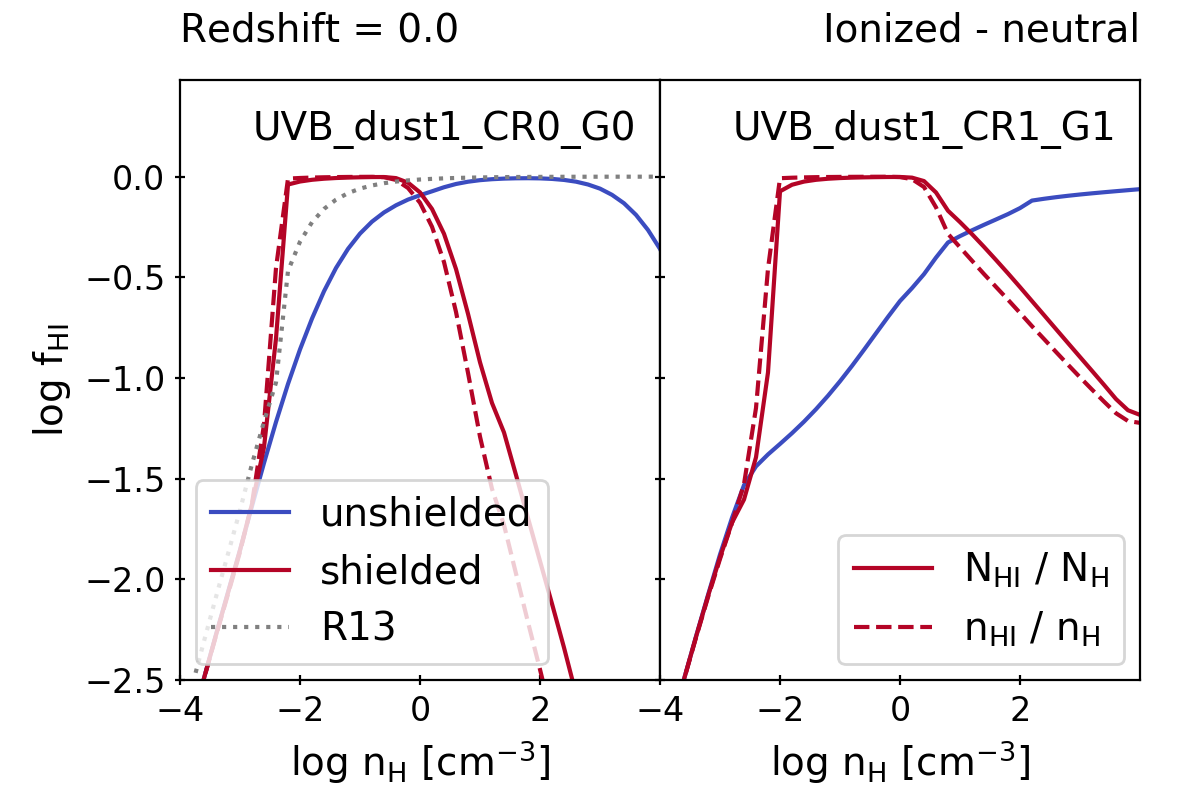}
		\caption{Neutral hydrogen fraction for two radiation fields (left panel: only UVB, right panel:  UVB + CR + ISRF) along their respective thermal equilibrium temperatures for self-shielded (red lines) and unshielded (blue lines) solar metallicity gas at $z=0$. For self-shielded gas both the volume ($n$) and column ($N$) density $\ion{H}{I}$ fractions are shown (solid lines: $N_{\ion{H}{I}} /\NH$, dashed lines: $n_{\ion{H}{I}} /\nH$ at the end of the shielding column). The dotted line in the left panel is $f_{\ion{H}{I}}$ from the fitting function of \citet{rahmati_evolution_2013} for the same $T_{\mathrm{eq}}$ as the self-shielded gas.}
		\label{fig:NHItrans}
	\end{center}
\end{figure}

For each radiation field, the transition from mostly ionized to mostly neutral hydrogen marks where self-shielding becomes important. Assuming that most of the gas follows the thermal equilibrium temperature, the tables directly provide typical hydrogen species fractions for each gas density. Fig.~\ref{fig:NHItrans} shows $f_{\ion{H}{I}}$ for solar metallicity and redshift $z=0$ for tables UVB\_dust1\_CR0\_G0 (UVB only, left panel) and UVB\_dust1\_CR1\_G1 (UVB, CRs and ISRF, right panel). For each radiation field, $f_{\ion{H}{I}}$ is presented for both unshielded (blue) and self-shielded (red) gas. For self-shielded gas, the neutral atomic hydrogen fraction is displayed both as a column density fraction $f_{\ion{H}{I}} = N_{\ion{H}{I}} /\NH$ (solid lines) and as a volume density fraction at the end of the shielding column $f_{\ion{H}{I}} = n_{\ion{H}{I}} /\nH$ (dashed lines).

For the UVB only tables (UVB\_dust1\_CR0\_G0), we compare $f_{\ion{H}{I}}$ from this work to the widely used fitting function from \citetalias{rahmati_evolution_2013}, for which $n_{\ion{H}{I}}/\nH$ is a function of gas density, temperature and photoionization rate $\Gamma_{\mathrm{phot}}$. The dotted line in the left panel of Fig.~\ref{fig:NHItrans} indicates $(n_{\ion{H}{I}}/\nH)_{\mathrm{R13}}$ for the same (equilibrium) temperatures. The characteristic density above which $f_{\ion{H}{I}}$ for self-shielded gas deviates from unshielded gas is the same in the shielded table and the \citetalias{rahmati_evolution_2013} fitting function.
While the transition is very steep in the self-shielded table presented in this work, the fitting function from \citetalias{rahmati_evolution_2013} has a more gradual increase. The difference here may be that $f_{\ion{H}{I}}$ from the tables assumes that all gas follows the thermal equilibrium temperature and a fixed $\nH-\NH$ relation, while $f_{\ion{H}{I}}$ from \citetalias{rahmati_evolution_2013} is a fit to cosmological radiative transfer simulations.
Note, that in the tables $f_{\ion{H}{I}}$ can decrease again towards higher densities as hydrogen becomes molecular. Molecular hydrogen is not included in \citetalias{rahmati_evolution_2013}.

\paragraph*{Atomic - molecular hydrogen: }  

\begin{figure} 
	\begin{center}
		\includegraphics*[width=\linewidth, trim = 0cm 0cm 0cm 0cm,clip]{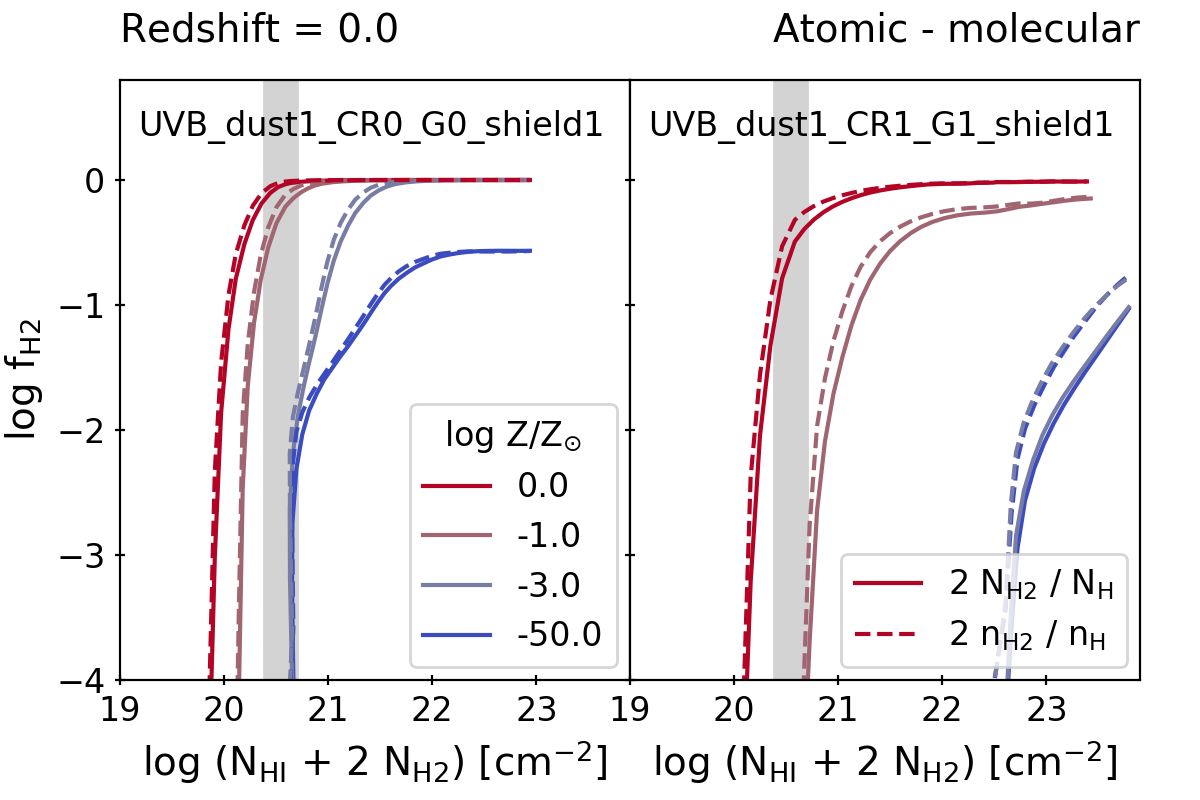}	
		\caption{Molecular hydrogen fraction $f_{\mathrm{H2}}$, defined as $2 N_{\mathrm{H2}}/ (N_{\ion{H}{I}} + 2 N_{\mathrm{H2}}) $ (column density fraction, solid lines) or $2 n_{\mathrm{H2}}/ (n_{\ion{H}{I}} + 2 n_{\mathrm{H2}}) $ (volume density fraction at the end of the shielding column, dashed lines) assuming thermal equilibrium temperatures for different radiation fields (different panels) and different metallicities (different line colours) for the self-shielded tables. The grey band brackets the column densities for the observed $\ion{H}{I}$-$\mathrm{H}_{\mathrm{2}}$ transition in the solar neighbourhood, from $\log (N_{\ion{H}{I}} + 2 N_{\mathrm{H2}}) [\scm] = 20.38$ (FUSE halo survey, \citealp{gillmon_fuse_2006}) to $\log (N_{\ion{H}{I}} + 2 N_{\mathrm{H2}}) [\scm] = 20.70$ (FUSE disk survey, \citealp{gillmon_fuse_2006}; Copernicus, \citealp{savage_survey_1977}).}
		\label{fig:NH2trans}
	\end{center}
\end{figure}

The transition between atomic and molecular hydrogen is summarised in Fig.~\ref{fig:NH2trans} for two self-shielded tables (ending with ``shield1"). The left panel does not include an interstellar radiation field, while the right panel includes both CRs and the ISRF. In each panel, the H$_{\mathrm{2}}$ fraction is shown for the thermal equilibrium temperature. The x-axis shows the column density of atomic plus molecular hydrogen through the gas cloud 

\begin{equation}
	N_{\ion{H}{I}} + 2 N_{\mathrm{H2}} = 2 \times N_{\mathrm{sh}} \cdot \left ( \frac {N_{\ion{H}{I}}}{\NH}  +2  \frac {N_{\mathrm{H2}}}{\NH}   \right )
\end{equation}

\noindent
where $N_{\mathrm{sh}}$ as well as the $\ion{H}{I}$ and H$_{\mathrm{2}}$ fractions are taken from the tables along the thermal equilibrium temperature (hdf5 group \texttt{/ThermEq/}). 

For solar metallicity, the results can be compared to observations of H$_{\mathrm{2}}$ fractions in the Milky Way in sight-lines towards stars (e.g. with Copernicus, \citealp{savage_survey_1977}) or AGN (e.g. the FUSE halo survey, \citealp{gillmon_fuse_2006}). Inside the Milky Way Galaxy, the transition from atomic to molecular hydrogen (from $\log f_{\mathrm{H2}}\equiv \log 2 N_{\mathrm{H2}}/(N_{\ion{H}{I}} +  2 N_{\mathrm{H}}) \lesssim -3$ to $\log f_{\mathrm{H2}} \gtrsim -1$) has been measured to occur around $\log (N_{\ion{H}{I}} + 2 N_{\mathrm{H2}})[\ccm] = 20.70$ \citep{savage_survey_1977}, while for high-latitude sight-lines, hydrogen becomes molecular at slightly lower column densities ($\log (N_{\ion{H}{I}} + 2 N_{\mathrm{H2}})[\ccm] = 20.38$, FUSE halo survey, \citealp{gillmon_fuse_2006}).

Using these tables in hydrodynamical simulations, we find that the gas temperature of individual resolution elements scatters up from the thermal equilibrium temperature but rarely goes below $T_{\mathrm{eq}}$. Higher temperatures for constant density lead to an increase in $N_{\mathrm{ref}}$ and typically to a decrease in $f_{\mathrm{H2}}$. In addition, the tables assume a constant density for each column density, while a sightline in a simulation can go through a variety of gas volume densities. We find that the atomic to molecular transition typically does not scatter to lower column densities in galaxy-scale simulations (Ploeckinger et al. in prep). Therefore, table UVB\_dust1\_CR1\_G1\_shield1 matches the column density where the \ion{H}{I}-H$_{\mathrm{2}}$ transition is observed to occur.

\subsection{Ion fractions}

\begin{figure*} 
	\begin{center}
		\includegraphics*[height=5cm, trim = 0cm 0cm 0.3cm 0cm,clip]{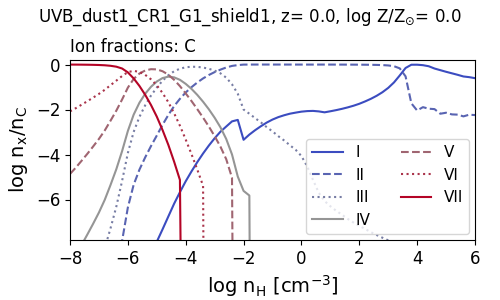}
		\includegraphics*[height=5cm, trim = 0cm 0cm 0.3cm 0cm,clip]{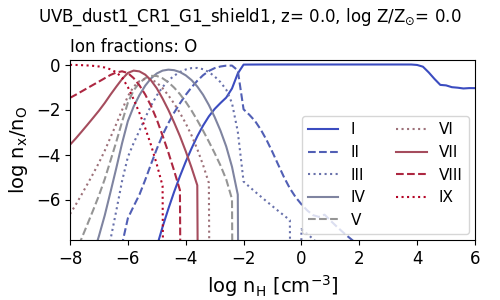}
		\caption{Ion fractions at the end of each shielding column for carbon (left) and oxygen (right) for table UVB\_dust1\_CR1\_shield1 at $z=0$ and for solar metallicity assuming thermal equilibrium. For densities of $\log \nH [\ccm]\gtrsim 4.5$ most C and O atoms are in CO. At $\log \nH [\ccm] \approx -2$ the equilibrium temperature decreases steeply with density (compare: red solid line in Fig~\ref{fig:Teqoverview}), which explains the features (most noticeable in the \ion{C}{I} fraction) around this density. }
		\label{fig:IonFracCTeq}
	\end{center}
\end{figure*}

Each table contains the gas-phase ion fractions of 11 selected elements (see Table~\ref{tab:alldatasets}).
Examples for table UVB\_dust1\_CR1\_G1\_shield1 at $z=0$ and for solar metallicity gas in thermal equilibrium are shown for carbon and oxygen (Fig.~\ref{fig:IonFracCTeq}). Ion fractions of all 11 tabulated elements can be found at the project webpages (see Sec.~\ref{sec:intro} for links).

As hydrogen is the most abundant element and it efficiently absorbs photons with energies higher than 13.6 eV, elements with minimum ionization energies at which the hydrogen photoionization cross section is large (e.g. helium: $E_i = 24.59 \,\mathrm{eV}$, nitrogen: $E_i = 14.53  \,\mathrm{eV}$, oxygen: $E_i = 13.62  \,\mathrm{eV}$, and neon: $E_i = 21.56  \,\mathrm{eV}$), are pre-dominantly neutral in self-shielded gas (see right panel of Fig.~\ref{fig:IonFracCTeq} for oxygen). Other elements, such as carbon ($E_i = 11.26  \,\mathrm{eV}$) and magnesium ($E_i = 7.65  \,\mathrm{eV}$) are singly ionised over a large range of volume densities (see left panel of Fig.~\ref{fig:IonFracCTeq} for carbon). 

Including dust depletion, the ion volume densities (here illustrated for \ion{C}{II}) can be calculated as

\begin{equation}\label{eq:iongas}
	n_{\ion{C}{II}}= \nH \left ( \frac {n_{\ion{C}{II}}}{n_{\mathrm{C}}} \right )_{\mathrm{gas}} \left (  \frac{n_{\mathrm{C}}}{\nH} \right )_{\mathrm{total}}  ( 1 - f_{\mathrm{dust}} ) \, ,
\end{equation}

\noindent
with the gas-phase ion fraction $(n_{\ion{C}{II}} / n_{\mathrm{C}})_{\mathrm{gas}}$ (dataset: \texttt{IonFractions}), the dust depletion fraction $f_{\mathrm{dust}}$ (dataset: \texttt{Depletion}), and the element abundance $n_{\mathrm{C}}/\nH$ (dataset: \texttt{TotalAbundances}).

\subsection{H2 and CO abundance}

\begin{figure} 
	\begin{center}
		\includegraphics*[width=\linewidth, trim = 0.7cm 0.5cm 0.3cm 0.3cm,clip]{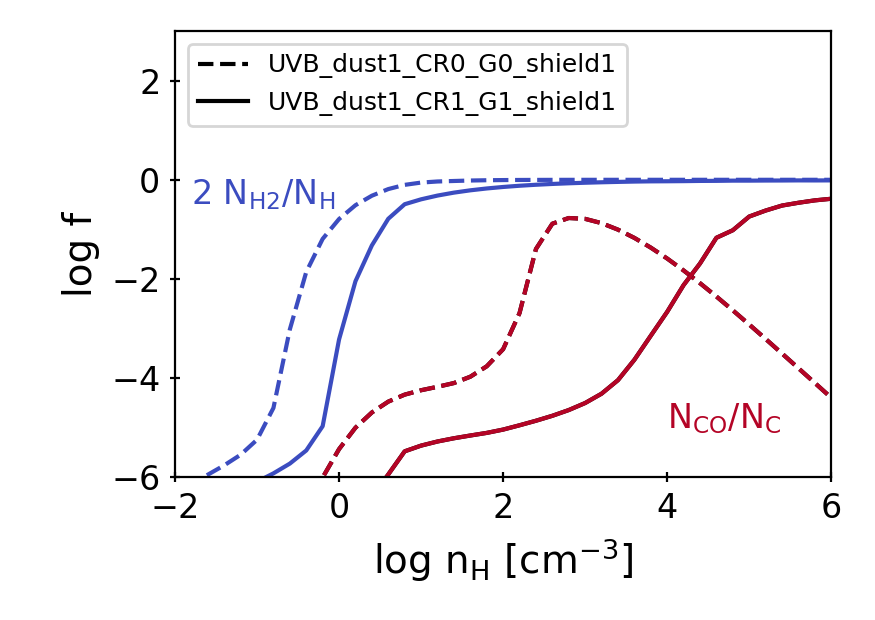}
		\caption{CO and H$_{\mathrm{2}}$ abundances for gas in thermal equilibrium at redshift $z=0$ and solar metallicity. Column density fraction of H2 ($2\NHtwo/\NH$, blue lines) and CO ($N_{\mathrm{CO}}/N_{\mathrm{C}}$, red lines) for two tables that include self-shielding (no ISRF and no CRs: dashed lines, including ISRF and CR: solid lines). The fraction of CO decreases with density for $\log \nH [\ccm] \gtrsim 3$ for the weaker radiation field (dashed lines) as CO is increasingly depleted onto dust grains (see Fig.~\ref{fig:molecules}).}
		\label{fig:COH2fractions}
	\end{center}
\end{figure}

Each tables contains the abundances of CO and H$_{\mathrm{2}}$, both in the last \textsc{cloudy} zone as well as integrated over the full shielding column. Fig.~\ref{fig:COH2fractions} shows these abundances for solar metallicity gas at $z=0$ at its thermal equilibrium temperature for self-shielded tables with and without ISRF and CR rates: UVB\_dust1\_CR0\_G0\_shield1 and UVB\_dust1\_CR1\_G1\_shield1. While the H2 abundance decreases when the ISRF and CRs are added, at high densities ($\log \nH [\ccm]>4$)  the CO abundance is largest when including an ISRF. 

\begin{figure} 
	\begin{center}
		\includegraphics*[width=\linewidth, trim = 0.3cm 0.3cm 0.3cm 0.3cm,clip]{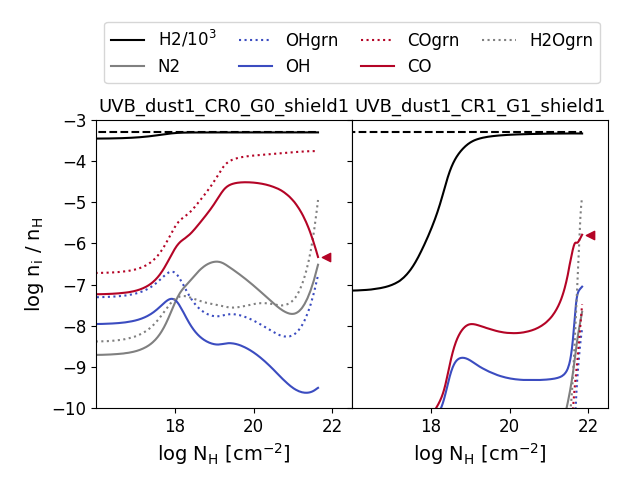}
		\caption{Abundances of selected molecules (and their depletion onto dust grains, indicated by the suffix ``grn") along the shielding column at an individual grid point of two tables (left panel: UVB\_dust1\_CR0\_G0\_shield1, right panel: UVB\_dust1\_CR1\_G1\_shield1) for the respective equilibrium temperatures at the end of the shielding columns ($\log T [K] = 1.00$, left and 1.41, right) of $z=0$, solar metallicity gas with a volume density of $\log\nH[\ccm] =4$. The H$_{\mathrm{2}}$ abundance $\nHtwo/\nH$, has been reduced by a factor of $10^3$ to fit on the plot. For guidance, the horizontal dashed line indicates the abundance where all H is in H$_{\mathrm{2}}$. As seen in Fig.~\ref{fig:COH2fractions}, the CO abundance at the last zone (indicated by a red triangle) for this density is comparable for both radiation fields, but the distribution throughout the column is very different.}
		\label{fig:molecules}
	\end{center}
\end{figure}

A closer look at the molecule abundances is presented in Fig.~\ref{fig:molecules}. Here, the dependence of their abundances on the depth into the gas cloud (column density $\NH$) is visualised for the two tables from Fig.~\ref{fig:COH2fractions}. For this, an individual grid point (at $\log \nH [\ccm] = 4$, see figure caption) is rerun with the additional output\footnote{\textsc{cloudy} command: \texttt{save molecules}} for the volume densities of 95 molecule species in each zone. Note that all these molecules are part of the chemistry for all grid points, but this additional output is not stored for the full table.
The left panel of Fig.~\ref{fig:molecules} (no ISRF, no CR) explains why gas phase CO is less abundant at the centre of the cloud (here at $\NH \approx 10^{22}\,\scm$): the majority of CO molecules are condensed into dust grains (``COgrn"). This is also the case for the OH and H$_2$O molecules (``OHgrn" and ``H2Ogrn"). For the stronger radiation field (right panel) molecule fractions peak at much higher column densities and therefore close to the centre of the gas cloud.

The resulting fraction $N_{\mathrm{CO}}/N_{\mathrm{C}}$ for the full shielding column is comparable for both radiation fields at a density of $\log \nH [\ccm] = 4$ (see Fig.~\ref{fig:COH2fractions} and triangles in Fig.~\ref{fig:molecules}) but without an ISRF, most of the CO is depleted onto dust grains, which explains the reduced gas phase CO fraction for high densities $\log \nH [\ccm] \gtrsim 4$ for the table without the ISRF (UVB\_dust1\_CR0\_G0\_shield1).

\section{Example results: Emission lines}\label{sec:emission}

\begin{figure} 
	\begin{center}
		\includegraphics*[width=\linewidth, trim = 0.cm 0.cm 0cm 0.cm, clip]{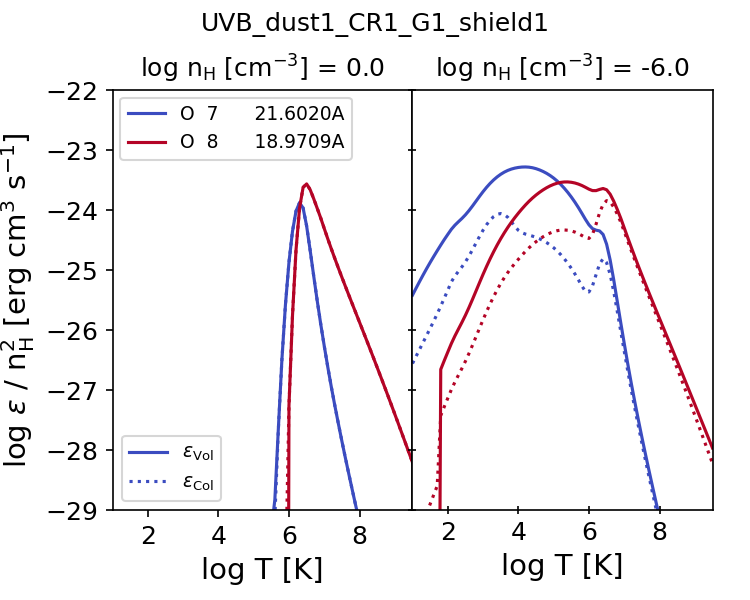}
		\caption{The normalized emergent emissivities, $\epsilon/\nH^2 \,[\mathrm{erg}\,\mathrm{cm}^{3}\,s^{-1}]$ for selected soft X-ray emission lines for constant hydrogen number density (left panel: $\log\nH [\ccm] = 0$, right panel: $\log\nH [\ccm] = -6$), solar metallicity and redshift $z=0$. The solid lines are the emissivities at the last \textsc{cloudy} zone ($\epsilon_{\mathrm{Vol}}$, dataset: \texttt{EmissivitiesVol}) and the dotted lines are the average emissivities of the full shielding column ($\epsilon_{\mathrm{Col}}$, dataset: \texttt{EmissivitiesCol}). For diffuse gas (right panel) with temperatures below $\log T [\K] \lesssim 6$, photo-ionization dominates over collisional ionization and the line emissivity therefore depends on density (compare different panels). The differences between $\epsilon_{\mathrm{Vol}}$ and $\epsilon_{\mathrm{Col}}$ for $\log\nH [\ccm] = -6$ (right panel) are unexpected as the gas is unshielded at this density and the variations within the small shielding column are negligible. This is likely an artefact in \textsc{cloudy} (see text for details).}
		\label{fig:emisXray}
	\end{center}
\end{figure}

\begin{figure} 
	\begin{center}
		\includegraphics*[width=\linewidth, trim = 0.cm 0.cm 0cm 0.cm, clip]{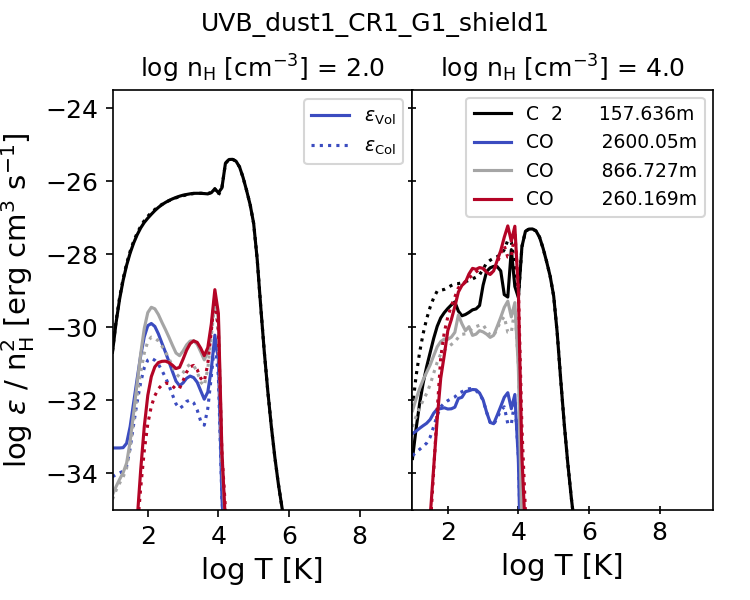}
		\caption{As Fig.~\ref{fig:emisXray} but for selected (sub-)mm / far infrared emission lines for high density gas (left panel: $\log\nH [\ccm] = 2$, right panel: $\log\nH [\ccm] = 4$). As molecules form in self-shielded gas and their abundance and level population depend on the shielding column density, the emissivity at the last \textsc{cloudy} zone $\epsilon_{\mathrm{Vol}}$ (i.e. the most shielded part of the gas column, solid lines) of the selected CO lines is typically larger than the average line emissivity of the shielding column $\epsilon_{\mathrm{Col}}$ (dotted lines). For high densities (e.g. $\log\nH [\ccm] = 4$, right panel) the emissivity of \ion{C}{II} (``C  2") is lower at high shielding column densities and therefore the column averaged emissivity is larger than the emissivity at the last \textsc{cloudy} zone ($\epsilon_{\mathrm{Col}} > \epsilon_{\mathrm{Vol}}$). }
		\label{fig:emisFIR}
	\end{center}
\end{figure}

Out of the 183 emission lines listed in Table~\ref{tab:lines} we show a few examples of soft X-ray emission lines in Fig.~\ref{fig:emisXray} and far-infrared/(sub-)mm emission lines in Fig.~\ref{fig:emisFIR}. In both figures the normalized emergent emissivities, $\epsilon/\nH^2$, for the last \textsc{cloudy} zone of the shielding column ($\epsilon_{\mathrm{Vol}}$) as well as the emissivity calculated by dividing the line intensity of the full shielding column by the length of the shielding column ($\epsilon_{\mathrm{Col}}$) from the fiducial model ``UVB\_dust1\_CR1\_G1\_shield1"~are displayed. 

Fig.~\ref{fig:emisXray} shows the strongest soft X-ray emission lines of \ion{O}{VII} and \ion{O}{VIII} (as in the top panels of figure 1 in \citealp{bertone_metal-line_2010}). For densities $\log \nH [\ccm] \gtrsim -3$ (see left panel for  $\log \nH [\ccm] =0$) the emissivities $\epsilon_{\mathrm{Vol}}$ (solid lines) and $\epsilon_{\mathrm{Col}}$ (dotted lines) agree, as expected for unshielded gas. For very low densities (e.g. $\log \nH [\ccm] = -6$, right panel) $\epsilon_{\mathrm{Vol}}$ and $\epsilon_{\mathrm{Col}}$ disagree for photo-ionized gas ($\log T [\K] \lesssim 6$). This has been traced down to differences between the output produced by the \textsc{cloudy} commands ``save lines emissivity"~and ``save line list absolute"~and even persists for one-zone models (e.g.~``UVB\_dust1\_CR1\_G1\_shield0"). The unexpected behaviour of \textsc{cloudy} has been posted to the \textsc{cloudy} group (see \url{https://cloudyastrophysics.groups.io/g/Main/message/4301} for details). This is unlikely affecting any simulation results as this issue only appears at very low densities where the emissivities are very small.

The results for selected CO lines as well as the \ion{C}{II} emission line at $157\,\mu\mathrm{m}$ are presented in Fig.~\ref{fig:emisFIR} for densities of $\log \nH [\ccm] = 2$ (left panel) and $\log \nH [\ccm] = 4$ (right panel). For higher densities the CO fraction increases and therefore the normalized emissivity of \ion{C}{II} decreases. At $\log \nH [\ccm] = 4$ (right panel) CO is collisionally excited to higher energy levels, leading to brighter emission lines of higher level transitions (compare e.g. $J=1 \rightarrow 0$ at 2600~$\mu\mathrm{m}$ and $J=3 \rightarrow 2$ at 867~$\mu\mathrm{m}$).

As CO forms mainly deep into the shielding column (see e.g. right panel of Fig.~\ref{fig:molecules}), the emissivities calculated at the last zone ($\epsilon_{\mathrm{Vol}}$, solid lines) are not expected to be identical to the shielding column averaged emissivity ($\epsilon_{\mathrm{Col}}$, dotted lines) but they converge to the same values for temperatures $\log T [\mathrm{K}] \gtrsim 4$, where the shielding column becomes small.

\section{Summary}\label{sec:summary}

We use \textsc{cloudy} version 17.01 to tabulate the properties of unshielded and self-shielded gas as it is exposed to metagalactic and interstellar radiation fields. The tables cover a sufficiently large range in redshift ($z=0$ \-- $9$), temperature ($\log T [K] = 1$ \-- $9.5$), metallicity ($\log Z/\mathrm{Z}_{\odot} = -4$ \-- $+0.5$, $Z=0$), and density ($\log \nH [\ccm] = -8$ \-- $+6$) to enable users to use one set of tables for all gas phases (ionised, atomic, molecular) both during a simulation (i.e. cooling and heating rates) as well as for analysing the simulation output in post-processing (i.e. species fractions, line emissivities).

The gas column density is an important factor to estimate both the local star formation rate as well as the attenuation of the radiation field. 
The reference column density $N_{\mathrm{ref}}$ in this work is based\footnote{$N_{\mathrm{ref}}$ equals $N_{\mathrm{J}}$ for ISM gas but in addition includes a smooth transition to optically thin gas (see Eq.~\ref{eq:Nnorm}).} on the Jeans column density $N_{\mathrm{J}}$, which is a typical scale for self-gravitating gas. 
As gas surface density is observed to correlate with the star formation rate surface density (Eq.~\ref{eq:KS}), we add an interstellar radiation field (ISRF) and cosmic rays  proportional to $N_{\mathrm{ref}}^{1.4}$ to the redshift-dependent UV/X-ray background. The UV (and X-ray) background is based on the model from \citetalias{faucher-giguere_cosmic_2020} but before hydrogen and helium reionization are complete, we attenuate the \citetalias{faucher-giguere_cosmic_2020} spectra to better match the effective rates inferred from observations by \citetalias{faucher-giguere_cosmic_2020} and therefore the observed electron scattering optical depth of $\tau_e = 0.054$ \citep{planck_collaboration_planck_2018}.

Dust grains are an important catalyst for the formation of $\mathrm{H}_{\mathrm{2}}$ and other molecules and also contribute to self-shielding. A constant dust-to-metals ratio is assumed for star formation rate surface densities above the solar neighbourhood value, while for lower $N_{\mathrm{ref}}$, the dust-to-metal ratio scales like the ISRF ($\propto N_{\mathrm{ref}}^{1.4}$, Eq.~\ref{eq:DGratio}). The gas phase abundances of individual metals that are depleted onto dust grains are reduced accordingly. 

At the centre of a gas cloud, the radiation field is attenuated by gas with a shielding column of $\NS = 0.5 N_{\mathrm{ref}}$ and the dust within. This is modelled by passing the incident radiation field through a gas shielding column, $\NS$. A large number of gas properties (see Table~\ref{tab:alldatasets} for a full list) is stored in hdf5 datasets at the unattenuated (unshielded; tables with \_shield0) edge of the gas clump, as well as at the gas cloud centre (self-shielded;  tables with \_shield1). Selected properties (e.g CO and H species fractions, line emissivities) are also available as averages over the shielding column.
In the tables for self-shielded gas, all elements (also all metals) contribute to the self-shielding and in turn respond to the attenuated radiation field.
Some datasets (e.g. cooling and heating rates in Figs.~\ref{fig:netcool2d} and \ref{fig:coolheat}, ion fractions in Figs.~\ref{fig:IonFracCTeq}) are presented for individual slices through the multi-dimensional hyper cuboid and we provide an easy to use graphical user interface to explore the full range.

As an example application, we showed characteristic densities for the most important phase transitions ($\mathrm{H}_{\mathrm{2}}$ to \ion{H}{I}, Fig.~\ref{fig:NHItrans}, and \ion{H}{I} to \ion{H}{II}, Fig.~\ref{fig:NH2trans}) and compared the results to values from the literature. 
For the transition from ionised to neutral hydrogen, we compared our results to the fitting function from \citetalias{rahmati_evolution_2013}. Fig.~\ref{fig:NHItrans} illustrates that self-shielding becomes important at very similar densities, although the increase in $f_{\ion{H}{I}}$ with density is steeper in this work than in \citetalias{rahmati_evolution_2013}.

The predicted transition between neutral and molecular hydrogen was compared to observations of individual sight lines within the MW disk and towards the MW halo, measuring both the \ion{H}{I} and $\mathrm{H}_{\mathrm{2}}$ column densities (Fig.~\ref{fig:NH2trans}). For our fiducial ISRF and CR rates (UVB\_dust1\_CR1\_G1\_shield1) the observations are very well matched. 

Two tables (UVB\_dust1\_CR1\_G1\_shield0 and UVB\_dust1\_CR1\_G1\_shield1) include line emissivities for 183 selected lines.
A full overview of the data files released with this work is in Table~\ref{tab:products}. In addition to the hdf5 data files we also provide a set of C routines that interpolate the tables and return the total net cooling rate for a given redshift, gas density, temperature, and metallicity or abundance ratio, which can be implemented in a hydrodynamics code, and a set of python3 routines that read the tables and produce plots via an easy to use graphical user interface. All data files and links to the github repositories for the python and C packages can be found at \url{http://radcool.strw.leidenuniv.nl/} and \url{https://www.sylviaploeckinger.com/radcool}.

Finally, practical information about how to use the tables in simulations, either alone or coupled to a non-equilibrium network, such as \textsc{chimes} \citep{richings_non-equilibrium_2014-1,richings_non-equilibrium_2014} or \textsc{grackle} \citep{smith_grackle:_2017}, as well as how to reproduce our results with the public code \textsc{cloudy}v17.01 are provided in the appendix.

\begin{table}
	\caption{Summary of files released with this work at \url{http://radcool.strw.leidenuniv.nl/} and \mbox{\url{https://www.sylviaploeckinger.com/radcool}}. }
	\begin{tabular}{lL{5cm}}
	File name & Description \\
	\hline
	$<$modelname$>$.hdf5            & Hdf5 files with cooling and heating rates, ion fractions, free electron fractions and molecule abundances for each model listed in Table~\ref{tab:tables}. The content of each datafile is listed in Table~\ref{tab:alldatasets} and described in Sec.~\ref{sec:datasets}.\\
	$<$modelname$>$\_lines.hdf5  & Hd5 files for models UVB\_dust1\_CR1\_G1\_shield0 and UVB\_dust1\_CR1\_G1\_shield1 containing the emissivities for the emission lines listed in Table~\ref{tab:lines} and described in Sec.~\ref{sec:lines}. \\
	modFG20\_cloudy.ascii		  & The modified \citetalias{faucher-giguere_cosmic_2020} spectra in the \textsc{cloudy} UVB format. See Appendix~\ref{sec:reprod} for how to use this UVB with \textsc{cloudy}.\\
	\end{tabular}\label{tab:products} 
\end{table}

\begin{table*} 
	\caption{Overview of the cooling processes stored in the hdf5 tables, with their index in the cooling datasets (column 1), their string identifiers (column 2), and their scaling (column 3, see text for details). Column 4 lists the \textsc{cloudy} labels for the individual cooling channels and additional information can be found in column 5. The star in column 1 indicates processes that are combined in ``TotalPrim"~while the remaining ones are combined in ``TotalMetal".}
	\begin{tabular}{lllll}
		i	 	&	Identifier		& 	Scaling 			& \textsc{cloudy} label			&	Comment \\
		\hline
		0$\star$ 	&	Hydrogen		& 	$n_{\mathrm{H}}$	& H, H-fb 				& H-fb: H + e $\rightarrow$ H$^{-}$ + h$\nu$\\
		1$\star$ 	&	Helium		& 	$n_{\mathrm{He}}$	& He					& 	\\ 
		2 		&	Carbon		& 	$n_{\mathrm{C}}$	& C					& 	\\
		3	 	&	Nitrogen		& 	$n_{\mathrm{N}}$	& N					&  	\\
		4 		&	Oxygen		& 	$n_{\mathrm{O}}$	& O					&	\\
		5 		&	Neon		& 	$n_{\mathrm{Ne}}$	& Ne					&	\\
		6 		&	Magnesium	& 	$n_{\mathrm{Mg}}$	& Mg					&	\\
		7 		&	Silicon		& 	$n_{\mathrm{Si}}$	& Si					&	\\
		8 		&	Sulphur		& 	$n_{\mathrm{S}}$	& S					&	\\
		9 		&	Calcium		& 	$n_{\mathrm{Ca}}$	& Ca					&	\\
		10		&	Iron			& 	$n_{\mathrm{Fe}}$	& Fe					&  	\\
		11 		&	OtherA		& 	Z				&  Li, Be, B, F, Na,  Al, 	& all other atoms\\
				&				&					&  P, Cl, Ar, K, Sc, Ti, V, 	&  	\\
				&				&					&  Cr, Mn, Co, Ni, Cu, Zn	&	\\
		12$\star$ 	&	H2			& 	$n_{\mathrm{H2}}$	&  H2cX				& H2 cooling from: collisions, \\
		 		&				&					&  H2ln, H2+			& H2 lines, photo continuum cooling\\
		13		&	molecules 	& 	Z				& molecule 			& cooling from all molecules other than H2 and HD \\
		14$\star$ 	&	HD			& 	$n_{\mathrm{HD}}$	& HD				& HD rotation cooling (see Sec.~\ref{sec:HD} for details) \\
		15$\star$ 	&	NetFFH		& 					& FFcm\_H			& net free-free brems cooling from H and He ions\\
		16	 	&	NetFFM		& 	Z				& FFcm\_M			& net free-free brems cooling from metal ions\\
		17$\star$ 	&	eeBrems		& 	$n_{\mathrm{e}}^2$	& eeff				& electron - electron bremsstrahlung\\
		18$\star$ 	& 	Compton		& 					& Compton			& Compton cooling 	\\
		19		&	Dust			& 	Z				& dust				& dust cooling					\\
		\hline
		20$\star$ 	&	TotalPrim		& 	prim				&					& 0, 1, 12, 14, 15, 17, 18\\
		21		&	TotalMetal		& 	metal			&					& 2, 3, 4, 5, 6, 7, 8, 9, 10, 11, 13, 16, 19\\
		\hline
	\end{tabular}\label{tab:cool}
\end{table*}

\begin{table*} 
	\caption{As Table~\ref{tab:cool} but for the individual heating contributions.}
	\begin{tabular}{lllll}
			ID Heat		&	Identifier		& Scaling 				& \textsc{cloudy} label			&	Comment \\
		\hline
	0$\star$ 	&	Hydrogen		& 	$n_{\mathrm{H}}$	& H, H-,  				& H$^{-}$ heating, \\ 
			&				&					& Hn=2				& Hn=2: net hydrogen photoionization heating of all excited states\\
	1$\star$ 	&	Helium		& 	$n_{\mathrm{He}}$	& He, He3l			& He3l: helium triplet line heating\\ 
	2 		&	Carbon		& 	$n_{\mathrm{C}}$	& C					& 	\\
	3 		&	Nitrogen		& 	$n_{\mathrm{N}}$	& N					&  	\\
	4 		&	Oxygen		& 	$n_{\mathrm{O}}$	& O					&	\\
	5 		&	Neon		& 	$n_{\mathrm{Ne}}$	& Ne					&	\\
	6 		&	Magnesium	& 	$n_{\mathrm{Mg}}$	& Mg					&	\\
	7 		&	Silicon		& 	$n_{\mathrm{Si}}$	& Si					&	\\
	8 		&	Sulphur		& 	$n_{\mathrm{S}}$	& S					&	\\
	9 		&	Calcium		& 	$n_{\mathrm{Ca}}$	& Ca					&	\\
	10 		&	Iron			& 	$n_{\mathrm{Fe}}$	& Fe, Fe 2				& Fe 2 line heating \\
	11 		&	OtherA		& 	Z				&  Li, Be, B, F, Na,  Al, 	& all other atoms\\
			&				&					&  P, Cl, Ar, K, Sc, Ti, V, 	&  	\\
			&				&					&  Cr, Mn, Co, Ni, Cu, Zn	&	\\
	12$\star$ 	&	H2			&	$n_{\mathrm{H2}}$	&  H2vH				& H2 heating from: collisions,	\\
			&				&					&  H2dH, H2+			& photodissociation, H2+ heating\\
			&				&					&  H2ph				& photoionization heating	\\
	13		&	COdiss	 	&	$n_{\mathrm{CO}}$	& COds				& CO dissociation heating\\
	14$\star$ 	&	CosmicRay	& 	CR rate			& CR H				& cosmic ray heating 	\\
	15		&	UTA			& 	Z				& UTA				& unresolved transition array (UTA) heating\\
	16		&	line			& 	Z				& line				& heating due to induced line absorption of continuum	\\
	17$\star$ 	&	Hlin			& 					& Hlin				& iso-sequence line heating\\
	18		&	ChaT		& 	Z				& ChaT				& heating due to charge transfer: \\
			&				&					& 					&- ionization of heavy element, recombination of hydrogen\\
			&				&					&					&- recombination of heavy element, ionization of hydrogen)\\
	19$\star$ 	&	HFF			& 					&  H FF				& free-free heating (if Bremsstrahlung has a net heating effect)\\
	20$\star$ 	& 	Compton		& 					& Compton			& Compton heating	\\
	21		&	Dust			& 	Z				& GrnP, GrnC			& grain photoionization, grain collisions				\\
	\hline
	22$\star$ 	&	TotalPrim		& 	prim				&					& 0, 1, 12, 14, 17, 19, 20 \\
	23		&	TotalMetal		& 	Z				&					& 2, 3, 4, 5, 6, 7, 8, 9, 10, 11, 13, 15, 16, 17, 18, 21\\
		\hline
	\end{tabular}\label{tab:heat}
\end{table*}

\section*{Acknowledgements}
We gratefully acknowledge helpful discussions with Gargi Shaw and Gary Ferland about the H$_{\mathrm{2}}$ model used in \textsc{cloudy}, Claude-Andr\'{e} Faucher-Gigu\`{e}re about the details of his UVB model, Andr\'{e}s Felipe Ramos Padilla about the line emissivities and members of the COLIBRE team about the application of this work in hydrodynamical simulations. Within the COLIBRE team we would like to highlight the contributions of Alejandro Benitez-Llambay, who pointed out the lack of HD cooling in Cloudy v17.01 and Alexander Richings, who provided in-depth knowledge about individual chemical processes as well as comparisons with CHIMES, his non-equilibrium network, which were crucial in identifying issues in the earlier stages of this work. SP was in part supported by European Research Council (ERC) Advanced Investigator grant DMIDAS (GA 786910, PI C. S. Frenk).\\
This work used the DiRAC@Durham facility managed by the Institute for Computational Cosmology on behalf of the STFC DiRAC HPC Facility (www.dirac.ac.uk). The equipment was funded by BEIS capital funding via STFC capital grants ST/K00042X/1, ST/P002293/1 and ST/R002371/1, Durham University and STFC operations grant ST/R000832/1. DiRAC is part of the National e-Infrastructure.

\section*{Data availability}
The data underlying this article are available in the Harvard Dataverse, at \url{https://dataverse.harvard.edu/dataverse/radcool} (rates and fractions data files: \url{https://doi.org/10.7910/DVN/GR3L5N}, line emissivities data files: \url{https://doi.org/10.7910/DVN/CRJ7GT}). More information as well as the routines to use and visualise the tables can be found at \url{http://radcool.strw.leidenuniv.nl/} and \url{https://www.sylviaploeckinger.com/radcool}.



\bibliographystyle{mnras}
\bibliography{CoolingTables} 


\appendix

\section{How to use the tables}\label{sec:howto}

The tables released with this work cover a large range of redshifts, temperatures, metallicities and densities. For any dataset listed in Table~\ref{tab:alldatasets}, its value can be obtained at any point in this multi-dimensional space by a 4D interpolation. Due to the relatively fine spacing, the exact interpolation scheme or whether different weights are applied for different dimensions, is unimportant. We recommend a linear interpolation in log and weighting the different dimensions equally, but other interpolation schemes are possible. An interpolation in log space can lead to species fractions not summing up to 1, even if the fractions at the individual grid points do. In this case, the fractions can be re-normalised to 1 after the interpolation. 

For solar abundance ratios the total cooling rate $\Lambda_{\mathrm{cool, total}}$ is

\begin{align}
	\Lambda_{\mathrm{cool,total}} (z, T, Z, \nH)  =& \Lambda_{\mathrm{cool,TotalPrim}} (z, T, Z, \nH)  \\ \nonumber
									  + & \Lambda_{\mathrm{cool,TotalMetal}} (z, T, Z, \nH)
\end{align}

\noindent
but in the more general case, where the abundance ratios differ from the solar values, the total cooling rate is calculated as

\begin{align} 
	\Lambda_{\mathrm{cool,total}} (z, T, Z, \nH)  = &\sum_{i} \frac{ (n_i/\nH)}{(n_i/\nH)_{\mathrm{table}}} \Lambda_{\mathrm{cool,i}} (z, T, Z, \nH)  \\ \nonumber
									& +  \Lambda_{\mathrm{cool,OtherA}} (z, T, Z, \nH) \\ \nonumber
									& +  \Lambda_{\mathrm{cool,H2}} (z, T, Z, \nH) \\ \nonumber
									& +  \Lambda_{\mathrm{cool,molecules}} (z, T, Z, \nH) \\  \nonumber
									& +  \Lambda_{\mathrm{cool,HD}} (z, T, Z, \nH) \\  \nonumber
									& +  \Lambda_{\mathrm{cool,NetFFH}} (z, T, Z, \nH) \\  \nonumber
									& +  \Lambda_{\mathrm{cool,NetFFM}} (z, T, Z, \nH) \\ \nonumber					
									& +  \Lambda_{\mathrm{cool,eeBrems}} (z, T, Z, \nH) \\  \nonumber
									& +  \Lambda_{\mathrm{cool,Compton}} (z, T, Z, \nH) \\  \nonumber
									& +  \Lambda_{\mathrm{cool,Dust}} (z, T, Z, \nH) 
\end{align}

\noindent where $i$ loops over all individually traced elements (see Table~\ref{tab:solarabundances}). Analogously, the total heating rate $\Lambda_{\mathrm{heat,total}}$ can be calculated by summing up either only the components $\Lambda_{\mathrm{heat,TotalPrim}}$ and $\Lambda_{\mathrm{heat,TotalMetal}}$ (for solar abundances) or the full list of heating channels (Table~\ref{tab:heat}) for general abundance ratios.

In Sec.~\ref{sec:extrapolation} we discuss how the cooling and heating rates can also be calculated for values outside the table ranges. In addition, Sec.~\ref{sec:noneq} demonstrates how the tables can be combined with a reduced non-equilibrium network. 

\subsection{Extrapolation outside the table boundaries}\label{sec:extrapolation}
\paragraph*{Redshift}

The redshift dimension in the tables goes from $z_{\mathrm{min}}=0$ to $z_{\mathrm{max}}=9$ .
For redshifts larger than $z_{\mathrm{max}}$, the Compton cooling should not be used from the tables as it is highly redshift dependent. 
Instead, the change in cooling rate due to inverse Compton cooling off the CMB can be calculated analytically with \citep[e.g.][]{mo_galaxy_2010}:

\begin{eqnarray}
	\Lambda_{\mathrm{cool,Compton}} / \nH^2 =& \frac{4 k_{\mathrm{B}} }{m_{\mathrm{e}} c}\, \sigma_{\mathrm{T}}\, \frac{n_{\mathrm{e}}}{\nH^2} \,a \,\left [ T_{\mathrm{CMB,0}} (1+z) \right ]^4 \cdot\\
	& [T - T_{\mathrm{CMB,0}} (1+z)] \quad  \mathrm{erg\,cm^3\,s^{-1}}
\end{eqnarray}

\noindent
with the Boltzmann constant $k_{\mathrm{B}} = 1.38066 \times 10^{-16}\, \mathrm{cm}^{2} \mathrm{g} \,\mathrm{s}^{-2} \K^{-1}$, the electron mass $m_{\mathrm{e}}=9.11 \times 10^{-28}\,\mathrm{g}$, the speed of light $c = 2.998\times10^{10}\,\mathrm{cm}\,\mathrm{s}^{-1}$, the Thomson cross-section $\sigma_{\mathrm{T}} = 6.65 \times 10^{-25}\,\mathrm{cm}^2$, the radiation constant $a = 7.57\times10^{-15}\,\mathrm{erg\,cm^{-3}\,K^4}$, the present day CMB temperature $T_{\mathrm{CMB,0}} = 2.73 \K$, the electron number density $n_{\mathrm{e}}$, the hydrogen number density $\nH$, the gas temperature $T$, and the current redshift $z$. The electron density $n_{\mathrm{e}}$ can be taken from the redshift bin $z = z_{\mathrm{max}}$ for $z > z_{\mathrm{max}}$.

\paragraph*{Temperature}
We impose a minimum temperature of $10\,\K$ and advise to impose a similar temperature floor in all simulations that make use of this tables.
In case the temperature exceeds the maximum temperature $T_{\mathrm{max}} = 10^{9.5}\,\K$, the contributions from ``eeBrems", ``NetFFH", and ``Compton'' cooling processes can be linearly extrapolated using the slope between $\log T[\K] = 9$ and $\log T[\K] = 9.5$ for $\log \Lambda_{\mathrm{cool,i}}(T)$ with $i$ is ``eeBrems", ``NetFFH", and ``Compton''.  The sum of these three extrapolations is the total cooling rate. All other cooling contributions, as well as all heating processes that we compute are negligible for $T>T_{\mathrm{max}}$.

\paragraph*{Metallicity}
For all heating and cooling processes $i$ that scale with either an element abundance or with metallicity (as indicated in Tables~\ref{tab:cool} and \ref{tab:heat}), the cooling (heating) rate for metallicities above the maximum metallicity of $Z_{\mathrm{max}} = 10^{0.5} \Zsol$ can be approximated by

\begin{eqnarray}
	\Lambda_{\mathrm{cool,i}} = \frac{Z} {Z_{\mathrm{max}}} \Lambda_{\mathrm{cool, i, Zmax}} \\
	\Lambda_{\mathrm{heat,i}} = \frac{Z} {Z_{\mathrm{max}}} \Lambda_{\mathrm{heat, i, Zmax}}	
\end{eqnarray}

\noindent
for metallicity scaling and 

\begin{eqnarray}
	\Lambda_{\mathrm {cool,i}} = \frac{n_{\mathrm {x}} / \nH} {(n_{\mathrm {x}} / \nH)_{\mathrm{max}}} \Lambda_{\mathrm{cool,i,Zmax}} \\
	\Lambda_{\mathrm {heat,i}} = \frac{n_{\mathrm {x}} / \nH} {(n_{\mathrm {x}} / \nH)_{\mathrm{max}}} \Lambda_{\mathrm{heat,i,Zmax}}
\end{eqnarray}

\noindent 
for element abundance scaling. Here $\Lambda_{\mathrm{cool, i, Zmax}}$ and $\Lambda_{\mathrm{heat, i, Zmax}}$ are the cooling and heating rates for the highest tabulated metallicity bin for the considered redshift, density, and temperature. $(n_{\mathrm {x}} / \nH)_{\mathrm{max}}$ is the maximum abundance of element $x$.

Cooling or heating channels labelled with $\star$ (primordial) in Tables~\ref{tab:cool} and \ref{tab:heat} do not scale with metallicity and should therefore be approximated by the maximum metallicity table entry:

\begin{eqnarray}
	\Lambda_{\mathrm{cool,i}} =  \Lambda_{\mathrm{cool,i, Zmax}} \\
	\Lambda_{\mathrm{heat,i}} =  \Lambda_{\mathrm{heat, i, Zmax}}	
\end{eqnarray}

\noindent
neglecting the shielding from additional metals.

The metallicity bins are spaced logarithmically with an additional metallicity bin for primordial abundances and therefore $Z = 0$. 
The primordial abundances can be used for $Z<Z_{\mathrm{min}}$, where $Z<Z_{\mathrm{min}}$ is the minimum (for $Z\ne0$) metallicity in the tables ($\log (Z/\Zsol)_{\mathrm{min}} = -4$).

\paragraph*{Density}
At high densities, where collisional ionization dominates over photo-ionization, most cooling rates scale as $\nH n_{\mathrm{e}}$ while most heating rates scale with $\nH$. For convenience, cooling and heating rates are both stored as $\left [ \Lambda_{\mathrm{cool}} / \nH^2 \right ]$ and $\left [ \Lambda_{\mathrm{heat}} / \nH^2 \right ]$, respectively. Assuming a constant free electron fraction for $\nH \ge n_{\mathrm{H,max}}$, the cooling and heating rates can be extrapolated as

\begin{equation}
	\Lambda_{\mathrm{cool}} (\nH > n_{\mathrm{H,max}}) = \left [ \frac {\Lambda_{\mathrm{cool}} }{n_{\mathrm{H,max}}^2} \right ] \, \nH^2 
\end{equation}

\noindent
and 

\begin{equation}
	\Lambda_{\mathrm{heat}} (\nH > n_{\mathrm{H,max}}) = \left [ \frac {\Lambda_{\mathrm{heat}} }{n_{\mathrm{H,max}}^2} \right ]\, \nH \, n_{\mathrm{H,max}} \; .
\end{equation}

\subsection{Coupling with a reduced non-equilibrium network} \label{sec:noneq} 

If the timescales for ionization and recombination are long compared to the dynamical timescale or the cooling / heating timescale, local ionization equilibrium is not a good assumption. 
As the number of species and reactions increase steeply with the number of elements, it is computationally very expensive to run a non-equilibrium network for a large number of elements. 

The smallest networks only calculate hydrogen and helium in non-equilibrium and add metal cooling from look-up tables. Other networks include more species and reactions to also 
trace the formation of CO. Independently of the elements included in a network, it can easily be combined with the tables presented in this work. 

\paragraph*{Cooling / heating rates:} 
The cooling and heating rates are tabulated according to the individual processes, as listed in Tables~\ref{tab:cool} and \ref{tab:heat}. If the (net) cooling rates for e.g. hydrogen and helium are calculated 
from the network, then any remaining cooling channels that are not included in the network can be obtained from the tables and added. 

\paragraph*{Electron densities:} 
The total number density of free electrons can be obtained in a similar way as explained above for the total cooling / heating rates. The dataset \texttt{ElectronFractions} contains the fraction of free electrons per hydrogen nucleus ($n_{\mathrm{e}}/\nH$) split into the contributions from each element. This allows one to individually add the electron density from each element that is not included in the network. 

Cooling processes typically include an interaction between an electron and another species, for example a carbon atom (i.e. collisional excitation, collisional ionization, recombination cooling, free-free emission), and scale therefore generally as $\Lambda_{\mathrm{cool}} \propto n_{\mathrm{e}} \nH$ (if each species number density scales with the hydrogen number density). As we tabulate the electron densities individually for each element, a non-equilibrium network can scale the metal cooling rates by the combined free electron density $n_{\mathrm{e,comb}} = (n_{\mathrm{e,H}} + n_{\mathrm{e,He}})_{\mathrm{neq}} + (n_{\mathrm{e}})_{\mathrm{eq}}$. In this example, the free electrons from hydrogen ($n_{\mathrm{e,H}}$) and helium ($n_{\mathrm{e,He}}$) are taken from the non-equilibrium (``neq") calculations while the free electrons from metals ($n_{\mathrm{e}}$) are added from the equilibrium (``eq") tables from this work. 

If the metal cooling rates are scaled as \mbox{$\Lambda_{\mathrm{comb}}  = \Lambda_{\mathrm{table}} (n_{\mathrm{e,comb}} / n_{\mathrm{e,table}})$} the cooling rates $\Lambda_{\mathrm{comb}}$, while calculated assuming ionization equilibrium (as in $ \Lambda_{\mathrm{table}}$), capture many of the non-equilibrium effects for metal cooling. For the heating rates this additional scaling is not necessary as $\Lambda_{\mathrm{heat}}$ scales with $\nH$.

\section{Including the effective FG20 background radiation field}\label{sec:fg20eff}

Before and during reionization, the photo-ionization rates calculated from the redshift-dependent UV / X-ray spectra from \citetalias{faucher-giguere_cosmic_2020} differ from their tabulated ``effective"~photo-ionization rates that are modelled to match the observed electron scattering optical depth of $\tau_e = 0.054$ \citep{planck_collaboration_planck_2018}. As \textsc{cloudy} uses a full spectrum as input radiation field, we modify the spectra from \citetalias{faucher-giguere_cosmic_2020} to match their ``effective"~photo-ionization rates. Assuming that before \ion{H}{I} reionization, the \ion{H}{I} ionizing radiation is attenuated by neutral hydrogen gas with a column density $N_{\ion{H}{I}}$ (and before \ion{He}{II} reionization, the \ion{He}{II} ionizing radiation is attenuated by singly ionized helium gas with column density $N_{\ion{He}{II}}$), the new spectrum $J_{\nu}^{\mathrm{modFG20}}$ can be derived from the original \citetalias{faucher-giguere_cosmic_2020} spectrum $J_{\nu}^{\mathrm{FG20}}$ as

\begin{equation}
    J_{\nu}^{\mathrm{modFG20}} = J_{\nu}^{\mathrm{FG20}}\, e^{-\sigma_{\ion{H}{I}} (\nu) N_{\ion{H}{I}}} \, e^{-\sigma_{\ion{He}{II}} (\nu) N_{\ion{He}{II}}}
\end{equation}

The photo-ionization cross sections $\sigma_{\ion{H}{I}}$ and $\sigma_{\ion{He}{II}}$ are taken from \citet{verner_atomic_1996} and the column densities $N_{\ion{H}{I}}$ and $N_{\ion{He}{II}}$ are free parameters that can be chosen so that the photo-ionization rates calculated from $J_{\nu}^{\mathrm{modFG20}}$ are equal to the effective rates from \citetalias{faucher-giguere_cosmic_2020}. The effective \ion{He}{II} photo-ionization from \citetalias{faucher-giguere_cosmic_2020} decreases very steeply for $z>4$ which would lead to arbitrarily high column densities. We therefore use a maximum shielding column of $N_{\ion{H}{I}\mathrm{,max}} = 10^{20}\,\mathrm{cm}^{-2}$ and $N_{\ion{He}{II}\mathrm{,max}} = (n_{\mathrm{He}}/n_{\mathrm{H}})_{\mathrm{prim}} \times N_{\ion{H}{I}\mathrm{,max}}$, where $(n_{\mathrm{He}}/n_{\mathrm{H}})_{\mathrm{prim}} = 0.08246$ is the primoridal helium abundance, yielding a non-zero photo-ionization rate of \ion{He}{II} between $z=4$ and $z=8$ (see Fig.~\ref{fig:SP20rates}). 

\begin{figure} 
	\begin{center}
		\includegraphics*[width=\linewidth, trim = 0.2cm 0cm 1.5cm 1.cm,clip]{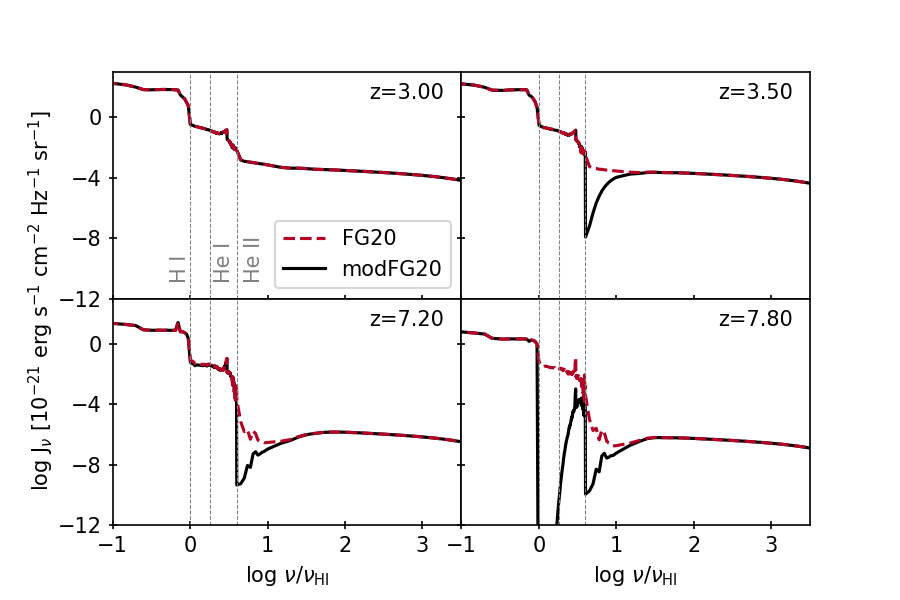}
		\caption{Background radiation field from \citetalias{faucher-giguere_cosmic_2020} (``FG20") and the modified \citetalias{faucher-giguere_cosmic_2020} radiation field used in this work (``modFG20") for selected redshifts after (left) and during (right) \ion{He}{II} (top) and \ion{H}{I} (bottom) reionization. The vertical lines indicate the \ion{H}{I},  \ion{He}{I} and \ion{He}{II} minimum ionization energies. }
		\label{fig:SP20spectrum}
	\end{center}
\end{figure}

\begin{figure} 
	\begin{center}
		\includegraphics*[width=\linewidth, trim = 0.5cm 0cm 1.2cm 0.5cm,clip]{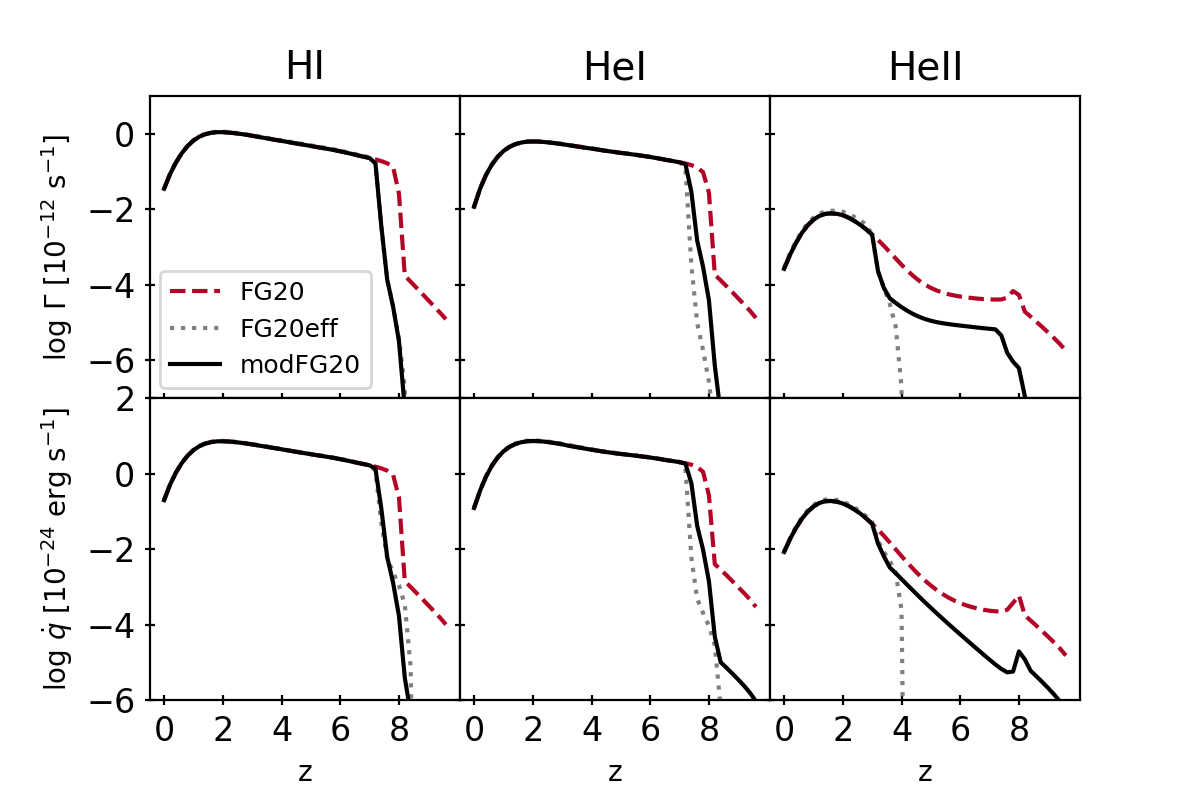}	
		\caption{Photo-ionization (top) and photo-heating (bottom) rates for \ion{H}{I} (left panels), \ion{He}{I} (middle panels), and \ion{He}{II} (right panels) calculated from the spectrum provided from \citetalias{faucher-giguere_cosmic_2020} (``FG20"), the effective rates from  \citetalias{faucher-giguere_cosmic_2020} (``FG20eff"), and calculated from the modified spectrum used in this work (``modFG20", see text for details). }
		\label{fig:SP20rates}
	\end{center}
\end{figure}

Fig.~\ref{fig:SP20spectrum} displays $J_{\nu}^{\mathrm{modFG20}}$ (black solid lines) used in this work and  $J_{\nu}^{\mathrm{FG20}}$ (red dashed lines) for four different redshifts. 
The top panels show the spectrum after (left, $z=3.0$) and during (right, $z=3.5$) \ion{He}{II} reionization and the bottom panels illustrate the UVB spectrum after (left, $z=7.2$) and during (right, $z=7.8$) \ion{H}{I} reionization. For $z\le3$ ($z_{\mathrm{rei},\ion{He}{II}} - \Delta z_{\mathrm{rei},\ion{He}{II}}=3$), both spectra are identical, as the \citetalias{faucher-giguere_cosmic_2020} effective rates match the rates from the spectra in \citetalias{faucher-giguere_cosmic_2020} and no additional attenuation is necessary. 

The photo-ionization and photo-heating rates calculated from $J_{\nu}^{\mathrm{modFG20}}$ are shown in Fig.~\ref{fig:SP20rates}. By design, the photo-ionization rate of \ion{H}{I} and \ion{He}{II} from ``modFG20"~(black solid lines) match the effective rates from FG20 (``FG20eff", grey dotted lines) better than is the case for the original FG20 spectrum (``FG20", red dashed lines). Additional attenuation for \ion{He}{I} is not necessary to obtain a good match in both the photo-ionization and photo-heating rates, as for energies close to the minimum ionization energy of \ion{He}{I}, the spectrum is already attenuated by the \ion{H}{I} shielding (see Fig.~\ref{fig:SP20spectrum}).

\section{\textsc{cloudy} commands for reproducibility}\label{sec:reprod}

\textsc{cloudy} was slightly modified to efficiently run a large grid of individual simulations with various scalings. This only involves additional command line keywords\footnote{E.g. a new keyword on the line \texttt{table ISM} normalizes the ISRF as in Eq.~\ref{eq:radG0}.} while the core calculations remain untouched. We explain in the following how standard \textsc{cloudy} v17.01 can be used to reproduce our results for individual grid points. 
Table~\ref{tab:c17} shows an example of a \textsc{cloudy} run with all included processes switched on (i.e. table UVB\_dust1\_CR1\_G1\_shield1). Not all command lines are present in each run (e.g. ``table ISM"\, is only included for tables with the ISRF). 

\noindent
\begin{table} 
	\caption{\textsc{cloudy} commands for reproducing an individual grid point at redshift $z$, temperature $\log T$, metallicity $\log Z/\Zsol$ and density $\log\nH$ from one of the tables. Column 2 lists the respective values for the arguments \texttt{[value]} on some command lines, e.g for the line \texttt{stop column}, the value from dataset \texttt{ShieldingColumn} for the selected grid points would be used for \texttt{[value]}. }
	\begin{tabular}{ll}
		\textsc{cloudy} v17.01 command									&	value \\
	\hline
		\texttt{constant density}											&		\\
		\texttt{constant temperature, t = [value] K log} 							&	$\log T$	\\
		\texttt{hden [value]} 												&	$\log \nH$	\\
		\texttt{stop column [value] log} 										&	ShieldingColumn	\\
		\texttt{abundances "solar\_GASS10.abn"\,no grains} 						&		\\
		\texttt{(or for log Z/Z$_{\odot}$ = -50:  }								&		\\
		\texttt{\quad abundances "primordial.abn")}									&		\\
		\texttt{element abundance helium [value] }								&	AbundanceHe\\
		\texttt{metals deplete }											&		\\
		\texttt{grains orion PAH [value] log no qheat } 							& 	DGratio - $\log (\mathrm{D/G})_{\mathrm{0}}$ 	\\
		\texttt{metals [value] log }											&	$\log Z/\Zsol$\\
		\texttt{CMB redshift [value] }										&	$z$	\\
		\texttt{Table HM12 redshift [value] }									&	$z$	\\
		\texttt{table ISM [value] }											&	RadField	\\
		\texttt{cosmic rays background [value] log }							&	CosmicRayRate - $\log \zeta_{\mathrm{0}}$\\
		\texttt{iterate to convergence }										&		\\
		\texttt{stop temperature off} 										&		\\
		\texttt{Database H2 }												&		\\
	\hline
	\end{tabular}\label{tab:c17}
\end{table}

There is no direct analogue in \textsc{cloudy} v17.01 to scale the metal depletion, as is done in this work. But the individual depletion factors can be set manually at the bottom of file \texttt{abund.cpp} in the \textsc{cloudy} source code  to $f_{\mathrm{gas, i}} = 1 - f_{\mathrm{dust,i}}$, with $f_{\mathrm{dust,i}}$ from the dataset \texttt{Depletion}.

For the UVB used here (modified  \citetalias{faucher-giguere_cosmic_2020}), replace the \texttt{hm12\_galaxy.ascii} file in \textsc{cloudy}'s data folder with the provided \texttt{modFG20\_cloudy.ascii} file, but keep the original filename (\texttt{hm12\_galaxy.ascii}).
As explained in Sec.~\ref{sec:CR} the dissociation rate of H$_{\mathrm{2}}$ by CRs is re-normalized to better match the UMIST values. This has been done by changing one line in the \textsc{cloudy} file \texttt{mole\_reactions.cpp} from
\mbox{\texttt{newreact("H2,CRPHOT=>H,H","h2crphh",1.,0.,0.)}} to \mbox{\texttt{newreact("H2,CRPHOT=>H,H","h2crphh",1.746e-2,0.,0.)}}.


\bsp	
\label{lastpage}
\end{document}